\documentclass[11pt]{article}
\usepackage[utf8]{inputenc}
\usepackage{amsmath}
\usepackage[margin = 1in]{geometry}
\usepackage{natbib}
\usepackage{amsthm}
\usepackage{amsfonts}
\usepackage{import}
\usepackage{url} 
\usepackage{setspace}
\usepackage{enumitem}
\usepackage{marginnote}
\usepackage{tikz}
\usetikzlibrary{positioning, arrows.meta, decorations.pathreplacing}

\usepackage[backref=page]{hyperref} 
\hypersetup{	colorlinks=true,	linkcolor=blue,	filecolor=magenta,	urlcolor=blue,	allcolors=blue,}
\usepackage{algorithm,algorithmicx,algpseudocode}

\usepackage{titlesec}
\titlespacing{\section}{0pt}{1ex}{0ex}
\titlespacing{\subsection}{0pt}{1ex}{0ex}
\titlespacing{\subsubsection}{0pt}{0.5ex}{0ex}

\usepackage{subcaption}

\usepackage{amsthm}
\usepackage{amsfonts}

\usepackage{bm}

\newcommand{\cm}[1]{\ignorespaces}

\usepackage{booktabs,multirow,tabularx}
\usepackage{graphics}
\usepackage{graphicx}

\usepackage{xcolor}
\definecolor{mypink}{RGB}{219, 48, 122}

\definecolor{mypurple}{RGB}{75,0,130}

	\newtheorem{theorem}{Theorem}
	\newtheorem{lemma}{Lemma} 
	\newtheorem{proposition}{Proposition} 
	\newtheorem{definition}{Definition}

	\newtheorem{corollary}{Corollary}
	\newtheorem{assumption}{Assumption}

\newcommand{\pscale}{p}
\newcommand{\pfactor}{p}
\newcommand{\Bsupport}{\mathcal{S}}

\newcommand{\bR}{\mathbb{R}}
\newcommand{\cS}{\mathcal{S}}
\newcommand{\cC}{\mathcal{C}}
\newcommand{\cU}{\mathcal{U}}
\newcommand{\cE}{\mathcal{E}}

\newcommand{\bE}{\mathbb{E}}
\newcommand{\bP}{\mathbb{P}}

\newcommand{\sgn}{\mathrm{sgn}}

\newcommand{\Var}{\mathrm{Var}}

\newcommand{\br}[1]{\left\{ #1 \right\} }
\newcommand{\mbr}[1]{\left[ #1 \right] }
\newcommand{\sbr}[1]{\left( #1 \right) }

\newcommand{\nbr}[1]{\left\| #1 \right\|}

\newcommand\numberthis{\addtocounter{equation}{1}\tag{\theequation}}

\def\logit{\text{logit}}
\def\nie{\text{NIE}}

\def\nde{\text{NDE}}

\def\dd{\text{d}}

\def\vox{\Delta s} 
\def\cvox{{s}} 
\def\leb{\lambda} 
\def\sthresh{\nu} 
\def\mpdf{M} 
\def\mcdf{\mathcal{M}} 
\def\bfmcdf{\bm{\mathcal{M}}} 
\def\normal{\mN} 
\def\bfa{\mathbf a}
\def\bfb{\mathbf b}
\def\bfc{\mathbf c}

\def\bfs{\mathbf s}

\def\bfg{\mathbf g}

\def\bfA{\mathbf A}

\def\bfC{\mathbf C}

\def\bfI{\mathbf I}
\def\bfX{\mathbf X}
\def\bfY{\mathbf Y}

\def\bfD{\mathbf D}

\def\bfM{\mathbf M}

\def\bfV{\mathbf V}
\def\bfU{\mathbf U}
\def\bfW{\mathbf W}

\def\bfm{\mathbf m}

\def\bfalpha{\boldsymbol \alpha}

\def\bftheta{\boldsymbol \theta}

\def\bfdelta{\boldsymbol \delta} 
\def\bfgamma{\boldsymbol \gamma}

\def\bfmu{\boldsymbol\mu}
\def\bfSigma{\boldsymbol\Sigma}

\def\bfpi{\boldsymbol\pi}
\def\bfzeta{\boldsymbol\zeta}
\def\bfxi{\boldsymbol\xi}
\def\bfmu{\boldsymbol\mu}
\def\bfzero{\boldsymbol 0}

\def\cA{\mathcal A}
\def\cB{\mathcal B}
\def\cN{\mathcal N}
\def\cS{\mathcal S}

\def\cP{\mathcal P}

\def\cR{\mathcal{R}}

\def\cGP{\cG\cP}
\def\cF{\mathcal F}

\def\mbR{\mathbb R}

\def\mN{\mathrm{N}}

\def\iid{\scriptsize \mbox{iid}}
\def\ind{\scriptsize \mbox{ind}}

\def\cGP{\mathcal{GP}}
\def\cSTGP{\mathcal{STGP}}

\def\bfTheta{{\ensuremath\boldsymbol{\Theta}}}
\def\bfrho{\boldsymbol{\rho}}

\def\diag{\mathrm{diag}}

\def\cA{\mathcal{A}}
\def\cGP{\mathcal{GP}}

\def\rT{\mathrm T}
\def\R{\mathbb R}

\newif\ifhidetext
\hidetextfalse 

\newcommand{\hidetext}[1]{%
  \ifhidetext
  \else
    #1
  \fi
}

\begin{document}

\def\spacingset#1{\renewcommand{\baselinestretch}%
{#1}\small\normalsize} \spacingset{1}


\title{\bf Bayesian Image Mediation Analysis}
 \author{Yuliang Xu $^1$\thanks{Most of this work was completed during my PhD training in the Department of Biostatistics at the University of Michigan.}, Timothy D Johnson$^2$,  Mary Heitzeg$^3$ and Jian Kang$^2$\thanks{To whom correspondence should be addressed: Jian Kang (jiankang@umich.edu)} \\[4mm]
    Department of Statistics, University of Chicago$^1$\\
    Department of Biostatistics, University of Michigan$^2$\\
    Department of Psychiatry, University of Michigan$^3$} 

    \date{}
  \maketitle

\bigskip
\begin{abstract}
Mediation analysis aims to separate the indirect effect through mediators from the direct effect of the exposure on the outcome. It is challenging to perform mediation analysis with neuroimaging data which involves high dimensionality, complex spatial correlations, sparse activation patterns and relatively low signal-to-noise ratio.  To address these issues, we develop a new spatially varying coefficient structural equation model for Bayesian Image Mediation Analysis (BIMA). We define spatially varying mediation effects within the potential outcomes framework, employing a soft-thresholded Gaussian process prior for functional parameters. We establish posterior consistency for spatially varying mediation effects along with selection consistency on important  regions that contribute to the mediation effects. We develop an efficient posterior computation algorithm scalable to analysis of large-scale imaging data. Through extensive simulations, we show that BIMA can improve the estimation accuracy and computational efficiency for high-dimensional mediation analysis over existing methods. We apply BIMA to analyze behavioral and fMRI data in the Adolescent Brain Cognitive Development (ABCD) study with a focus on inferring the mediation effects of the parental education level on the children's general cognitive ability that are mediated through the working memory brain activity. 
\end{abstract}

\noindent%
{\it Keywords:} spatially varying mediation effects; Soft thresholded Gaussian process; Posterior consistency.
\vfill

\newpage

\spacingset{1.2} 
\section{Introduction}
\label{sec:intro}

Mediation analysis is an important statistical tool that decomposes the total effects of an exposure or treatment variable on an outcome variable into direct effects and indirect effects through mediator variables~\citep{mackinnon2012introduction}. Mediation analysis has been widely adopted to gain insights into mechanisms of exposure-outcome effects in many research areas including epidemiology, environmental science, genomics, and neuroimaging. Recent advances in neuroimaging have presented great opportunities and challenges for mediation analysis with large-scale complex neuroimaging data. In many neuroimaging studies, it is of great interest to identify important brain image mediators that mediate the effect of an exposure variable, such as age, social economic status, medical treatment, or substance use, to an outcome variable, such as cognitive status or disease status. 

Recent studies \citep{cermakova2023parental,halabicky2023low} have demonstrated that parental education levels are significantly associated with children's cognitive abilities, including both general cognitive ability and specific cognitive functions such as working memory.  In particular, parental education has been shown to influence neural development pathways that contribute to cognitive functions, which are captured by fMRI during tasks like the working memory task.

Our work is motivated by the brain image mediation analysis in the Adolescent  Brain  Cognitive  Development  (ABCD) study, the largest long-term study of brain development and child health in the United States. Our objective is to investigate  how  parental education levels impact a child's general cognitive ability that is mediated through brain function development measured by working memory task fMRI.  We use general cognitive ability as the outcome because it provides a comprehensive assessment of a child's overall cognitive function, which encompasses not only working memory but also other critical skills such as reading, spelling, and math abilities \citep{alloway2008working}. These abilities are often correlated and influenced by common neural processes, making general cognitive ability a robust and representative measure for examining broader cognitive development. By using this summary outcome, we can account for the cumulative effect of parental education on multiple facets of cognitive performance, providing a more holistic view of the relationship between socioeconomic factors and brain development.

We consider voxel-level task fMRI contrast maps as the image mediators which pose several challenges for mediation analysis. First, the number of voxel-level image mediators can be up to 200,000 in a standard brain template, potentially requiring large computational resources for implementing the statistical algorithm. Second, brain image mediators exhibit complex correlation patterns such as the correlations among neighboring voxels and the correlation between brain regions with complementary functions. Ignoring or inappropriately accounting for the correlation may introduce bias or loss of statistical efficiency in estimating the mediation effects.   Third, due to the low signal-to-noise ratio of brain imaging data, the voxel-level image mediators may have weak or zero effects on the outcome variable. The standard mediation analysis approach may suffer from low power and high false positive rates when detecting active mediators.

Recent work on high-dimensional mediation analysis provides different angles to tackle these challenges, with different statistical models tailored to specific application domains, such as penalized high-dimensional survival analysis \citep{luo2020high} or  DNA methylation markers \citep{zhang2016estimating, guo2022high}. 
For imaging applications, \cite{lindquist2012functional} first extended the mediation analysis framework into functional data analysis and proposed a model based on least-squares estimation and penalized regression, without considering correlation among individual-level noise in the mediators.  Built upon this work, \cite{chen2018high} proposed a method based on principle component analysis, where high-dimensional correlated mediators are mapped to uncorrelated mediators through orthogonal transformations. The orthogonal maps are sequentially estimated from maximizing the likelihood of the joint model on each direction of the mediators separately. However, interpreting the estimated coefficients relies on untestable assumptions that mediators are randomly assigned to individuals, making the functional causal effect inseparable from the individual-level noise.

Aside from the sequential mediator modeling idea, \cite{zhao2022pathway} proposed a marginal mediator model with correlated errors, introducing a convex Pathway Lasso penalty to directly penalize the product term in the indirect effect. This method showed improved computational efficiency and accuracy over sequential mediator models but did not account for sparsity in functional coefficients. In a different approach, \cite{zhao2020sparse} addressed high-dimensional mediation by using sparse principal component analysis to map correlated mediators onto an independent space, applying penalized regression techniques like the elastic net to enforce sparsity in the outcome model. Expanding on these ideas,  \cite{zhao2023mediation} developed a method for high-dimensional outcomes and mediators, using independent screening to identify significant outcome-mediator pairs. \cite{Nath2023-mv} took a machine learning-based approach, mapping high-dimensional imaging mediators to a single latent variable for use in traditional mediation models, though this reduced the interpretability of the mediation effects.

Focusing on temporal mediation effects, \cite{zhao2019granger} proposed Granger mediation analysis, a novel framework for causal mediation analysis of multiple time series, inspired by an fMRI experiment. The framework combines causal mediation analysis and vector autoregressive (VAR) models to address challenges in time-series data, improving estimation bias and statistical power compared to existing approaches.

For Bayesian analysis of mediation effects, \cite{yuan2009bayesian} presented a pioneering work in both single-level and multi-level models, demonstrating that Bayesian mediation analysis can improve estimation efficiency by incorporating prior knowledge. \cite{daniels2012bayesian} proposed a Bayesian nonparametric method to model conditional densities for a single continuous mediator with binary outcome. In the high-dimensional mediation setting, \cite{song2020cors} proposed Bayesian mixture models to account for a large set of correlated mediators with application to biomarker identification. In particular, to deal withsparsity and correlation in a high dimensional parameter, a membership parameter was used to indicate whether the signal at a certain location is zero or not, and correlation structure is assumed for this membership parameter. In \cite{song2020ptg} and \cite{song2020shrinkage}, different types of Bayesian mixture models were proposed with less focus on the correlation among different locations. In Section \ref{subsec:comparison} we provide more details to these methods and compare them with our proposed method through simulation studies.

To the best of our knowledge, there is a lack of a Bayesian mediation analysis method for high-dimensional imaging data that can incorporate flexible spatial correlation structure, individual-level spatial noise, and sparsity in the functional coefficients. To fill this gap,  we propose a new structural equation model with spatially varying coefficients and adopt the soft-thresholded Gaussian processes~\citep[STGP]{kang2018} as priors for Bayesian Image Mediation Analysis (BIMA).  Under the potential outcomes framework, the proposed BIMA framework consists of two spatially varying coefficient models: a scalar-on-image regression model for the joint effect from the exposure and the image mediator on the outcome (the outcome model), and an image-on-scalar regression model for the effect of the exposure on the image mediator (the mediator model). By assigning the STGP priors, we ensure large prior support for the piecewise smooth and sparse spatially varying coefficients in both models, based on which we formally define the spatially varying mediation effects under the potential outcomes framework. To accommodate population heterogeneity in imaging data, we introduce spatially varying random effects for each individual in the mediator model, improving the efficiency of estimating the mediation effects. For posterior computation, we develop a modified Metropolis-adjusted Langevin algorithm (MALA) that boosts the computational efficiency via block updating and is scalable to high-dimensional imaging data analysis with many observations. 

We perform rigorous theoretical analyses of BIMA. We establish posterior consistency of all the spatially varying coefficients in the mediator and outcome models under the $L_2$ empirical norm, leading to  posterior consistency of the spatially varying mediation effects under the $L_1$ empirical norm. Different from previous theoretical work on Bayesian scalar-on-image models \citep{kang2018}, the image mediation analysis requires us to address the randomness of the functional mediator in the scalar-on-image outcome model while considering the mediator model as the generative model. Hence we proposed a new formulation for functional mediation where the mediator is treated as a random signed measure in the outcome model, and as a random function in the mediator model. This new formulation provides a coherent definition of the natural indirect effect with existing mediation literature while keeping the image mediator bounded in probability in the outcome model. 

The rest of the article is structured as follows. In Section \ref{sec:bima}, we introduce the BIMA framework with definitions, models, and prior specifications. In Section \ref{sec:theory}, we perform a theoretical analysis of the proposed methods, where we establish model identifiability and posterior consistency of the spatially varying mediation effects.
Then, we develop the posterior computation algorithm in Section \ref{sec:computation} and perform extensive simulations in Section \ref{sec:simulation}. 
Finally, we apply BIMA to the analysis of the fMRI and cognitive data in the ABCD study in Section \ref{sec:realdata} and conclude the paper in Section \ref{sec:conclusion}.

\section{Bayesian Image Mediation Analysis}
\label{sec:bima}

\subsection{General Notation}

Let $\R^d$ denote a $d$-dimensional Euclidean vector space. Let $\cS \subset \R^d$ be a compact support. Let $\normal(\mu,\sigma^2)$ represent a normal distribution with mean $\mu$ and variance $\sigma^2$. Let $L^2(\cS)$ be the space of square-integrable functions supported on $\cS$. Let $\br{s_1,\dots,s_p}$ be a set of $p$ fixed design points in $\cS$.
For any function $f(s)$ in $L^2(\Bsupport)$, let $\|f\|_{q,\pscale} = \br{\pscale^{-1} \sum_{j=1}^{\pscale}|f(\cvox_j)|^q}^{1/q}$ be the $L_q$ empirical norm on the fixed grid with $p$ voxels. For any vector $\bfa = (a_1,\ldots,a_d)^\top \in \bR^d$, let $\|\bfa\|_q= \br{\sum_{i=1}^d|a_i|^q}^{1/q}$ be the $L_q$ vector norm. For any functions $f,g\in L^2(\cS)$, define the inner product $\langle f,g\rangle :=\int_{\Bsupport} f(s)g(s)\leb (\dd s)$ where $\leb$ is a Lebesgue measure. The empirical inner product is defined as $\langle f,g\rangle_{\pscale}:=\pscale^{-1}\sum_{j=1}^p f(\cvox_j)g(\cvox_j)$.
Let $\cC^{\rho}(\Bsupport)$ be the order-$\rho$ H\"{o}lder space on $\Bsupport$ for a positive integer $\rho$. For a set $\cB$, $\bar\cB$ is used to denote the closure of the set, while $\partial\cB$ denotes its boundary. Let $\cGP(\nu,\kappa)$ denote a Gaussian Process with mean function $\nu(\cdot)$ and covariance matrix $\kappa(\cdot,\cdot)$. 

\subsection{Spatially-Varying Coefficient Structural Equation Models}\label{sec:model}

Suppose the data consists of $n$ individuals. For individual $i (i = 1,\ldots, n)$, let $Y_i\in \R$ denote the outcome variable, $X_i \in \R$ denote the exposure variable, $\bfC_i = (C_{i,1},\dots,C_{i,q})^\top\in\R^q$ be a vector of $q$ potential confounding variables. Suppose the imaging data are observed on a compact support $\cS$. Let $\br{\vox_1,\dots, \vox_p}$ be an evenly-spaced partition of $\cS$, representing voxels in the fMRI data.
 Let $\cvox_j$ be the center of the voxel $\vox_j$ for $j = 1,\ldots, p$.  Let $\bfM_i = \sbr{M_i(s_j),\ldots, M_i(s_p)}^\top$ be a vector of observed image intensities, where $\mpdf_i(s)$ represent the image intensity function at location $s \in \cS$.

To perform image mediation analysis, we consider spatially varying coefficient structural equation models which consist of scalar-on-image regression as the outcome model~\eqref{eq:model1} and image-on-scalar regression as the mediator model~\eqref{eq:model2}. For $i = 1,\ldots, n$, we assume
\begin{align}
Y_i &= \sum_{j=1}^{p} \beta(\cvox_j) \mcdf_{i}(\vox_j)  + \gamma X_i + \bfxi^{\top} \bfC_i + \epsilon_{Y,i},\quad \epsilon_{Y,i}\overset{\iid}{\sim} \normal(0,\sigma_Y^2),\numberthis\label{eq:model1}\\
\mpdf_{i}(\cvox_j) &= \alpha(\cvox_j) X_i + \bfzeta^\top(\cvox_j)\bfC_{i} + \eta_i(\cvox_j)+ \epsilon_{M,i}(\cvox_j),\quad \epsilon_{M,i}(\cvox_j)\overset{\iid}{\sim} \mN(0,\sigma_M^2)\numberthis\label{eq:model2}
\end{align}
where $\mcdf_i(\vox)=\int_{\vox} M_i(s)\leb(\dd s)$ is the total intensity measure over the small partition $\vox$.  Throughout this paper, we assume that the volume of the whole brain $\leb(\cS)=1$ and one partition is $\leb(\vox_j) = p^{-1}$ for any $j=1,\dots,p$, and $\mcdf_i(\vox)$ can be approximated as $p^{-1}\mpdf_i(s)$ in practice when $n$ and $p$ are finite. In theory where $n$ and $p$ can go to infinity, we refer to section \ref{sec:wiener} and use \eqref{eq:model2-3} as the approximation for $\mcdf_i(\vox)$. In fact, model \eqref{eq:model1} uses the finite approximation $\int_{\cS}\beta(s)\mcdf_i(\dd s) \approx\sum_{j=1}^{p} \beta(\cvox_j) \mcdf_{i}(\vox_j)$. To see this, for a compact $\cS$ with an evenly-spaced partition $\cS = \oplus_{j=1}^p\vox_j$, where $\oplus$ represents the union of mutually exclusive sets,  $\int_{\cS} \beta(s) \mcdf_i(\dd s) = \int_{\oplus_{j=1}^p\vox_j} \beta(s) \mcdf_i(\dd s) = \sum_{j=1}^p \int_{\vox_j}\beta(s) \mcdf_i(\dd s)\approx \sum_{j=1}^{p} \beta(\cvox_j) \mcdf_{i}(\vox_j)$. More details are discussed in Section \ref{sec:wiener}.

In the outcome model~\eqref{eq:model1}, $\beta(s)$ represents the spatially-varying effects of the image mediator on the outcome variable.  The scalar coefficient $\gamma$ is the direct effect of $X_i$ on $Y_i$. The vector coefficient $\bfxi\in \mbR^q$ represents the confounding effects. The random noise terms,  $\epsilon_{Y,i}$, are independent and identically distributed according to a normal distribution with mean zero and variance $\sigma^2_Y$. 

In the mediator model~\eqref{eq:model2}, $\alpha(s)$ is the spatially-varying functional parameter ofinterest. $\bfzeta(s) = \{\zeta_{1}(s),\ldots, \zeta_{q}(s)\}^\top$ is a vector of the coefficients for the confounders; $\eta_i(s)$ is the spatially-varying individual effect that capture the individual variations unexplained by the exposure variable $X_i$ and the observed confounders $\bfC_i$; and $\epsilon_{M,i}(\cvox_j)$ is the spatially independent noise term across locations and subjects with constant variance $\sigma^2_M$. 

One of our main contributions is to utilize spatial information of the mediator by using Gaussian Process (GP) and Soft-thresholded Gaussian Process (STGP) priors to model the spatial structures of different functional effects. The following subsection provides the prior specification for the above functional parameters. 

\subsection{Prior Specifications}\label{sec:prior}
Due to the sparsity of brain signals in the ABCD working memory task fMRI data \citep{zhao2023bayesian, lin2024latent}, we enforce a sparsity assumption on $\alpha(s)$ and $\beta(s)$. To model the sparsity and the spatial smoothness in the spatially varying mediation effects $\mathcal{E}(s)$, we adopt the soft-thresholded Gaussian process (STGP) proposed in \cite{kang2018} for $\alpha(s)$ and $\beta(s)$, separately. For the individual effects $\eta_i(s)$ and confounding effects $\zeta_{k}(s)$, we assign regular Gaussian process priors. Let $T_\sthresh:\mathbb{R} \mapsto \mathbb{R}$ be a soft-thresholded operator defined as
$T_\sthresh(x) := \{x-\sgn(x)\sthresh\}I(|x|>\sthresh)$ for any $\sthresh \geq 0$. 
\begin{definition}[\cite{kang2018}]\label{def:STGP}
Let $\tilde f(s)$ be a Gaussian process (GP) with mean zero and the covariance kernel $\kappa_f$, denoted as $\tilde f\sim \cGP(0,\kappa_f)$. For any $\sthresh \geq 0$, set $f(s) = T_{\sthresh}\{\tilde f(s)\}$. 
Then $f(s)$ is a STGP with covariance kernel $\kappa_f$ and threshold parameter $\sthresh$, denoted as $f\sim\cSTGP(\sthresh_f,\kappa_f)$.
\end{definition}
 In summary, we have the following prior specifications,  
\begin{align}
    &\beta \sim  \cSTGP(\sthresh_{\beta},\sigma^2_\beta\kappa), \ \  \alpha \sim \cSTGP(\sthresh_{\alpha},\sigma^2_\alpha\kappa),\ \ \zeta_{k} \sim \cGP(0,\sigma^2_\zeta\kappa),  \ \  \eta_i \sim \cGP(0,\sigma^2_\eta\kappa),\label{eq:priors}
\end{align}
for $i = 1,\ldots, n$ and $k = 1,\ldots, q$. As explained in Section 3.2 in \cite{kang2018}, given a positive threshold value $\sthresh>0$,  STGP is flexible to fit a wide range of sparsity levels. The specific values for the thresholding parameters $\nu_\alpha$ and $\nu_\beta$ in practice are chosen within a reasonable range according to the effect size of $\alpha$ and $\beta$.

The choice of $\kappa$, the kernel function for the latent Gaussian process, controls the smoothness of the functional parameters. To utilize the spatial information in the mediator $M_i(s)$, the key insights are to (i) use STGP for $\alpha$ and $\beta$ so that the spatial structure is accounted for by the latent Gaussian kernel, and (ii) use GP on $\eta_i$ so that the individual level spatial-varying effects in the mediator are also accounted for. For the remaining  parameters, a normal prior with mean zero are assigned to $\gamma, \bfxi$, and inverse-gamma priors are assigned to the variance parameters $\sigma_Y^2$, $\sigma_M^2$, $\sigma_\beta^2$, $\sigma_\alpha^2$ and $\sigma_\eta^2$.

\subsection{Connection to the Wiener process}\label{sec:wiener}
 When the support $\cS$ is one-dimensional, the finite summation $\sum_{j=1}^p\beta(\cvox_j)\mcdf_i(\vox_j)$ in model \eqref{eq:model1} is an approximation to the continuous integral $\int_{\cS} \beta(s) \mcdf_i(\dd s)$. In fact, when $\cS=[0,1]\in \R$, the continuous version of model \eqref{eq:model1} and \eqref{eq:model2} can be represented as 
\begin{align*}
    Y_i &= \int_\cS \beta(s) \mcdf_{i}(\dd s)  + \gamma X_i + \bfxi^{\top} \bfC_i + \epsilon_{Y,i}, \\
     \mcdf_{i}(\dd s) &= \br{\alpha(s) X_i   + \bfzeta^\top(s)\bfC_{i} + \eta_i(s) }  \leb(\dd s)+ \sigma_M\dd W_{i,M}(s), \numberthis\label{eq:Wiener}
\end{align*}
where $\epsilon_{Y,i}\overset{\iid}{\sim} \mN(0,\sigma_Y^2)$ and $W_{i,M}(s)$ is the Wiener process \citep{durrett2019probability}. 

\begin{figure}%
\centering
\begin{minipage}{0.4\textwidth}
  \centering
  \begin{tikzpicture}[
  plainnode/.style={draw=none, fill=none, minimum size=0mm, text=black}, 
  roundnode/.style={circle, draw=green!60, fill=green!30, very thick, minimum size=7mm}, 
  squarednode/.style={rectangle, draw=blue!60, fill=blue!30, very thick, minimum size=7mm},
  every edge/.style={draw, -{Stealth[length=3mm, width=2mm]}, ultra thick}, 
  bracket/.style={decorate,decoration={brace,amplitude=10pt,raise=5pt}, thick} 
]

\node[roundnode]      (X)   at (0, 0)   {$X_i$};          
\node[squarednode]      (M)   at (3, 0)   {$M_i(s)_{s\in\cS}$};       
\node[roundnode]      (Y)   at (6, 0)   {$Y_i$};          
\node[roundnode]      (C)   at (3, -2)   {$C_i$};         
\node[plainnode]      (eta)   at (0.3, -1)   {$\eta_i$}; 

\draw[->] (X) -- node[above] {$\alpha$} (M);     
\draw[->] (M) -- node[above] {$\beta$} (Y);     
\draw[->, bend left=45] (X) to node[above] {$\gamma$} (Y); 
\draw[->] (C) -- node[left] {$\xi_M$} (M);            
\draw[->] (C) -- node[right] {$\xi_Y$} (Y);              
\draw[->] (eta) -- (M);                                  

\draw[bracket] (1.3, 0.2) -- (4.7, 0.2) node[midway, above=14pt] {$\mathcal{E}$};

\end{tikzpicture}
  \caption{Graph representation of the mediation model. $M_i(s)_{s\in\cS}$ is a random function supported on a spatial domain $s\in \cS$. $\alpha,\beta,\cE,\xi_M,\eta_i$ are all functional parameters. }
  \label{fig:structure}
\end{minipage}
\quad\quad
\begin{minipage}{.4\textwidth}
  \centering
  \def\svgwidth{\columnwidth}
\begingroup%
  \makeatletter%
  \providecommand\color[2][]{%
    \errmessage{(Inkscape) Color is used for the text in Inkscape, but the package 'color.sty' is not loaded}%
    \renewcommand\color[2][]{}%
  }%
  \providecommand\transparent[1]{%
    \errmessage{(Inkscape) Transparency is used (non-zero) for the text in Inkscape, but the package 'transparent.sty' is not loaded}%
    \renewcommand\transparent[1]{}%
  }%
  \providecommand\rotatebox[2]{#2}%
  \newcommand*\fsize{\dimexpr\f@size pt\relax}%
  \newcommand*\lineheight[1]{\fontsize{\fsize}{#1\fsize}\selectfont}%
  \ifx\svgwidth\undefined%
    \setlength{\unitlength}{267.24617305bp}%
    \ifx\svgscale\undefined%
      \relax%
    \else%
      \setlength{\unitlength}{\unitlength * \real{\svgscale}}%
    \fi%
  \else%
    \setlength{\unitlength}{\svgwidth}%
  \fi%
  \global\let\svgwidth\undefined%
  \global\let\svgscale\undefined%
  \makeatother%
  \begin{picture}(1,0.52325452)%
    \lineheight{1}%
    \setlength\tabcolsep{0pt}%
    \put(0,0){\includegraphics[width=\unitlength,page=1]{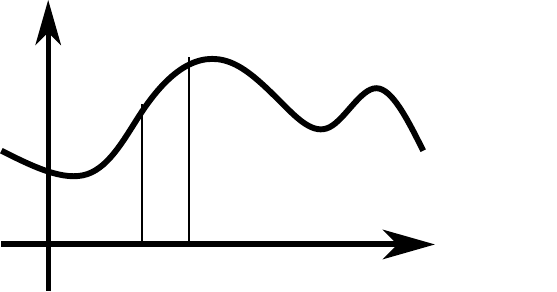}}%
    \put(0.25494789,0.0395472){\makebox(0,0)[lt]{\lineheight{1.25}\smash{\begin{tabular}[t]{l}$\Delta s$\end{tabular}}}}%
    \put(0,0){\includegraphics[width=\unitlength,page=2]{curve2.pdf}}%
    \put(0.58228905,0.38212966){\makebox(0,0)[lt]{\lineheight{1.25}\smash{\begin{tabular}[t]{l}$\mpdf(\cvox)$\end{tabular}}}}%
    \put(0.77093809,0.09483794){\makebox(0,0)[lt]{\lineheight{1.25}\smash{\begin{tabular}[t]{l}$\Bsupport$\end{tabular}}}}%
    \put(0.34275258,0.20793127){\color[rgb]{0.34117647,0.34117647,0.89411765}\makebox(0,0)[lt]{\lineheight{1.25}\smash{\begin{tabular}[t]{l}$\mcdf(\vox)$\end{tabular}}}}%
  \end{picture}%
\endgroup%
\\
  \caption{Illustration of the definitions of the intensity measure $\mathcal{M}(\Delta s)$ and the intensity function $M(s)$ in one-dimensional support $\Bsupport$.}
  \label{fig:curve}
\end{minipage}%

\end{figure}

 In neuroimaging applications, we can only observe $\mpdf_i(s)$ on fixed grids $\br{j=1,\dots,p}$, without loss of generality, we can approximate the values of $\mpdf_i(s),\alpha(s),\bfzeta(s)$ and $\eta_i(s)$ within each $\vox_j$ by the functional values at its center $\cvox_j$, using the approximation for any $s\in\vox_j,\alpha(s)\equiv \alpha(\cvox_j)$, similar for $\bfzeta,\eta_i$. Hence \eqref{eq:Wiener} can be approximated by
\begin{align*}
    \mcdf_{i}(\vox_j) &=  \br{\alpha(\cvox_j) X_i + \bfzeta^\top(\cvox_j)\bfC_{i} + \eta_i(\cvox_j)}\leb(\vox_j) + \varepsilon_{M,i}(\vox_j),
     \numberthis \label{eq:model2-3}
\end{align*}
where $\varepsilon_{M,i}(\vox_j)\sim\mN\{0,\sigma_M^2 \leb(\vox_j)\}$. Note that the general definition $\mcdf_i(\vox) = \int_{\vox} M_i(s)\leb(\dd s)$ is consistent with Equation \eqref{eq:model2-3}, by plugging in the outcome model \eqref{eq:model2} (see details in Section \ref{supp_sec:approx_mediator}). The relationship between the function-valued $\mpdf_i(s)$ and the integrated image intensity $\mcdf_{i}(\vox_j)$ over $\vox_j$ in the one-dimensional case is illustrated in Figure \ref{fig:curve}.
The advantage of using $\sum_{j=1}^p\beta(\cvox_j)\mcdf_i(\vox_j)$ in the scalar-on-image model \eqref{eq:model1} compared to other existing formulations \citep{kang2018, lindquist2012functional} can be explained in two ways. First, the summation $\sum_{j=1}^{p} \beta(\cvox_j) \mcdf_{i}(\vox_j)$ in 
 \eqref{eq:model1} is a natural approximation to the inner-product on $L^2(\cS)$, hence in mediation analysis, as explained in the next section, we can naturally express the total indirect effect as $\sum_{j=1}^p \beta(\cvox_j)\alpha(\cvox_j)\leb(\vox_j)$. Other formulations such as $\beta(s_j)M_i(s_j)/\sqrt{p}$ in \cite{kang2018} do not have this property. Second, the variance of $\varepsilon_{M,i}(\vox_j)$ is by design proportional to $\leb(\vox_j)$ instead of $\sbr{\leb(\vox_j)}^2$. This plays a key role in constructing a test function when showing posterior consistency in model \eqref{eq:model1}, and ensures that we have enough variability in the design matrix in \eqref{eq:model1} to be able to estimate $\beta(s)$. In fact, the $M_i(s_j)/\sqrt{p}$ as used in \cite{kang2018} also has variance proportional to $1/p$, but they assume the mean part of $M_i(s)$ is zero for all $s\in \cS$, so that $\beta(s_j)\mathbb{E}\br{M_i(s_j)}/\sqrt{p}$ will not explode as $p\to \infty$, but this assumption is not practical in mediation analysis. \cite{lindquist2012functional} also uses an inner product formulation, but they only assume that all the functional parameters can be represented by finitely many basis functions, and the number of basis does not increase with $n$ nor $p$, whereas in our case, we study all sparse, piece-wise smooth function in $L_2(\cS)$.

\subsection{Causal Mediation Analysis}\label{subsec:causal_mediation}

We define the main mediation parameter of interest first.
\begin{definition}\label{def:mediation_func}
Let $\mathcal{E}(s)= \alpha(s)\beta(s)$ be the spatially-varying mediation effect (SVME). 
\end{definition}

Under the causal inference framework~\citep{rubin1974estimating}, for individual $i$, we define $Y_{i,(x,\bfm)}$ as the potential outcome variable that would have been observed when the image mediator $\bfM_i = \bfm$ and the exposure variable $X_i = x$. Let $M_{i,(x)}(s_j)$ represent the image intensity value at location $s_j$ when the individual $i$ receives exposure $x$. Let $\bfM_{i,(x)} = (M_{i,(x)}(s_1),\ldots, M_{i,(x)}(s_p))^\rT$ be the potential image mediator on $\cS$.
When the exposure variable $X_i$ changes from $x$ to $x'$,
combining equations \eqref{eq:model1} and \eqref{eq:model2-3}, we represent the natural indirect effect (NIE) and the natural direct effect (NDE) as follows: \begin{align}
\nie(x,x') = \bE\left[ Y_{i,\{x, \bfM_{i,(x)}\}} - Y_{i,\{x, \bfM_{i,(x')}\}} \mid \bfC_i\right] 
&= \sum_{j=1}^p \mathcal{E}(\cvox_j) \leb(\vox_j)(x - x') \label{eq:NIE}\\
\nde(x,x') = \bE\left[  Y_{i,\{x, \bfM_{i,(x')}\}} - Y_{i,\{x', \bfM_{i,(x')}\}} \mid \bfC_i\right] &= \gamma(x-x').\label{eq:NDE}
\end{align}
The detailed derivation of NIE and NDE based on the causal assumptions and our structural equation models is provided in Supplementary Section \ref{sec:app_causal_interpretation}.

Following the line of work in functional mediation analysis \citep{lindquist2012functional,wang2023high, song2020cors}, we impose the stable unit treatment value assumption (SUTVA)~\citep{rubin1980randomization} and the following set of causal assumptions to ensure the causal identification of NIE and NDE:

\noindent\textbf{Causal Assumptions:} For any $i$, $x$ and $\bfm$, we assume: \textbf{[A1]} $Y_{i,(x,\bfm)}\perp X_i\mid \bfC_i$, \textbf{[A2]} $Y_{i,(x,\bfm)}\perp \bfM_i\mid \{\bfC_i,X_i\}$, \textbf{[A3]} $\bfM_{i,(x)}\perp X_i\mid \bfC_i$, \textbf{[A4]} $Y_{i,(x,\bfm)} \perp \bfM_{i,(x')}\mid \bfC_i$.

The causal assumptions ensure that: (i) NIE and NDE can be identified, and (ii) NIE and NDE can be estimated from observable data. See \cite{vanderweele2014mediation} for a detailed interpretation of the above assumptions. We provide a detailed interpretation of causal assumptions under the ABCD study setting in Section \ref{sec:realdata} and Supplementary Section \ref{sec:app_causal_interpretation}.

The current causal framework only allows one to causally identify NIE and NDE, instead of the mediation effect of individual voxels. Under this framework, BIMA, along with other spatially varying mediation approaches~\citep{wang2023high, song2020cors, jiang2023causal}, aims to investigate the contribution of different spatial regions to the natural indirect effect (NIE). Conceptually, $\mathcal{E}(\cdot)$ is treated as a single functional mediator, rather than a collection of multiple mediators each exerting their own mediation effects. Our objective is to identify the spatial decomposition of nonzero regions within $\mathcal{E}(\cdot)$ that contribute to the NIE, analogous to the temporal decomposition of the NIE proposed in~\cite{lindquist2012functional}.

 In image mediation analysis, we are interested in which locations contribute to the NIE or the mediation effects.  From~\eqref{eq:NIE}, it is straightforward to see that  $\mathcal{E}(\cvox_j)$ represents the contribution of location $\cvox_j$ to the $\nie(x,x')$ for any $x\neq x'$, which is the motivation of Definition \ref{def:mediation_func}.  For any location $s\in \Bsupport$, $\mathcal{E}(s)$ characterizes the impact of the location $s$ on the NIE. Both $\mathcal{E}(s)$ and $\pfactor^{-1}\sum_{j=1}^p \mathcal{E}(\cvox_j)$ are the parameters of main interest.
It is generally believed that not all brain locations contribute to the mediation effects, and $\mathcal{E}(s)$ is naturally a sparse function when either $\alpha(s)$ or $\beta(s)$ is sparse.

\section{Theoretical Properties}
\label{sec:theory}
This section we establish posterior consistency  for the spatially varying mediation effects  $\mathcal{E}(s)$ under the empirical $L_1$ norm. To achieve this goal, we first show posterior consistency for  $\beta(s)$ in the outcome model~\eqref{eq:model1} and  $\alpha(s)$ in the mediator model~\eqref{eq:model2}, respectively. All the derivations and proofs are provided in the Supplementary Material.

\subsection{Notation and Assumptions} \label{subsec:notation_and_assumption}
To perform the theoretical analysis, we introduce additional notation. Let $\bfY = (Y_1,\ldots, Y_n)\in \R^n$, $\bfX = (X_1,\dots, X_n)^\top\in \R^n$, $\bfM = (\bfM_1,\ldots, \bfM_n)^\top\in \R^{n\times p}$ and $\bfC = (\bfC_1,\dots,\bfC_n)^\top \in \R^{n\times q}$.
Let $\alpha_0(s)$, $\beta_0(s)$, $\eta_{i,0}(s)$ and $\bfzeta_{0}(s)$ represent the corresponding true spatially varying coefficients in the BIMA models~\eqref{eq:model1} and \eqref{eq:model2} that generate the observed data $\bfY$ and $\bfM$ given $\bfX$ and $\bfC$. Let $\mathcal{E}_0(s) = \alpha_0(s)\beta_0(s)$ represent the true spatially varying mediation effects.  We assume that all of the true spatially varying coefficients are square-integrable in $L^2(\cS)$. For matrix $A$, $\det(A)$ denotes the determinant of $A$, $\sigma_{\min}(A),\sigma_{\max}(A)$ denote the smallest and the largest singular value of $A$ respectively.

Next, we define a functional space for the sparse and piecewise smooth spatially varying coefficients. 

\begin{definition}[Sparse functional space]\label{def:sparse_param}
Define the sparse functional space $\Theta^{SP}=\{f(s):s\in \Bsupport\}$ as the collection of spatially-varying coefficient functions that satisfy the three conditions.  a) (Continuous) $f(s)$ is a continuous function on $\Bsupport$; b) (Sparse) Assume there exist two disjoint nonempty open sets $\cR_{-1}$ and $\cR_{1}$, and $\partial \cR_{-1}\cap \partial \cR_1=\emptyset$ such that $\forall s\in\cR_{1},~f(s)>0$; $\forall s\in\cR_{-1}, ~f(s)<0$. $\cR_{0} = \Bsupport-(\cR_{1}\cup \cR_{-1})$, and assume $\cR_0$ has nonempty interior; and c) (Piecewise smooth) For any $s\in \bar \cR_1\cup \bar \cR_{-1}$, $f(s)\in \cC^\rho(\bar \cR_1\cup \bar \cR_{-1})$, $\rho \geq 1$. 
\end{definition}
This definition has been adopted for specifying the true parameter space of  scalar-on-image regression, see Definition 2 in \cite{kang2018}. In BIMA, $\alpha(s)$ and $\beta(s)$ are assumed to be in the sparse functional space in Definition \ref{def:sparse_param}, and later in the proof of Theorem \ref{thm:total_consistency}, we will show that $\mathcal{E}(s)$ as defined in Definition \ref{def:mediation_func} also belongs to the sparse functional space in Definition \ref{def:sparse_param} when both $\alpha(s)$ and $\beta(s)$ are in this sparse functional space.

Next, we will introduce the parameter space for models \eqref{eq:model1} and \eqref{eq:model2}.
\begin{definition}[Parameter space]\label{def:param_space}
    Let $\Theta_\alpha$, $\Theta_\beta$, $\Theta_\eta$ and $\Theta_{\zeta}$ be the parameter space in $L^2(\cS)$ for $\alpha$, $\beta$, $\br{\eta_i}_{i=1}^n$ and $\br{\zeta_{k}}_{k=1}^q$ respectively. Let $\br{\psi_l(s)}_{l=1}^\infty$ be a set of basis of $L^2(\cS)$, we specify the following constraints for each parameter space: (a) $\Theta_\alpha\subset \Theta^{SP}$; (b) $\Theta_\beta\subset \Theta^{SP}$, and for any $\beta\in \Theta_\beta$, define $\theta_{\beta,l} = \int_\cS\beta(s)\psi_l(s)\leb(\dd s)$, there exists $L_n = n^{\nu_1}$ where $\nu_1\in (0,1)$ and $\nu_2>0$ such that $\sum_{l=L_n}^\infty \theta_{\beta,l}^2\leq L_n^{-\nu_2}$; (c) $\Theta_\eta, \Theta_{\zeta} \subset C^\rho(\cS)$; (d) There exists a constant $K>0$ such that for any $f, g$ in $\Theta_\alpha$, $\Theta_\beta$, $\Theta_\eta$, $\Theta_{\zeta}$ and $\br{\psi_l(s)}_{l=1}^\infty$, the fixed grid approximation error $|\int_\cS f(s)g(s)\leb (\dd s) - p^{-1}\sum_{j=1}^p f(\cvox_j)g(\cvox_j)| \leq K p^{-2/d}$.
\end{definition}

\noindent \textbf{Remark.}
In the case of region partition $\cS = \cup_{r=1}^R \cS_r$, we can construct the basis based on each region. Let $\br{\psi_{l,r}(s)}_{l=1}^\infty$ be the basis of $L^2(\cS_r)$, and construct $\psi_l(s) = \sum_{r=1}^R \psi_{l,r}(s)I(s\in \cS_r)$. The basis decomposition for $f(s)\in L^2(\cS)$ can be written as $\theta_{f,l} = \int_{\cS} \psi_l(s)f(s)\leb(ds) = \sum_{r=1}^R \int_{\cS_r} \psi_l(s)f(s)\leb(ds) = \sum_{r=1}^R \theta_{f,r,l}$. The decay rate condition in Definition \ref{def:param_space} stays the same for $\theta_{f,l}$ because of the finite summation.

In Definition \ref{def:param_space}, (a)-(c) define the smoothness and sparse feature of the parameter space, where $\alpha(s)$, $\beta(s)$ are assumed to be piecewise-smooth, sparse and continuous functions, and the individual effect $\eta_i(s)$ and the confounding effects $\zeta_{k}(s)$ in model \eqref{eq:model2} are only required to be smooth but not necessarily sparse. Definition \ref{def:param_space}(d) sets an upper bound for the fixed grid approximation error. Assumption \ref{asm:true_func} below specifies the smoothness of the underlying Gaussian processes and the rate of $p$ as $n\to\infty$.

\begin{assumption}\label{asm:true_func}
Given the dimension $d$ of $\cS$ and a constant $\tau$ satisfying $d>1+1/\tau$, $\tau\geq 1$, assume that 
a) (Smooth Kernel) for each $s$, the kernel function $\kappa(s,\cdot)$ introduced in the priors \eqref{eq:priors} has continuous partial derivatives up to order $2\rho+2$ for some positive integer $\rho$, i.e. $\kappa(s,\cdot)\in \cC^{2\rho+2}(\Bsupport)$, and $d+3/(2\tau) < \rho$; b) (Dimension Limits) $p\geq O(n^{\tau d})$.
\end{assumption}
\noindent The Assumption \ref{asm:true_func}(a) is the standard condition~\citep{ghosal2006} to ensure the sufficient smoothness of the latent Gaussian processes $\tilde\beta(s)$, $\tilde\alpha(s), \zeta_{k}(s)$ and $\eta_i(s)$. The Assumption \ref{asm:true_func}(b) is to specify the order of the number of voxels as the sample size increases, implying that our method can handle high resolution images. The total number of voxels $p$ needs to be sufficiently large for the posterior of the function-valued $\beta(s)$ to concentrate around the true function $\beta_0(s)$. The number of confounders $q$ is finite and does not grow with $n$.

As the mediator model \eqref{eq:model2} involves spatially varying coefficients $\eta_i(s)$ as individual effect parameters, the model identifiability is not trivial and requires some mild conditions on the observations of exposure variables and confounding factors.

\begin{assumption}\label{asm:alpha_iden}
(a) Each element in $(\bfX,\bfC)$ has a finite fourth moment with sub-Gaussian tails, and $\sigma_{\min}\br{(\bfX,\bfC)}>\sqrt{n}$ almost surely;
 (b)  Conditioning on $(\bfX,\bfC)$, there exists a matrix $\bfW = (W_{i,k})\in \R^{n\times (q+1)}$ such that $\det\{\bfW^\top (\bfX,\bfC)\}\neq 0$; and (c) there exists a constant vector $\bfb=(b_1,\ldots, b_q)^\top $ such that for any $s\in\cS$ and $k=1,\ldots,q+1$,  $\sum_{i=1}^n W_{i,k} \eta_i(s) = b_k$.  
\end{assumption}
\noindent Assumption \ref{asm:alpha_iden}(a) 
is a reasonable assumption in linear regression with the design matrix $(\bfX,\bfC)$ \citep{armagan2013posterior}. For (b) and (c), one example that can satisfy the above assumption is to set $b = 0\in \R^{q+1}$, $\bfW = (\bfX,\bfC)$, and if we express $\eta_i(s) = \sum_{l=1}^\infty\theta_{\eta,i,l}\psi_l(s)$ as infinite sums of basis in the Hilbert space, then each $\sbr{\theta_{\eta,i,l}}_{i=1}^n\in \R^n$ is generated from a subspace orthogonal to $\text{span}\{\bfX,\bfC_1,\ldots,\bfC_q\}$. We enforce this assumption in the sampling algorithm by updating $\sbr{\theta_{\eta,i,l}}_{i=1}^n$ from a constrained multivariate normal distribution.

With Assumption \ref{asm:alpha_iden}, we can establish the model identifiability in \eqref{eq:model2} and show that if the spatially varying coefficients are different from the true value, the mean function of $M_i(s)$, denoted as $\mu_{M,i}(s) := \alpha(s)X_i + \bfzeta^\top(s)\bfC_{i} + \eta_i(s)$, will also be deviated from the true mean function $\mu_{M,i,0}(s):=\alpha_0(s)X_i + \bfzeta^\top_{ 0}(s)\bfC_{i} + \eta_{i,0}(s)$.

Let $\Theta_M = \Theta_\alpha \times \Theta_{\bfzeta} \times \left(\prod_i \Theta_{\eta,i}\right)$ be the joint parameter space for all parameters in the mean function $\mu_{M,i}(s)$. For any $\epsilon>0$ and some constant $c_0>0$, define the following two subsets of $\Theta_M$ as $ \cU^c_M = \left\{\Theta_M: \|\alpha-\alpha_0\|^2_{2,\pscale} + \sum_{k=1}^{q}\|\zeta_{k}-\zeta_{k,0}\|^2_{\pscale}  + n^{-1}\sum_{i=1}^n\|\eta_i-\eta_i\|^2_{2,\pscale} >\epsilon^2\right\}$ and $\cU^c_{M,\mu} = \left\{\Theta_M: n^{-1}\sum_{i=1}^n\|\mu_{M,i} - \mu_{M,i,0}\|^2_{2,\pscale}>c_0\epsilon^2\right\}$. 

\begin{proposition}\label{prop:iden}
Under Assumptions \ref{asm:alpha_iden}, (a) the mediator model \eqref{eq:model2} is identifiable; and (b) $\cU^c_M \subset \cU^c_{M,\mu}$ almost surely with respect to $(\bfX, \bfC)$.
\end{proposition}

\subsection{Posterior consistency}
First, we show joint posterior consistency of all the spatially varying coefficients in the mediator model~\eqref{eq:model2} as the number of images $n\to\infty$ and  the number of voxels $\pscale\to \infty$.

The following empirical $L_2$ norm consistency result is proved by verifying conditions in Theorem A.1 in \cite{Ghosal2004}. For the proof of existence of test, we borrow techniques from Proposition 11 in \cite{vandervaart2011}.  
\begin{theorem}
\label{thm:alpha_consistency}

Suppose Assumptions \ref{asm:true_func}-\ref{asm:alpha_iden} hold in the mediator model (\ref{eq:model2}). For any $\epsilon>0$, as $n\to \infty$, 
$\Pi(\cU^c_M \mid \bfM, \bfX, \bfC) \to 0$
in $P^{n}_0$- probability. This further implies that $\Pi ( \|\alpha-\alpha_0\|_{2,\pscale}>\epsilon \mid \bfM, \bfX, \bfC) \to 0$ and $
    \Pi (n^{-1}\sum_{i=1}^n \|\eta_i - \eta_{i,0}\|^2_{2,\pscale}>\epsilon^2 \mid \bfM, \bfX, \bfC) \to 0$
in $P^{n}_0$- probability.
\end{theorem}


Next, we establish the $L_2$ consistency on $\beta(s)$ with the following assumptions. 

For any $f\in L^2(\Bsupport)$, given the basis $\br{\psi_l(s)}_{l=1}^\infty$ in Definition \ref{def:param_space}, $f(s) = \sum_{l=1}^\infty \theta_{f,l}\psi_l(s)$, 
where $\sum_{l=1}^\infty\theta_{f,l}^2 <\infty$. Let $r_L(s) = \sum_{l=L}^\infty \theta_{f,l}\psi_l(s)$ be the remainder term after choosing a cutoff $L$ as the finite sum approximation.  Note that the remainder term $\int_{\cS}r_L(x)^2\leb (\dd s) = \sum_{l=L}^\infty \theta_{f,l}^2\to 0$ as $L\to\infty$ (Appendix E in \cite{vandervaart_2017}). We employ the basis expression to show the posterior consistency in model \eqref{eq:model1}, especially for studying the role of $\mcdf_i(\vox_j)$.

Denote $\tilde\bfgamma = (\gamma,\bfxi^\top)^\top\in \R^{q+1}$, $\tilde\bfX_i = (X_i,\bfC_i^\rT)^\rT\in \R^{q+1}$. Let $\beta(s) = \sum_{l=1}^\infty \theta_{\beta,l}\psi_l(\cvox_j)$. Let $\tilde{\mcdf}_{i,l} = \sum_{j=1}^p \psi_l(\cvox_j)\mcdf_i(\vox_j)$, and define the $n\times L_n$ matrix $\bm{\tilde{\mcdf}}_n:=(\tilde{\mcdf}_{i,l})_{i=1,\ldots, n, l=1,\ldots, L_n}$. Further, denote $\tilde\bfW_n = (\bm{\tilde{\mcdf}}_n,\tilde\bfX) \in \R^{n\times(L_n+1+q)}$ as the design matrix.

We state the following assumption for constructing the consistency test in Theorem \ref{thm:beta_consistency}.

\begin{assumption}\label{asm:W}
    The least singular value of $\tilde\bfW_n$ satisfies $0 < c_{\min} < \liminf_{n\to\infty}\sigma_{\min}(\tilde\bfW_n)/\sqrt{n}$
with probability $1-\exp(-\tilde c n)$ for some constant $\tilde c ,c_{\min}>0$.
\end{assumption}
A similar assumption has been made in \cite{armagan2013posterior}. One extreme example that satisfies Assumption \ref{asm:W} is when $\tilde\bfW_n$ has mean-zero i.i.d. subgaussian entries. We will also give an example in the Supplementary Material \ref{subsec:verify} that satisfies Assumption \ref{asm:W} and follows the generative model \eqref{eq:model2} under some conditions. 

\noindent\textbf{Remark.} Assumption \ref{asm:W} demonstrates the variability in the design matrix $\tilde \bfW_n$: the posterior consistency of $\beta(s)$ can only be guaranteed when the variability of the design matrix is sufficiently large, implying that the level of complexity of the functional parameter $\beta(s)$ we can possibly estimate is determined by the complexity of the input imaging data.

\begin{theorem}
\label{thm:beta_consistency}
Suppose Assumptions \ref{asm:true_func} - \ref{asm:W} hold in the outcome model \eqref{eq:model1} and the priors on $\tilde\bfgamma$ satisfy that  $\Pi(\|\tilde\bfgamma-\tilde\bfgamma_0\|_2^2<\epsilon)>0$ for any $\epsilon>0$. 
 Then for any $\epsilon>0$, we have,  as $n\to \infty$, 
$\Pi \left(\|\beta - \beta_0\|_{2,\pscale} + \|\tilde\bfgamma - \tilde\bfgamma_0\|_2>\epsilon \mid \bfY,\bfM, \bfX,\bfC \right) \to 0$
in $P_0^{n}$- probability. This implies that $$\Pi \left(\|\beta - \beta_0\|_{2,\pscale}>\epsilon \mid \bfY,\bfM, \bfX,\bfC \right) \to 0$$
in $P^{n}_0$- probability.
\end{theorem}

In the proof of Theorem \ref{thm:beta_consistency}, especially in constructing the test for $H_0: \beta(s)=\beta_0(s) \text{ v.s. } H_1:\|\beta-\beta_0\|_{2,\pscale}>\epsilon$ through the basis approximation of $\beta(s)$, verifying conditions 
in the Supplementary Material for $M_i(s)$ in model \eqref{eq:model2} provides insight into the relationship between models \eqref{eq:model1} and \eqref{eq:model2}: sufficient variability in $M_i(s)$ ensures posterior consistency of $\beta(s)$.

\begin{theorem}
\label{thm:total_consistency} (Posterior consistency of SVME) 
Under Assumptions \ref{asm:true_func} - \ref{asm:W},
for any $\epsilon>0$, as $n\to \infty$, 
$\Pi\left( \|\mathcal{E}-\mathcal{E}_0\|_{1,\pscale}<\epsilon \mid \bfY,\bfM, \bfX,\bfC\right)\to 1$
in $P^{n}_0$-probability.
\end{theorem}

This theorem implies that the posterior distribution of SVME concentrates on an arbitrarily small neighborhood of its true value with probability tending to one when the sample size goes to infinity. Here the sample size refers to the number of images $n$. By Assumption~\ref{asm:true_func}, in this case. the number of voxels $p$ also goes to infinity. This theorem also implies the consistency of estimating NIE using posterior inference by BIMA in the following corollary. 

\begin{corollary}
(Posterior consistency of NIE) For any $\epsilon > 0$, as $n \to \infty$, 
\begin{align*}
\Pi\left( \pscale^{-1}\left|\sum_{j=1}^{\pscale} \mathcal{E}(s_j) - \sum_{j=1}^{\pscale} \mathcal{E}_0(s_j) \right|<\epsilon \ \Bigg|\  \bfY,\bfM, \bfX,\bfC\right)\to 1
\end{align*} in $P^{n}_0$-probability. 
\end{corollary}

From Theorem~\ref{thm:total_consistency}, we can further establish the posterior sign consistency of SVME. Consider a minimum effect size $\delta > 0$, define $\cR^+_\delta =  \br{s:\mathcal{E}_0(s)>\delta}$ and $\cR^{-}_\delta = \br{s:\mathcal{E}_0(s)<-\delta}$, which represent the true positive SVME region and the true negative SVME region respectively. Let $\cR_0 =\{s: \mathcal{E}_0(s)=0\}$ represent a region of which the true SVME is zero.  

\begin{corollary} \label{corol:sign_consistency}
 (Posterior sign consistency of SVME)  For any $\delta>0$, let $\cR_{\delta} = \cR^+_{\delta} \cup \cR^-_{\delta}\cup \cR_0$, Then as $n\to \infty$,
$\Pi\left[\mathrm{sign}\{\mathcal{E}(s)\}=\mathrm{sign}\{\mathcal{E}_0(s)\}, \forall s\in\cR_{\delta}\mid\bfY,\bfM, \bfX,\bfC\right]\to 1$ 
in $P^{n}_0$-probability, where $\mathrm{sign}(x) = 1$ if $x>0$, $\mathrm{sign}(x) = -1$ if $x<0$ and $\mathrm{sign}(0) = 0$. 
\end{corollary}

This corollary ensures that with a large posterior probability BIMA can identify the important regions with significant positive and negative SVMEs that contributes the NIE.

\section{Posterior Computation}
\label{sec:computation}
The posterior computation for BIMA is challenging due to the complexity of the nonparametric inference, the high-dimensional parameter space and the non-conjugate prior specifications for the spatially-varying coefficients in the model. To address these challenges, we next construct an equivalent model representation. 
\subsection{Model representation and approximation}\label{subsec:computation_region_approx}
We approximate the STGPs and GPs using a basis expansion approach. By Mercer's theorem~ \citep{williams2006gaussian}, the correlation kernel function in \eqref{eq:priors} can be decomposed by infinite series of orthonormal basis functions $\kappa(s,s') = \sum_{l=1}^\infty\lambda_l \psi_l(s)\psi_{l}(s')$, and the corresponding GP $g(s)\sim \cGP(0,\sigma_g^2\kappa)$ can be expressed as $g(s) = \sum_{l=1}^\infty \theta_{g,l}\psi_l(s)$ where $\theta_{g,l}\overset{\ind}{\sim}N(0,\lambda_l\sigma_g^2)$.

In our implementation, we allow region partition to speed up the computation, and assume a region-independence prior kernel structure for the spatially varying parameters $\beta,\alpha,\zeta_k, \eta_i$. In real data analysis, the brain anatomic region parcellation defines the region partition. Assume there are $r = 1,\dots, R$ regions that form a partition of the support $\cS$, denoted as $\cS_1,\dots, \cS_R$. The kernel function $\kappa(s_j, s_k) = 0$ for any $s_j\in \cS_r, s_k\in\cS_{r'}, r\neq r'$, and the prior covariance matrix on the fixed grid has a block diagonal structure. For the whole brain analysis as one region, one can choose $R=1$.

For the $r$-th region, let $p_r$ be the number of voxels in $\cS_r$, $Q_r = \sbr{\psi_l(s_{r,j})}_{l=1,j=1}^{L_r,p_r} \in \R^{p_r\times L_r}$ be the matrix with the $(j,l)$-th component $\psi_l(s_{r,j})$, $\br{s_{r,j}}_{j=1}^{p_r}$ forms the fixed grid in $\cS_r$. Because of the basis approximation with cutoff $L_r$, $Q_r$ is not necessarily an orthonormal matrix, hence we use QR decomposition to get an approximated orthonormal $Q_r$, i.e. $Q_r^T Q_r = I_{L_r}$, where $I_{L_r}$ is the identity matrix. With the region partition, the GP priors on the $r$-th region can be approximated as $g_r = \sbr{g(s_{r,1}),\dots,g(s_{r,p_r})}^T \approx Q_r\bftheta_{g,r}$, where $\bftheta_{g,r}\sim \cN\sbr{0,\sigma_g^2 D_r}$, $D_r$ is a diagonal matrix with eigen-values $(\lambda_{r,1},\dots,\lambda_{r, L_r})^\rT\in \R^{L_r}$.

After truncating the expansion at sufficiently large  $\br{L_r}_{r=1}^L$, STGPs and GPs in the prior specifications~\eqref{eq:priors}, which all share the same kernel, can be approximated by $\beta_r = T_{\sthresh}(\tilde\beta_r) \approx T_{\sthresh}\left( Q_r\bftheta_{\tilde\beta,r}\right)$, $\alpha_r = T_{\sthresh}(\tilde\alpha_r) \approx T_{\nu}\left( Q_r\bftheta_{\tilde\alpha,r}\right)$, $\zeta_{k,r} \approx Q_r\bftheta_{\zeta,k,r}$ and $\eta_{i,r} \approx  Q_r\bftheta_{\eta,i,r}$
where the corresponding basis coefficients follow independent normal priors: $\bftheta_{\tilde\beta,r}\sim \mN_{L_r}(0, \sigma^2_\beta D_r)$, $\bftheta_{\tilde\alpha,r}\sim \mN_{L_r}(0, \sigma^2_\alpha D_r)$, $\bftheta_{\zeta,k,r}\sim \mN_{L_r}(0, \sigma^2_\zeta D_r)$ and $\bftheta_{\eta,i,r}\sim \mN_{L_r}(0, \sigma^2_\eta D_r)$. 
We discuss the details for choosing $L_r$ in Section 4.2. Denote  $\mcdf_{i}(\cS_r) = \sbr{\mcdf_{i}(\vox_{r,j})}_{j=1}^{p_r}\in \R^{p_r}$, $\mpdf_{i}(\cS_r) = \sbr{\mpdf_{i}(\cvox_{r,j})}_{j=1}^{p_r}\in \R^{p_r}$,
Then the BIMA model can be approximated as follows: $Y_i = \sum_{r=1}^{R} T_\sthresh\sbr{Q_r\bftheta_{\tilde\beta,r}} \mcdf_{i}(\cS_r) + \gamma X_i + \bfzeta^{\rT}_Y \bfC_i + \epsilon_{Y,i}$ and 
$\mpdf_{i}(\cS_r) = T_\sthresh\sbr{Q_r\bftheta_{\tilde\alpha,r}}  X_i + \sum_{k=1}^q Q_r\bftheta_{\zeta,k,r} C_{i,k} + Q_r\bftheta_{\eta,i,r}+ \epsilon_{M_r,i}$, 
where $\epsilon_{Y,i}\sim \mN(0,\sigma^2_Y)$ and  $\epsilon_{M_r,i} \sim \mN_{p_r}(0,\sigma_M^2I_{p_r})$.
From the above model representation, both $\bftheta_{\zeta,k,r}$ and $\bftheta_{\eta,i,r}$ have conjugate posteriors, but $T_\sthresh$ is not a linear function, and $\bftheta_{\tilde\beta,r}$ and $\bftheta_{\tilde\alpha,r}$ do not have conjugate posteriors. To overcome this, the Metropolis-adjusted Langevin algorithm (MALA) is used to sample $\bftheta_{\tilde\beta,r}$ and $\bftheta_{\tilde\alpha,r}$. However, the first-order derivative of the soft-thresholded function $T_\sthresh(x)$ does not exist at the two change points $x=\pm \sthresh$. To approximate the first-order derivative, either the derivative of a smooth function approximation or an indicator function approximation: $d \hat T_\sthresh(x)= I(|x|\geq \sthresh)$ works in our case. The latter one provides better computational efficiency, and is implemented in our algorithm. 
For both models (1) and (2), most of MALA’s cost comes from computing the log-posterior gradient, particularly the log-likelihood component. In model (1), the gradient involves a term 
$Q_r^\top \hat{D}_r \bm{\mcdf}_r \hat{\epsilon}_{\bfY,r}$, 
yielding $O(L_r\,p_r\,n)$ complexity per region $r$, where $L_r$ is the number of basis functions, $p_r$ is the number of spatial locations, and $n$ is the sample size. Because $\hat{D}_\nu$ is diagonal (and often sparse), the actual cost is closer to $O(L_r\,m_{\beta,r}\,n)$, where $m_{\beta,r}$ counts the nonzero elements of $\beta$.
Model (2) has the same $O(L_r\,m_{\alpha,r}\,n)$ cost for updating $\alpha(s)$ but introduces extra parameters $\bftheta_{\zeta,k,r}$ and $\bftheta_{\eta,i,r}$. The Gibbs update for $\bftheta_{\zeta,k,r}$ is $O(L_r)$ due to $Q_r$’s orthonormality. In contrast, the hyperplane MVN algorithm for $\bftheta_{\eta,i,r}$ requires $O\bigl(\max\{(q+1)n^2,\,(n-q-1)^2n\}\bigr)$, dominated by $n$ when $q \ll n$. Thus, the complexity in region $r$ is
$O\bigl(\max\{L_r\,m_{\alpha,r}\,n,\;L_r\,(n-q-1)^2\,n\}\bigr).$
Hence, updating $\bftheta_{\eta,i,r}$ typically poses the main computational bottleneck, especially for sparse signals where $m_{\alpha,r} \ll n$. 

\subsection{Covariance kernel specifications and estimation }

We can choose different covariance kernels for the GPs in models \eqref{eq:model1} and \eqref{eq:model2}. Given the covariance kernel function $\kappa(\cdot,\cdot)$, to obtain the coefficients $\lambda_l$ and the basis functions $\psi_l(s)$, Sections 4.3.1 and 4.3.2 in \cite{williams2006gaussian}  provide the analytic solution for squared exponential kernel, and an approximation method for other kernel functions with no analytic solutions. In practice when $\psi_l(s)$ has no analytical solutions, such as the Mat\'ern kernel, we use eigen decomposition on the covariance matrix, and take the first $L$ eigenvalues as the approximated $\lambda_l$, the first $L$ eigenvectors as the approximated $\psi_l(s)$, then apply QR decomposition on the approximated basis functions to obtain orthonormal basis. The limitation of this method is that the covariance matrix is difficult to compute in high dimensions due to precision issues. Hence in high dimensions we split the entire space $\cS$ into smaller regions, and compute the basis functions on each region independently. This also aligns with the imaging application with the whole brain atlas. Another benefit is that by splitting the whole parameter space into smaller regions, the sampling space gets smaller and it becomes easier to accept the proposed vector $\beta(s)$ on each region with much less directions to explore. In practice, to choose the number of basis functions $L_r$ for region $r$ with $p_r$ voxels, we first compute the covariance matrix in $\R^{p_r\times p_r}$ with appropriately tuned covariance parameters, get the eigen-value of such covariance matrix, and choose the cutoff such that the summation $\sum_{l=1}^{L_r}\lambda_l$ is over $90\%$ of $\sum_{l=1}^{p_r}\lambda_l$, i.e. the eigenvalues before cutoff account for over $90\%$ of the total eigenvalues. We provide the detailed sensitivity analysis on choosing the covariance parameters in the Supplementary Material. 

\subsection{The MCMC algorithm}\label{sec:MCMC_algo}
We develop an efficient Markov chain Monte Carlo (MCMC) algorithm for posterior computation. 
To update parameters~$\br{\bftheta_{\tilde\beta,r},\bftheta_{\tilde\alpha,r}}_{r=1}^R$,
 we adopt the Metropolis-adjusted Langevin algorithm (MALA). The step size is tuned during the burn-in period to ensure an acceptance rate between 0.2 and 0.4. The target acceptance rate for each region is set to be proportional to the inverse of the number of basis functions in that region, in order to produce a relatively large effective sample size of the MCMC sample. 

To incorporate the block structure with MALA, in each iteration, the proposal $\bftheta_{\tilde\beta,r}$ or $\bftheta_{\tilde\alpha,r}$ for region $\cS_r$ is based on the target posterior density conditional on $\bftheta_{\tilde\beta,r'}$ or $\bftheta_{\tilde\alpha,r'}$ supported on all other regions where $r'\neq r$. The acceptance ratio is also computed region by region.

MALA has a considerable computational cost especially in high dimensional sampling, where the step size has to be very small to have an acceptance rate reasonably greater than 0. It is important to have a good initial value. To obtain the initial values, we consider a working model with the spatially varying coefficients $\beta(s)$ and $\alpha(s)$ following GP instead of STGP. With the basis expansion approach, we can straightforwardly use Gibbs sampling to obtain the approximated posterior samples of $\beta(s)$ and $\alpha(s)$ of the working model. The posterior mean values of $\beta(s)$ and $\alpha(s)$ estimated from the working model can be used to specify the initial value of the basis coefficients in the MALA algorithm. More detailed discussion on choosing the initial value can be found in Supplementary Material Section \ref{supp_sec:additional_rda}.

To impose identifiability Assumption \ref{asm:alpha_iden}, the posterior of $\theta_{\eta,i,l}$ is sampled from a constrained multivariate normal distribution, with the constraint $\tilde\bfX^\rT\bm{\theta_{\eta,l}}=\bm{0}$ where $\bm{\theta_{\eta,l}} = (\theta_{\eta,1,l},\dots,\theta_{\eta,n,l})^\rT$. The algorithm for sampling multivariate normal distribution constrained on a hyperplane follows Algorithm 1 in \citep{cong2017fast}.

For the rest of the parameters, with available conjugate full conditional posteriors, we use  Gibbs sampling to update. The algorithm is implemented in Rcpp \citep{Rcpp} with RcppArmadillo \citep{RcppArmadillo}. The implementation is wrapped as an R package \texttt{BIMA}. 
\hidetext{\footnote{Available on Github \url{https://github.com/yuliangxu/BIMA}}}

\section{Simulations}
\label{sec:simulation}

To demonstrate the performance of BIMA, we first compare it with existing Bayesian mediation methods in terms of selection and estimation accuracy as well ass computing time and algorithm stability in Section \ref{subsec:comparison}. In Section~\ref{subsec:comp_sens}, we focus on evaluating computational scalability and robustness of BIMA in high-dimensional settings where most existing Bayesian methods are not applicable.  We vary the sample size, noise variance, and image patterns, and conduct a sensitivity analysis on the performance of BIMA under different settings and prior specifications. We also compare BIMA with BI-GMRF~\citep{wang2023high} in one special high dimensional setting considered in~\cite{wang2023high}. The results are reported in the Section~\ref{sec:compare_BIGMRF} of the Supplementary Material.

\subsection{Comparison with existing Bayesian methods}\label{subsec:comparison}
 In this section, we compare BIMA with two recently proposed Bayesian methods: product threshold Gaussian prior~\citep[PTG]{song2020ptg} and Correlated Selection Model~\citep[CorS]{song2020cors}.

PTG constructs priors by thresholding bivariate Gaussian latent vectors and using their product to induce sparsity in the model parameters. While PTG effectively controls sparsity, it does not account for spatial correlations between locations, making it less suitable for spatially correlated data like brain imaging. CorS uses a four-component mixture model to specify different sparsity patterns in model parameters and incorporates spatial correlations in the prior specifications. This approach is more appropriate for spatial applications, as it models correlations between locations through the inclusion of spatially varying mixing weights. The detailed model specifications for PTG and CorS are provided in Section~\ref{sec:PTG_CorS_Settings} of the Supplementary Material. 

In this simulation, BIMA adopts a modified square-exponential kernel 
$\kappa(s,s'; a, b) = \mathrm{cor}\{\beta(s),\beta(s')\} =  \exp\{-a (s^2 + {s'}^2) - b \|s-s'\|^2 \}$ with
 $a=0.01$ and $b=10$. We split the input image into four regions. We use Hermite polynomials up to the 10th degree, resulting in 66 basis coefficients to approximate each region. The initial values for all parameters are obtained from Gibbs sampling with Gaussian process priors for $\alpha$ and $\beta$. The threshold parameter $\sthresh=0.5$ in STGP priors. For the outcome model \eqref{eq:model1}, a total of $10^5$ iterations are performed, with the acceptance probability tuned to be around 0.2 for each region during the first 80\% of burn-in iterations. The mediator model \eqref{eq:model2} follows the same setting, except with a total of 5000 iterations and a burn-in period comprising the first 90\%.

\begin{figure}[htb]
\begin{center}
\includegraphics[width=0.7\textwidth]{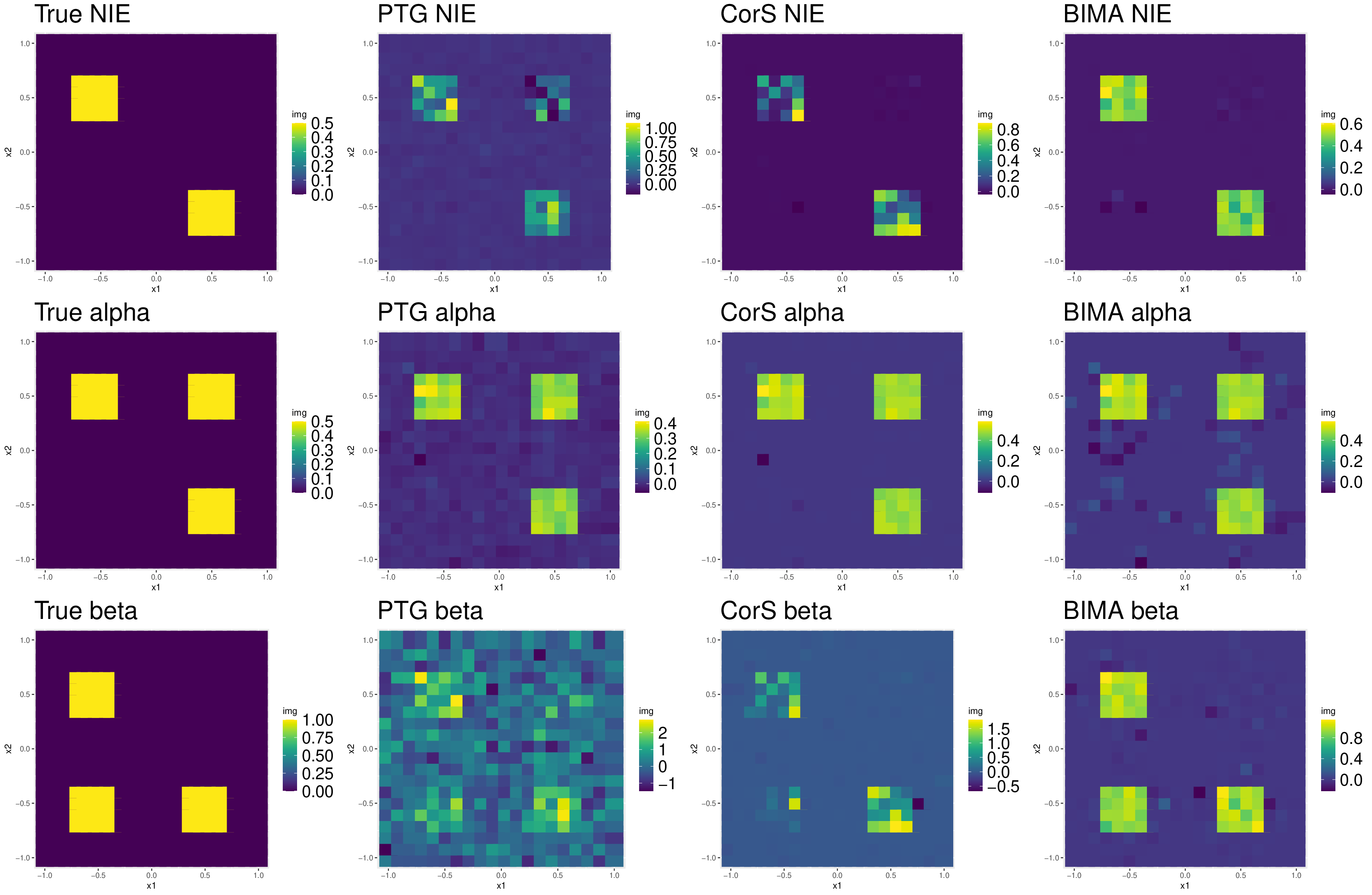}
\caption{Comparison on the posterior mean of the 3 methods with the true images. Rows from top to bottom represent functional NIE $\mathcal{E}(s)$, $\alpha(s)$, $\beta(s)$. Columns from left to right represent true images, posterior mean from PTG model, posterior mean from CorS model, posterior mean from BIMA model.}
\label{fig:comparison}
\end{center}
\end{figure}

Figure \ref{fig:comparison} shows the true image for $\alpha(s)$, $\beta(s)$, and $\mathcal{E}(s)$, i.e. NIE. Table \ref{tb:comparison} summarizes posterior samples of NIE using three methods with 100 replicated simulations. The final result of NIE is tuned using the inclusion probability of the sampled NIE for all 3 methods in the following way: for each location $s_j$, we estimate the empirical probability $\hat P(\nie(s_j) \neq0)$ from the MCMC sample of NIE, and set a threshold $t$ on  $\hat P(\nie(s_j) \neq0)$: if $\hat P(\nie(s_j) \neq 0)<t$, $\nie(s_j)=0$, otherwise $\nie(s_j)$ equals the posterior sample mean. By tuning $t$, we can control the FDR to be below 10\%. Although we set the target FDR to be 10\% for all 3 methods, it is still possible that FDR cannot be tuned to be less than 10\% with any $t<1$ when the sample is very noisy, in which case the largest possible $t$ is used, and the tuned FDR can be larger than 10\%. In the extreme case where the largest possible $t$ still maps all location to 0, we get the NAs as shown in Table \ref{tb:comparison}. These NA replication results are excluded from the summary statistics in Table \ref{tb:comparison}.

\renewcommand{\arraystretch}{0.85}
\begin{table}[ht]
\small
\centering
\caption{Comparison of posterior inferences on NIE among different methods including PTG, CorS and BIMA based on 100 replications. The standard errors are reported in the brackets}
\label{tb:comparison}
\begin{subtable}[ht]
{\columnwidth}
\centering
\caption{Selection accuracy including the overall accuracy (ACC),  false discovery rate (FDR) and true positive rate (TPR). All values are in percentage.} 
\begin{tabular}{c|ccc|ccc|ccc}
\toprule
\multicolumn{10}{c}{Selection Accuracy}      \\
\midrule                                                                                                  & \multicolumn{3}{c|}{PTG}          & \multicolumn{3}{c|}{CorS}                               & \multicolumn{3}{c}{BIMA}                                              \\
$(n, p)$    & FDR      & TPR    & ACC  & FDR  & TPR & ACC                                & FDR                                     & TPR       & ACC      \\\midrule
$(200,400)$ & 9 (15)  & 20 (19) & 93 (1) & 1 (2) & 80 (37) & 98 (3)  & 7 (3) & 95 (3) & 99 (0) \\
$(300,400)$ & 21 (21) & 16 (14) & 93 (1) & 1 (2) & 100 (0) & 100 (0) & 6 (3) & 93 (5) & 99 (0) \\
$(200,676)$ & 14 (14) & 11 (12) & 93 (1) & 0 (0) & 3 (2)   & 93 (0)  & 8 (2) & 96 (3) & 99 (0) \\
$(300,676)$ & 10 (14) & 17 (11) & 94 (1) & 1 (1) & 80 (36) & 98 (3)  & 7 (2) & 96 (3) & 99 (0) \\

\bottomrule
\end{tabular}
\end{subtable}

\bigskip

\begin{subtable}[ht]{\columnwidth}
\caption{Estimation and computation performance including mean squared errors (MSE) in the true activation region (multiplied by 100) and computation time in seconds. }
\begin{tabular}{c|ccc|cccc|cc}
\toprule
\multicolumn{10}{c}{Estimation and Computation time}    \\
\midrule
              & \multicolumn{3}{c|}{MSE (Activation)} & \multicolumn{4}{c|}{Time (Seconds)}                                                               & \multicolumn{2}{c}{\#of NA} \\
$(n, p)$        & PTG      & CorS     & BIMA       & PTG  & CorS  & BIMA \eqref{eq:model1} & BIMA \eqref{eq:model2} & PTG         & CorS          \\\midrule
$(200,400)$ & 24 (1) & 5 (10) & 2 (1) & 251 (7)   & 26 (3) & 27 (2)  & 28 (1) & 31 & 7  \\
$(300,400)$ & 24 (1) & 0 (0)  & 2 (1) & 385 (8)   & 25 (2) & 35 (3)  & 61 (1) & 22 & 0  \\
$(200,676)$ & 24 (0) & 25 (1) & 2 (1) & 663 (13)  & 75 (1)  & 54 (6)  & 35 (1) & 60 & 60 \\
$(300,676)$ & 24 (0) & 5 (9)  & 1 (1) & 1026 (21) & 76 (2)  & 64 (11) & 71 (2) & 21 & 11 \\
\bottomrule
\end{tabular}

\end{subtable}

\end{table}

From Table \ref{tb:comparison}, PTG performs the least ideal in the correlated image setting as shown in Figure \ref{fig:comparison}, especially in the estimation for the mediator effect $\beta(s)$. In general, $\beta(s)$ is more challenging to estimate than $\alpha(s)$ for two reasons: i) The mediator model \eqref{eq:model2} has $n\times \pscale$ observations to estimate $\pscale$ dimensional $\alpha(s)$, leading to a higher signal to noise ratio than $\beta$ in model \eqref{eq:model1}; ii) In the outcome model \eqref{eq:model1}, $M$ and $X$ are correlated through \eqref{eq:model2}, making it more difficult to separate the effect $\beta(s)$ from $\gamma$.

CorS model performs very well when $n$ is close to $p$. However in the higher dimensional setting, when $n$ is much smaller than $p$, CorS has a lower power than BIMA. BIMA performs well and is stable across all four settings, indicating that it is a suitable method especially for high-dimensional spatially correlated mediators, when $n$ is considerably less than $p$, such as in brain imaging application. Potential improvement can be made for BIMA when the kernel bases are tuned to accurately represent the smoothness of input mediators.

\begin{figure}[ht]
\begin{center}
\includegraphics[width=0.6\textwidth]{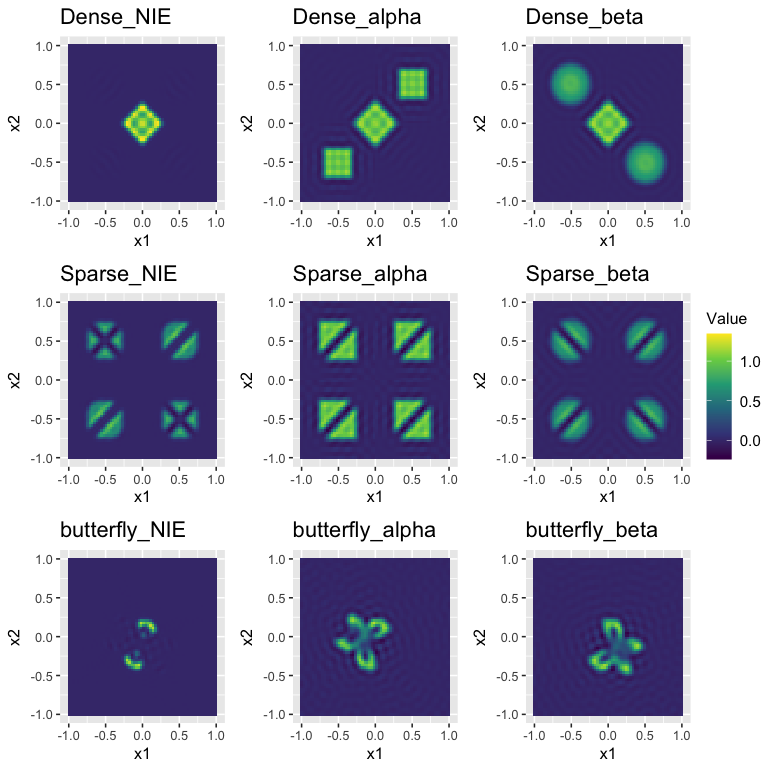}
\caption{Input image pattern for the simulation study. Rows from top to bottom represent dense, sparse, and butterfly patterns. Columns from left to right represent input image NIE, $\alpha(s)$, $\beta(s)$. $p=4096$.}
\label{fig:pattern}
\end{center}
\end{figure}

\subsection{Computational scalability and sensitivity analysis}\label{subsec:comp_sens}
To further illustrate the performance of our proposed method, we conduct simulation studies under various settings with three sets of patterns as shown in Figure \ref{fig:pattern}. For dense and sparse patterns, each image is split into 4 regions, each region being a $32\times 32$ grid. For the butterfly pattern, the entire image is one region of size $64\times 64$, with no region split. The threshold parameter $\sthresh=0.5$ in STGP priors. In this simulation, we use the Mat\'ern kernel in accordance with the sharp patterns in Figure \ref{fig:pattern}. 
\begin{align}\label{eq:matern}
    \kappa(s',s;u,\rho) = C_u(\|s'-s\|_2^2/\rho), ~ C_u(d):= \frac{2^{1-u}}{\Gamma(u)}\left( \sqrt{2u} d \right)^u K_u(\sqrt{2u d})
\end{align}
The number of basis for each region is set to be 20\% of the region size. The scale parameter $\rho=2$, and $u=1/5$. Due to the high dimension of mediators, we let the MALA algorithm update only $\beta(s)$ for the first 40\% of MCMC iterations to get $\beta(s)$ to a stable value, then jointly updating all other parameters in \eqref{eq:model1} using Gibbs Sampling. All other settings are the same as in Section \ref{subsec:comparison}, and the summary statistics for NIE in Table \ref{tb:simulation} are also tuned in the same way using inclusion probability. Table \ref{tb:simulation}(b) gives a sensitivity analysis result using different thresholds $\sthresh$ in the STGP priors to show that the estimation is not  too sensitive to the choice of $\sthresh$ within a small range.

\begin{table}[ht]
\begin{center}
\caption{Computational scalability and sensitivity analysis results. Selection accuracy  (multiplied by 100) includes  false discovery rate (FDR), true positive rate (TPR) and overall accuracy (ACC). Computational time (in minutes) are separately reported for fitting model \eqref{eq:model1} (T1) and model \eqref{eq:model2} (T2). 
The reported values are the average over 100 replications. The standard deviations are reported in the brackets.} 
\label{tb:simulation}
\begin{subtable}[ht]
{\columnwidth}
\centering
\caption{Peformance of BIMA in simulations for different sample sizes ($n$) and the random noise standard deviations in model~\eqref{eq:model1} ($\sigma_Y$).}
\begin{tabular}{*{8}{l}}
\toprule
\multicolumn{8}{c}{Under different generative model, $\sthresh=0.5$} \\
\cmidrule(r){1-8}
Pattern & $n$ & $\sigma_Y$ & FDR & TPR & ACC & T1 & T2 \\
\midrule
Dense & 1000 & 0.1 & 1 (1) & 98 (1) & 100 (0) & 13 (2) & 184 (16) \\
\textbf{Sparse} & 1000 & 0.1 & 3 (3) & 100 (0) & 100 (0) & 19 (4) & 212 (28) \\
Dense & \textbf{5000} & 0.1 & 1 (2) & 97 (4) & 100 (1) & 54 (15) & 1145 (140) \\
Dense & 1000 & \textbf{0.5} & 1 (1) & 98 (1) & 100 (0) & 17 (5) & 209 (37) \\
\textbf{Butterfly}, $\sthresh =0.2$ & \textbf{500} & 0.1 & 6 (6) & 98 (1) & 100 (0) & 20 (4) & 373 (38) \\
\bottomrule
\end{tabular}
\end{subtable}
\bigskip

\begin{subtable}[ht]
{\columnwidth}
\centering
\caption{Sensitivity analysis with different threshold values ($\sthresh$).}
\begin{tabular}{cccccc}
\toprule
\multicolumn{6}{c}{Under different sensitivity parameter $\sthresh$.}         \\
\multicolumn{6}{c}{Dense pattern, $n=1000$, $\sigma_Y=0.1$.}                  \\
\cmidrule(r){1-6}
$\sthresh$                    & FDR   & TPR     & ACC     & T1     & T2       \\
\midrule
\textbf{0.3} & 5 (1) & 99 (0)  & 99 (0)  & 20 (5) & 198 (18) \\
\textbf{0.6} & 0 (0) & 97 (10) & 100 (1) & 17 (4) & 222 (28) \\
\bottomrule
\end{tabular}
\end{subtable}
\end{center}
\end{table}

Table \ref{tb:simulation} demonstrates that our proposed method has stable performance across different settings. In Table \ref{tb:simulation}, the mediator model is fully updated and converged including all individual effects $\sbr{\eta_i}_{i=1}^n$. Fully updating $\sbr{\eta_i}_{i=1}^n$ can take much longer time for the entire model to converge compared to directly setting the individual effects all to 0. In the case all $\eta_i$ fixed at 0, the estimation for $\alpha$ and $\bfzeta$ are almost the same compared to updating the full model from the $p=4,096$ simulation studies that we have observed.
When $n=1,000$ and $p=4,096$,   the computational time of fitting BIMA with running $30,000$ MCMC  iterations is less than four hours for both models \eqref{eq:model1} and \eqref{eq:model2}. In comparison, the CorS method takes 9.8 hours when $n=1,000,p=2,000$, with $1.5\times 10^5$ iterations. Our approach has significantly better computational efficiency in this high-dimensional setting.

\section{Analysis of ABCD fMRI Data}
\label{sec:realdata}

In this section, we apply BIMA to analyze the fMRI data in the Adolescent Brain Cognitive Development (ABCD) study Release 1.0 \citep{casey2018ABCD}. The 2-back 3mm task fMRI contrast map ($61 \times 73\times 61$) is used, and the preprocessing steps are described in \cite{sripada2020toward}. After preprocessing and removing missing data, the final complete data set consists of $n= 1,861$ subjects.  We focus on the first 90 Automated Anatomical Labeling~\citep[AAL]{rolls2020automated} regions, as they primarily include cortical areas critical to cognitive functions such as working memory, while regions 91 to 116, which are subcortical and cerebellar, are less consistently implicated in working memory tasks. Thus, the number of voxels in the brain image mediator for our anlaysis is $p=47,636$. 

We aim at examining the natural indirect effect (NIE) of parental education level on children's general cognitive ability scores, mediated through brain imaging data. We explore the varying roles of different brain regions as mediators in the development of cognitive ability of a child. Hence the exposure is a binary variable indicating whether the parent has a college or higher degree. The outcome variable is the g-score that reflects a child's general cognitive ability~ \citep{sripada2020toward}. The confounders in our model include age, gender, race and ethnicity, and household income. For the multi-level variables race and ethnicity (Asian, Black, Hispanic, Other, White), household income (less than 50k, between 50k and 100k, greater than 100k), we use binary coding for each level. Table \ref{tb:summary_ABCD} provides the summary statistics of the ABCD data.

 In the context of the ABCD study, the causal inference assumptions [A1]--[A4] in Section \ref{subsec:causal_mediation} can be interpreted as follows: [A1] given the observed confounders, no unobserved factors influence the relationship between parental education and children's general cognitive ability scores; [A2] after accounting for observed confounders and parental education, no additional confounders affect the relationship between children's brain image intensity and general cognitive ability scores; [A3] given the observed confounders, no unmeasured factors influence the relationship between children's brain image intensity and parental education level; and [A4] assuming [A2] holds, no value of parental education can alter the relationship between children's working memory task activity and their general cognitive ability, once observed confounders are considered.  Please refer to additional discussions of these causal assumptions including the SUTVA assumption and its interpretation in the Supplementary Section \ref{sec:app_causal_interpretation}.

In Section \ref{supp:sec:SA} of the the Supplementary Material, we provide a sensitivity analysis algorithm, along with the result of NIE and NDE when a single binary unmeasured confounder has different levels of effect on the outcome and the mediator, where these unmeasured confounding effects are assumed to be spatially constant. Treating unobserved confounders in high-dimensional mediation problems is still an open research area. We refer readers to a further discussion on other approaches to account for unobserved confounders in this section.

In this analysis, we use the Mat\'ern kernel where the hyper-parameters $u$ and $\rho$ are specified for each region according to the estimated covariance matrices. The number of voxels for each region varies from $62$ to $1,510$. To determine the number of basis, we select up to $500$ locations within a certain range of the centroid for each region. Using these locations, we compute the empirical covariance matrix for each region. The cutoff for the number of basis is then chosen in such a way that it accounts for 90\% of the total sum of all the singular values of the estimated covariance matrix. Because the hyper-parameter $\sthresh$ in the STGP prior and the kernel parameters $u, \rho$ in each region are all prefixed, we provide a detailed description of selecting these parameters via testing MSE in the Supplementary Material. The final threshold $\sthresh_\beta$ for $\beta(s)$ is set to be 0.05, and the final threshold $\sthresh_\alpha$ for $\alpha(s)$ is set to be 0.1. The choice of $\nu$ is also based on testing MSE. Detailed sensitivity analysis can be found in the Supplementary Material.

\begin{figure}
\centering
\begin{tabular}{c}
Posterior inclusion probability (color range $[0.1,0.5]$)\\
\includegraphics[width=0.9\textwidth]{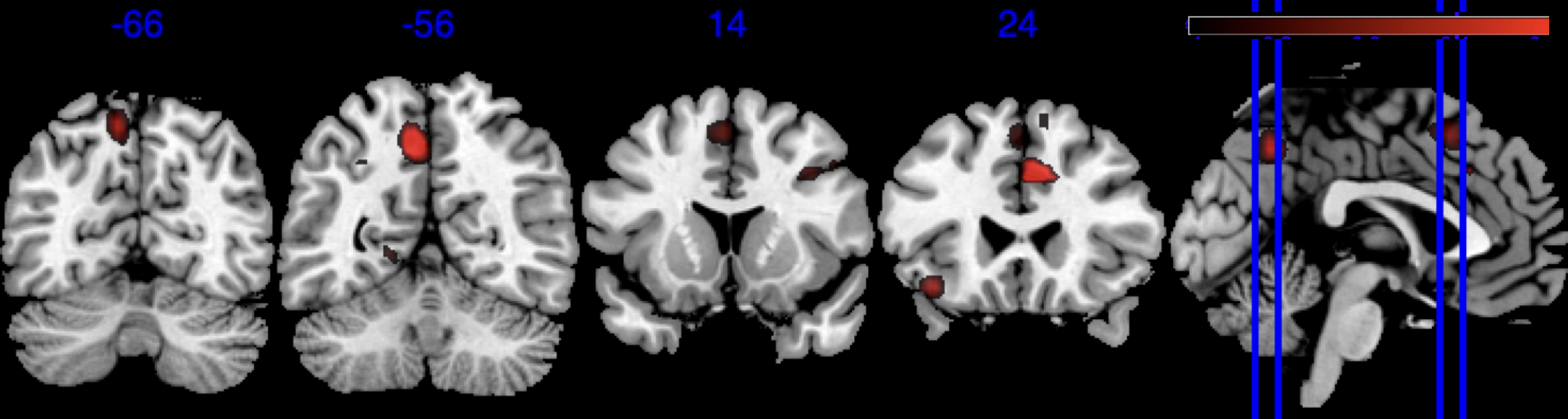}\\
Positive posterior mean of the spatial mediation effects (color range $[10^{-5}, 10^{-3}]$)\\
\includegraphics[width=0.9\textwidth]{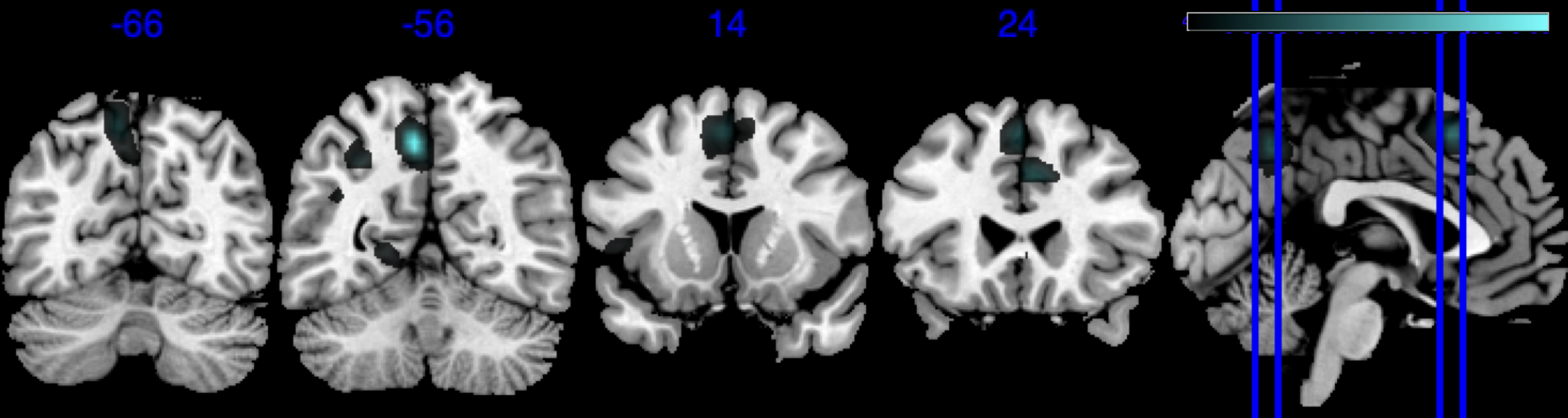}\\
Negative posterior mean of the spatial mediation effects (color range $[-10^{-4}, -10^{-5}]$)\\
\includegraphics[width=0.9\textwidth]{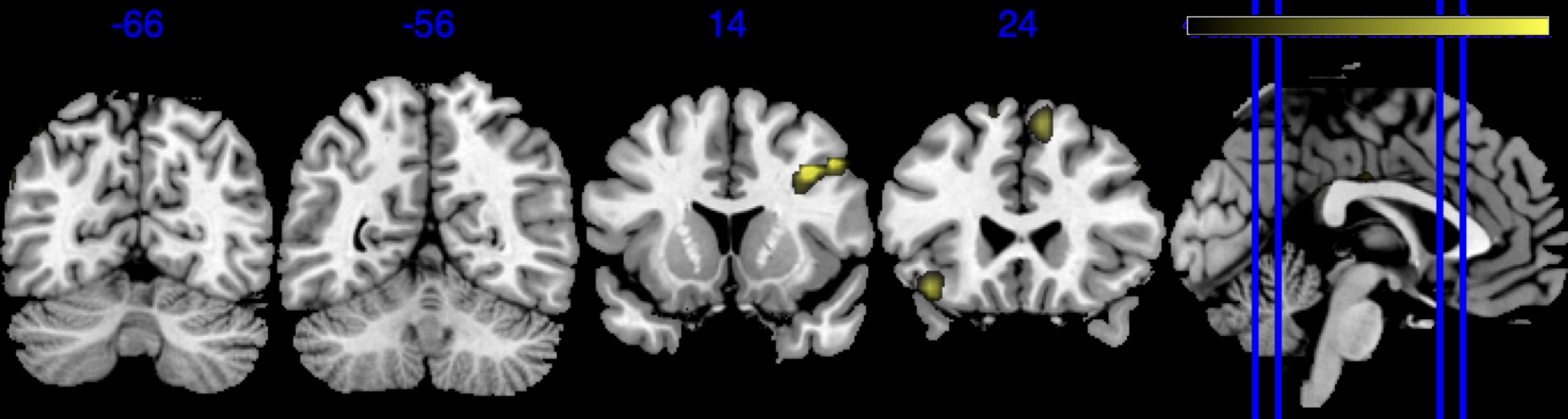}

\end{tabular}

\caption{Posterior inference on spatially varying indirect effects of parental education on the general cognitive ability that are mediated through the working memory brain activity. The Coronal view slides cutting through 3 of the top 10 regions with largest number of active pixels: the left Precuneus (Precuneus\_L), left Inferior parietal gyrus (Parietal\_Inf\_L) and the left Supplementary motor area (Supp\_Motor\_Area\_L).}
\label{fig:total_mean}
\end{figure}

We performed 100,000 iterations for the outcome model \eqref{eq:model1}, discarding the first 50\% as burn-in and thinning to retain 1,000 posterior samples. For the mediator model \eqref{eq:model2}, we ran 40,000 iterations with a 30,000 burn-in, thinning every 10 iterations to obtain 1,000 posterior samples.  Table \ref{tb:top10} gives a summary of both the overall NIE and NDE and the top seven regions identified with the largest number of active voxels. The definition of NIE in each region is $p^{-1}\sum_{s\in \cS_r}\beta(s)\alpha(s)$, where $\cS_r$ is the collection of all voxels in region $r$. The rule for selecting the active voxels is based on cutting the posterior inclusion probability (PIP) at 50\%, and the three regions with active voxels are reported in Table \ref{tb:top10}. Due to the very small effect sizes and low signal-to-noise ratio, we also include regions with voxels' PIP greater than 10\%.
The posterior of NDE $\gamma$ has a mean of 0.27 with the 95\% credible interval $(0.20,0.36)$. The posterior of NIE $\mathcal{E}$ has a mean of 0.0885 with the 95\% credible interval $(0.066,0.111)$. The total effect of parental education level on general cognitive ability score is 0.36, with 95\% credible interval (0.29,0.45). This suggests that parents with college degrees have a positive impact on children's cognitive abilities, and about 25\% of the effect is mediated through brain cognitive development. 
Figure \ref{fig:total_mean} shows the estimated activation regions and the NIE in coronal view slides. Among the top identified activation regions, the most interesting is the left precuneus, which plays a key role in episodic memory, visuospatial processing, and self-consciousness \citep{lou2004parietal, wallentin2006parallel}. This region has been consistently implicated in cognitive processes related to memory retrieval and spatial awareness, which are crucial components of children's cognitive development. In addition, other identified regions, such as the left inferior parietal region and the left postcentral gyrus, are associated with the interpretation of sensory information \citep{radua2010neural, diguiseppi2023neuroanatomy}. These regions are involved in integrating and processing sensory inputs, which are essential for tasks that require coordination between perception and cognition, such as working memory and executive function. These findings align with existing literature on the neural correlates of cognitive function, particularly in children. By identifying regions that have been consistently associated with cognitive processes, our results not only demonstrate the scientific validity of the BIMA approach but also provide meaningful insights into the brain areas that underlie cognitive abilities as captured by the ABCD study.

\begin{table}[h!]
\caption{Top regions ordered by the number of active voxels with $PIP > 50\%$ or $PIP > 10\%$. Columns 2 to 5 are timed by 100. NIE(+) and NIE(-) are defined as $p^{-1}\sum_{s\in\mathcal{r}_r}\mathcal{E}(s)I(\mathcal{E}(s)>0)$ and $p^{-1}\sum_{s\in\mathcal{r}_r}\mathcal{E}(s)I(\mathcal{E}(s)<0)$ for each region $r$. Average IP is the averaged inclusion probability over all voxels in the entire region. }
\centering
\resizebox{\columnwidth}{!}{%
\begin{tabular}{lllllll}
\toprule
\textbf{}                        & \textbf{NIE}               & \textbf{NIE(+)}               & \textbf{NIE(-)}                & \textbf{NDE}               & \textbf{\begin{tabular}[c]{@{}l@{}}Time (hours)\\ model \eqref{eq:model1}\end{tabular}} & \textbf{\begin{tabular}[c]{@{}l@{}}Time (hours) \\ model \eqref{eq:model2} \end{tabular}} \\
\textbf{Overall}               &8.85 & 10.57 & -1.72 & 27.37            & 1.60                                                               & 85.93      \\  \midrule      
\multicolumn{7}{c}{\textbf{PIP\textgreater{}50\%}}                                                                                                                                                                                                                                                                       \\
\textbf{Region Name (AAL Atlas)} & \textbf{NIE}               & \textbf{NIE(+)}               & \textbf{NIE(-)}            & \textbf{\begin{tabular}[c]{@{}l@{}}Average \\ PIP\end{tabular}}           & \textbf{\begin{tabular}[c]{@{}l@{}}\# of active\\  voxels\end{tabular}}        & \textbf{\begin{tabular}[c]{@{}l@{}} Region   \\ Size \end{tabular}}       \\
Cingulum\_Mid\_R        & 0.18  & 0.18   & 0.00   & 0.90                                                   & 55                                                                                              & 605                                                                                              \\
Precuneus\_L            & 0.35  & 0.35   & 0.00   & 0.50                                                   & 34                                                                                              & 1079                                                                                             \\
Parietal\_Inf\_L        & 0.28  & 0.28   & 0.00   & 0.57                                                   & 17                                                                                              & 696                                                                                              \\
\midrule
\multicolumn{7}{c}{\textbf{PIP\textgreater{}10\%}}               \\
\textbf{Region Name (AAL Atlas)} & \textbf{NIE}               & \textbf{NIE(+)}               & \textbf{NIE(-)}            & \textbf{\begin{tabular}[c]{@{}l@{}}Average \\ PIP\end{tabular}}           & \textbf{\begin{tabular}[c]{@{}l@{}}\# of active\\  voxels\end{tabular}}        & \textbf{\begin{tabular}[c]{@{}l@{}} Region   \\ Size \end{tabular}}       \\
Precuneus\_L          & 3.53  & 3.53 & -0.01 & 4.98 & 109 & 1079 \\
Parietal\_Inf\_L      & 2.83  & 2.83 & 0.00  & 5.67 & 99  & 696  \\
Postcentral\_L        & 0.21  & 0.21 & 0.00  & 1.98 & 71  & 1159 \\
Cingulum\_Mid\_R      & 1.82  & 1.82 & 0.00  & 8.98 & 67  & 605  \\
Supp\_Motor\_Area\_L  & 1.14  & 1.16 & -0.02 & 2.38 & 52  & 656  \\
Frontal\_Inf\_Oper\_R & -0.46 & 0.00 & -0.47 & 1.83 & 27  & 421  \\
Frontal\_Inf\_Orb\_L  & -0.12 & 0.02 & -0.13 & 1.98 & 21  & 503 \\
\bottomrule
\end{tabular}
}

\label{tb:top10}
\end{table}

\section{Conclusions}
\label{sec:conclusion}

In this paper, we develop a new spatially varying coefficient structural equation model, BIMA, for high-dimensional neuroimaging mediation analysis. BIMA addresses key challenges in analyzing neuroimaging data, including the high dimensionality of brain images, complex spatial correlations, sparse activation patterns, and relatively low signal-to-noise ratios. By leveraging a soft-thresholded Gaussian process (STGP) prior for spatially varying functional parameters, we not only establish posterior consistency for the mediation effects but also demonstrate selection consistency in identifying key brain regions that contribute to these effects. Our efficient posterior computation algorithm allows image mediation analysis to scale to large datasets, the fMRI data in the ABCD study, in a fully Bayesian framework.

Similar to all mediation analysis frameworks, BIMA relies on certain causal assumptions, including the Stable Unit Treatment Value Assumption (SUTVA) and assumptions underlying the identification of natural indirect and direct effects. In the context of the ABCD study, SUTVA implies that one child's parental education level does not affect another child's cognitive ability—a reasonable assumption given the study design. However, the assumption of no unmeasured confounders is more challenging, particularly in neuroimaging studies. It is likely that unmeasured confounders influence both the mediator (e.g., brain activity) and the outcome (e.g., cognitive ability). While we did not explicitly model such confounders in BIMA, addressing unmeasured confounding factors remains an important area for future research in image mediation analysis.

\bigskip
\bibliographystyle{asa}
\bibliography{ref}

\appendix

\section{Proof}

\subsection{Proof of Proposition \ref{prop:iden}}

\begin{proof}[Proof of Proposition \ref{prop:iden}]

In this proof we omit the notations $\mu_{M,i}$ to $\mu_i$ for simplicity.
First we show the identifiability of model \eqref{eq:model2}, namely part (a) in Proposition \ref{prop:iden}.

Consider two parameter sets $\Theta_M = \br{\alpha, \br{\zeta_k}_{k=1}^q, \br{\eta_i}_{i=1}^n}$ and $\Theta_M'= \br{\alpha', \br{\zeta'_k}_{k=1}^q, \br{\eta'_i}_{i=1}^n}$
Suppose the probability distributions of $\mathbf{\mpdf}$ given $\bfX$ and $\bfC$ under $\Theta_M$ and $\Theta_M'$ are equal, i.e., 
$$\pi(\mathbf{\mpdf} \mid \bfX, \bfC,\Theta_M) = \pi(\mathbf{\mpdf} \mid \bfX, \bfC, \Theta_M'),$$
where $\bfX$ and $\bfC$ satisfy the Assumption \ref{asm:alpha_iden}. Note that $\mathbf{\mpdf} = \{\mpdf_i(s)\}$. The  joint distributions of two multi-dimensional random variables are the same implies that the corresponding marginal distributions of any element of the two random variables are also the same. Hence we have for any $i \in \{1,\ldots, n\}$ and any $s\in \cB$, 
\[\pi(\mpdf_i(s) \mid \bfX,\bfC, \Theta_M) = \pi(\mpdf_i(s) \mid \bfX, \bfC, \Theta_M'). \]

Since $\mpdf_i(s)$ follows a normal distribution, for $i \in \{1,\ldots, n\}$ and any $s\in\cB$, 
\[\mu'_i(s) = \mu_i(s)\  \mbox{and} \  \sigma'^2_M = \sigma^2_M,\]
where $\mu_i(s) = \alpha(s) X_i + \eta_i(s) + \sum_{k=1}^q\bfzeta_k(s) C_{i,k}$ and $\mu'_i(s) = \alpha'(s) X_i + \eta'_i(s)+\sum_{k=1}^q\bfzeta_k(s) C_{i,k}$.  Consider the decomposition of $\mu_i(s)$, $\mu'_i(s)$, $\alpha(s), \alpha'(s)$, $\eta_i(s)$ and $\eta'_i(s)$. 
$$\mu_i(s) = \sum_{l=1}^\infty \theta_{\mu,i,l}\psi_l(s),\quad \alpha(s) = \sum_{l=1}^\infty \theta_{\alpha,l}\psi_l(s) , \quad \eta_i(s) = \sum_{l=1}^\infty \theta_{\eta,i,l}\psi_l(s),\quad \zeta_k(s) = \sum_{l=1}^\infty \theta_{\zeta,k,l}\psi_l(s)  $$
$$\mu'_i(s) = \sum_{l=1}^\infty \theta_{\mu',i,l}\psi_l(s),\quad \alpha'(s) = \sum_{l=1}^\infty \theta_{\alpha',l}\psi_l(s), \quad \eta'_i(s) = \sum_{l=1}^\infty \theta_{\eta',i,l}\psi_l(s),\quad \zeta'_k(s) = \sum_{l=1}^\infty \theta_{\zeta',k,l}\psi_l(s),$$
where the basis coefficients are satisfied with the following identities.
\begin{align*}
\theta_{\mu,i,l}= \theta_{\alpha,l} X_i + \theta_{\eta,i,l} +  \sum_{k=1}^q \theta_{\zeta,k,l}C_{i,k},\quad \mbox{and}\quad \theta_{\mu',i,l}= \theta_{\alpha',l} X_i + \theta_{\eta',i,l}  +\sum_{k=1}^q \theta_{\zeta',k,l}C_{i,k}. 
\end{align*}
Since $\mu_i(s) = \mu'_i(s)$ for any $i \in \{1,\ldots, n\}$ and any $s\in\mathcal{B}$, then for any $l \geq 1$, $\theta_{\mu,i, l} = \theta_{\mu',i, l}. $
Then we have $(\theta_{\alpha,l} - \theta_{\alpha',l} )  X_i + \theta_{\eta,1,l} - \theta_{\eta',1,l} +  \sum_{k=1}^q \sbr{\theta_{\zeta,k,l} - \theta_{\zeta',k,l}}C_{1,k} = 0$. 
According to the Assumption \ref{asm:alpha_iden}, for $t=1,\ldots,q+1$,  $
\sum_{i=1}^n W_{i,t} (\theta_{\eta,i,l} - \theta_{\eta',i,l}) = 0$. Let $\bfb_l = (\theta_{\alpha_1,l} - \theta'_{\alpha_1,l}, \theta_{\zeta,1,l} - \theta_{\zeta',1,l}, \ldots, \theta_{\zeta,q,l} - \theta_{\zeta',q,l},  \theta_{\eta,1,l} - \theta'_{\eta,1,,l},\ldots,  \theta_{\eta,n,,l} - \theta'_{\eta,n,,l})^\top$ for any $l\geq 1$ and 
$$\bfA= \left(\begin{array}{ccc}
         \bfzero_{(q+1)\times 1} & \bfzero_{(q+1)\times q} &  \bfW^\top\\
         \bfX & \bfC & \bfI_n 
    \end{array}\right),$$
where $\bfb_l$ is of dimension $(q+1+n)\times 1$ and $\bfA$ is of dimension $(n+q+1)\times (n+q+1)$.  Then we have the linear system: $\bfA\bfb_l = \bfzero_{(n+q+1)\times 1}$.

Denote $\bm{\tilde X} = (\bfX_{n\times 1}, \bfC_{n\times q})\in\R^{n\times(q+1)}$.
Note that $\mathrm{det}(\bfA) = \mathrm{det}(\bfzero - \bfW^\top \bfI^{-1}_n \bm{\tilde X})\det(\bfI_n) =  \det(\bfW^\rT \bm{\tilde X})\neq 0$ by Assumption \ref{asm:alpha_iden}. This implies that $\bfzero_{n+1+q}$  is the unique solution of $\bfA \bfb_l = \bfzero_{n+1+q}$.  Thus 
\begin{align*}
\theta_{\alpha,l} = \theta_{\alpha',l}, \qquad \theta_{\eta,i,l} = \theta_{\eta',i,l},\qquad \theta_{\zeta,k,l} = \theta_{\zeta',k,l}
\end{align*}
This further implies that for any $s$ and any $i$, 
\begin{align*}
\alpha(s) = \alpha'(s),\qquad \eta_i(s) = \eta'_i(s), \qquad \zeta_k(s) = \zeta'_k(s)
\end{align*}
This proves the identifiability of model \eqref{eq:model2}. Next, we show the statement in (b) in Proposition \ref{prop:iden}. Part (b) will be used in the proof of Theorem \ref{thm:alpha_consistency}.

By directly setting $\bfW = \bm{\tilde{X}}$, and $\sum_{i=1}^n W_{i,t} \eta_i(s) = 0$ for $t=1,\dots,q+1$, we know that $\sum_{i=1}^n \tilde X_{i,t} \eta_i(s)=0$ for $t=1,\dots,q+1$. For each $s$, let $\bm{\tilde\alpha}(s) = \{\alpha(s),\zeta_{1}(s),\dots,\zeta_{q}(s)\}^\rT \in \R^{q+1}$ and $\bm{\tilde\alpha}'(s) = \{\alpha'(s),\zeta'_{1}(s),\dots,\zeta'_{q}(s)\}^\rT \in \R^{q+1}$. Let $\tilde \bfb_l = (\theta_{\alpha_1,l} - \theta'_{\alpha_1,l},\theta_{\zeta,1,l} - \theta_{\zeta',1,l},\ldots, \theta_{\zeta,q,l} - \theta_{\zeta',q,l})^\rT$ and $\bfg_l = (\theta_{\eta,1,l} - \theta'_{\eta,1,,l},\ldots,  \theta_{\eta,n,,l} - \theta'_{\eta,n,,l})^\rT$. Then $ \tilde\bfX_i^\rT \{\tilde\bfalpha(s) - \tilde\bfalpha'(s)\} = \sum_{l=1}^\infty\tilde\bfX_i^\rT\tilde\bfb_l\psi_l(s)$ and $
    \tilde\eta_i(s) - \tilde\eta'_i(s) = \sum_{l=1}^\infty g_{l,i}\psi_l(s)$. 
Since $\int_{\cS}\{\mu_i(s) - \mu'_i(s)\}^2 \leb(\dd s)$ is finite, by Fubini's theorem,
\begin{align*}
    &\frac{1}{n}\sum_{i=1}^n\int_{\cS}\{\mu_i(s) - \mu'_i(s)\}^2 \leb(\dd s) \\
    &= \frac{1}{n} \int_{\cS}\sum_{i=1}^n  \br{ \bm{\tilde X}_i^\rT (\bm{\tilde\alpha}(s)  - \bm{\tilde\alpha}'(s) )  }^2 \leb(\dd s) + 
     \frac{1}{n}\int_\cS\sum_{i=1}^n \br{\eta_i(s) - \eta'_i(s)}^2 \leb(\dd s)\\
     &=\frac{1}{n}\int_\cS\sum_{i=1}^n\br{ \sbr{\sum_{l=1}^\infty \tilde\bfX^\rT\tilde\bfb_l\psi_l(s) }^2 + \sbr{ \sum_{l=1}^\infty g_{l,i}^2\psi_l(s) }^2 }\leb(\dd s)\\
    &=\frac{1}{n}\sum_{i=1}^n\br{ \sum_{l=1}^\infty (\tilde\bfX_i^\rT \tilde\bfb_l)^2 + \sum_{l=1}^\infty g_{l,i}^2 } \\
    &=\frac{1}{n} \sum_{l=1}^\infty \|\tilde\bfX\tilde\bfb_l\|_2^2 + \frac{1}{n}\sum_{l=1}^\infty\|\bfg_l\|_2^2.
\end{align*}
By Assumption \ref{asm:alpha_iden}(a) that $\sigma_{\min}(\tilde\bfX)>\sqrt{n}$,
$  \|\tilde\bfX\tilde\bfb_l\|_2^2\geq \sigma_{\min}^2(\tilde\bfX)\|\tilde\bfb_l\|_2^2 \geq n\|\tilde\bfb_l\|_2^2$. 
Hence 
\[\frac{1}{n}\sum_{i=1}^n\int_{\cS}\{\mu_i(s) - \mu'_i(s)\}^2 \lambda(ds) \geq \sum_{l=1}^\infty\|\tilde\bfb_l\|_2^2 + \frac{1}{n}\sum_{l=1}^\infty \|\bfg_l\|_2^2.\]

Note that the empirical norm $\|f\|_{2,\pscale}$ is a finite grid approximation of the Hilbert space inner product $\sqrt{\int_{\cS} f^2(s)\leb(\dd s)}$. By Definition \ref{def:param_space}(d), the approximation error is given by $err(f) = \left| \|f\|^2_{2,\pscale} - \int_{\cS} f^2(s)\leb(\dd s)\right|\leq Kp^{-2/d}$.
\begin{align*}
    &\|\alpha-\alpha'\|_{2,\pscale}^2 = \sum_{l=1}^\infty(\theta_{\alpha,l} - \theta_{\alpha',l})^2 + err(\alpha-\alpha')\\
    &\|\zeta_k-\zeta'_k\|_{2,\pscale}^2 = \sum_{l=1}^\infty(\theta_{\zeta_k,l} - \theta_{\zeta'_k,l})^2 + err(\zeta_k-\zeta'_k), ~ k=1,...,q\\
    &\|\eta_i-\eta'_i\|_{2,\pscale}^2 = \sum_{l=1}^\infty(\theta_{\eta_i,l} - \theta_{\eta'_i,l})^2 + err(\eta_i-\eta'_i), ~ i=1,...,n
\end{align*}
For $n$ large enough such that $Kp^{-2/d}<\frac{1}{q+3}\epsilon^2$, the following inequality
\[\|\alpha(s)-\alpha'(s)\|_{2,\pscale}^2+\sum_{k=1}^q\|\zeta_k(s)-\zeta'_k(s)\|_{2,\pscale}^2+\frac{1}{n}\sum_{i=1}^n\|\eta_i(s)-\eta'_i(s)\|_{2,\pscale}^2 >\epsilon^2\]
implies that there exists constant $c_1'$
$\sum_{l=1}^\infty\|\tilde\bfb_l\|_2^2 + n^{-1}\sum_{l=1}^\infty \|\bfg_l\|_2^2>c'_1\epsilon^2$
which further implies that there exists constant $c_0$, $$\frac{1}{n}\sum_{i=1}^n\|\mu_i(s) - \mu'_i(s) \|_{2,\pscale}^2 >c_0\epsilon^2$$
Hence Proposition \ref{prop:iden}(b) follows.

\end{proof}

\subsection{ Proof of Theorem \ref{thm:alpha_consistency} }

Theorem \ref{thm:alpha_consistency} is proved by checking the conditions in Theorem A.1 in \cite{Ghosal2004}. 

For simplicity, throughout the proof of Theorem \ref{thm:alpha_consistency}, we use the following notations: $\theta = \{\alpha,\br{\zeta_{k}}_{k=1}^q,\br{\eta_i}_{i=1}^n\}$, and the true parameters denoted as $\theta_0 =\{\alpha_0,\br{\zeta^0_{k}}_{k=1}^q,\br{\eta^0_i}_{i=1}^n\}$. In addition, let $\mu_i(s) = \alpha(s)X_i + \sum_{k=1}^q\zeta_k(s) C_{i,k} +\eta_i(s)$ be the mean function given $\{X_i,\{C_{i,k}\}_{k=1}^q\}_{i=1}^n$, and $\mu_i^0(s)$ be the mean function under the true parameters.

Conditional on $\{X_i,\{C_{i,k}\}_{k=1}^q\}_{i=1}^n$, for individual $i$ and location $s_j$, $\mpdf_i(s_j)$ follows independent  distributions across $i=1,\dots,n, j=1,\dots,p$, with density function $\pi(\mpdf_i(s_j);\theta) = \phi(\mu_i(s_j), \sigma^2)$, where $\phi(\mu_i(s_j), \sigma^2)$ is used to denote the normal density with mean $\mu_i(s_j)$ and variance $\sigma^2$.
Let $\Lambda_{i,j}(\theta_0,\theta) := \log\{\pi(\mpdf_i(s_j);\theta_0)/\pi(\mpdf_i(s_j);\theta)\}$. 

First, we verify the prior positivity condition as follows.

\begin{lemma}
(Prior positivity condition) There exists a set $B$, $\Pi(B)>0$ such that 
    \begin{enumerate}
        \item 
        $\liminf_{\br{n,p}\to\infty}\Pi\left\{\theta\in B:(n\pscale)^{-1}\sum_{i=1}^n\sum_{j=1}^{\pscale}\mathbb{E}_{\theta_0}\{\Lambda_{i,j}(\theta_0,\theta)\}<\epsilon\right\}>0 \text{ for all }\epsilon>0$; and 
        \item 
        $(n\pscale)^{-2}\sum_{i=1}^n\sum_{j=1}^{\pscale}\Var_{\theta_0}\{\Lambda_{i,j}(\theta_0,\theta)\}\to 0$, as $n\to \infty$ and $p\to \infty$,   for all $\theta\in B$.
    \end{enumerate}
\end{lemma}

\begin{proof}
Define 
\begin{align}\label{eq:theta_m_inf}
    \|\theta-\theta_0\|_\infty = \max\br{\sup_{s\in\cS}|\alpha(s)-\alpha_0(s)|, \max_k\sup_{s\in\cS}|\zeta_{k}(s)-\zeta_{k}^0(s)|, \max_i\sup_{s\in\cS}|\eta_i(s)-\eta_i^0(s)|}.
\end{align}

For constant $\delta>0$, consider 
\begin{align*}
    B_\delta&=\br{\theta\in\Theta:\|\theta-\theta_0\|_\infty<\delta }.
\end{align*}
Since the prior distributions for the above parameters are independent, to show $\Pi(B_\delta)>0$, we only need to show that the prior of each term in \eqref{eq:theta_m_inf} being upper bounded by a constant has a positive probability.

    By Theorem 4 in \cite{ghosal2006}, for any $i=1,\dots,n$, $k=1,\dots,q$,
    \[\Pi\left(\sup_{s\in \cS} |\eta_i(s)-\eta_{i}^0(s)|<\delta\right)>0, \quad  \Pi\left(\sup_{s\in \cS} |\zeta_k(s)-\zeta_{k}^0(s)|<\delta\right)>0.\]  
    
    By Lemma 2 in \cite{kang2018}, for any threshold $\sthresh>0$ and any true $\alpha_0(s)\in \Theta_{\alpha}$, there exists $\tilde\alpha(s)$ in the RKHS of $\kappa(\cdot,\cdot)$ such that $\alpha_0 = T_\sthresh(\tilde\alpha_0)$. Note that the soft-thresholding function $T_\sthresh(x)$ is a 1-Lipschitz continuous function of $x$, and by Theorem 4 in \cite{ghosal2006}, we have $\Pi\left(\sup_{s\in \cS} |\tilde\alpha(s)-\tilde\alpha_0(s)|<\delta\right)>0$, which implies $\Pi\left(\sup_{s\in\cS} |T_\sthresh(\tilde\alpha(s))-T_\sthresh(\tilde\alpha_0(s))|<\delta\right)>0$.  Hence for any $\theta\in B_\delta$, where $\Pi(B_\delta)>0$, we have
\begin{align*}
    \mathbb{E}_{\theta_0}\mbr{\Lambda_{i,j}(\theta_0,\theta)}
    =& \mathbb{E}\mbr{\mathbb{E}_{\theta_0}\br{\Lambda_{i,j}(\theta_0,\theta) \mid \bfX, \bfC }}\\
    =&- \frac{1}{2\sigma_M^2}\mathbb{E}\mbr{\mathbb{E}_{\theta_0}\br{(M_i(s_j)-\mu_i^0(s_j))^2 \mid \bfX, \bfC} }\\
    & +\frac{1}{2\sigma_M^2}\mathbb{E}\mbr{\mathbb{E}_{\theta_0}\br{(M_i(s_j) - \mu_i^0(s_j) +\mu_i^0(s_j)-\mu_i(s_j))^2 \mid \bfX, \bfC}}\\
    =& \mathbb{E}\mbr{\frac{1}{2\sigma^2_M}(\mu_i^0(s_j) - \mu_i(s_j))^2 }
\end{align*}
Note that
    \begin{align*}
        &\frac{1}{2\sigma_M^2}\{\mu_i^0(s_j)-\mu_i(s_j)\}^2 \\
        &\leq \frac{1}{2\sigma_M^2}\left[\{\alpha(s_j)-\alpha_0(s_j)\}X_i+\sum_{k=1}^q \{\zeta_k(s_j)-\zeta_{k}^0(s_j)\}C_{i,k}+\{\eta_i(s_j)-\eta_i^0(s_j)\}\right]^2\\
        &\leq \frac{2}{\sigma_M^2}\left[  X_i^2\{\alpha(s_j)-\alpha_0(s_j)\}^2 + \sum_{k=1}^q \{\zeta_k(s_j)-\zeta_{k}^0(s_j)\}^2 C_{i,k}^2 + \{\eta_i(s_j)-\eta_i^0(s_j)\}^2 \right]
    \end{align*}
    By choosing a constant $K_{\max}$ such that  $\max_i\{\mathbb{E}\br{|X_i|^2},\max_k\mathbb{E}\br{|C_{i,k}|^2}\} \leq K_{\max},$ then for any $\theta\in B_\delta$, $(n\pscale)^{-1}\sum_{i=1}^n\sum_{j=1}^{\pscale}\bE_{\theta_0}\br{\Lambda_{i,j}(\theta_0,\theta)}<2\sigma_M^{-2}K_{\max}(2+q)\delta^2$, hence for a small enough $\epsilon$ such that $0<\epsilon<\sigma_M^{-2}K_{\max}(2+q)\delta^2$,
    \begin{align*}
        &\liminf_{\br{n,p}\to\infty}\Pi\br{\theta\in B_\delta:\frac{1}{n\pscale}\sum_{i=1}^n\sum_{j=1}^{\pscale}\bE_{\theta_0}\sbr{\Lambda_{i,j}(\theta_0,\theta)}<\epsilon} \\
        &\geq \Pi\br{\|\theta-\theta_0\|_\infty\leq \sqrt{\sbr{\frac{2}{\sigma_M^2}K_{\max}(2+q)}^{-1} \epsilon } }> 0.
    \end{align*}
    To show the second condition, we only need to show that for any $i,j$ and any $\theta\in B_\delta$, the variance $\Var_{\theta_0}\br{\Lambda_{i,j}(\theta_0,\theta)}$ is bounded by some constant. 
    \begin{align*}
        \Var_{\theta_0}\br{\Lambda_{i,j}(\theta_0,\theta)}&= \mathbb{E}\br{\Var_{\theta_0}\br{\Lambda_{i,j}(\theta_0,\theta)\mid \bfX, \bfC}} + 
        \Var\br{\mathbb{E}_{\theta_0}\br{\Lambda_{i,j}(\theta_0,\theta)\mid \bfX, \bfC}} \\
        &= \mathbb{E}\br{\frac{1}{\sigma_M^2}\sbr{\mu_i^0(s_j)- \mu_i(s_j)}^2} + \Var\br{\frac{1}{2\sigma^2_M}(\mu_i^0(s_j) - \mu_i(s_j))^2}\\
        &\leq \max\br{\frac{4}{\sigma_M^2}K_{\max}(2+q)\delta^2,\frac{4}{\sigma_M^4}K_{\max,V}(2+q)\delta^4}<\infty,
    \end{align*}
    where $K_{\max,V}\geq \max_i\br{\Var(X_i^2),\max_k \Var(C_{i,k}^2)}$.

\end{proof}

Before the test construction, we add a useful lemma on the tail probability of the maximum of sub-Gaussian random variables.
\begin{lemma}\label{lem:max_subG_tail}
Let $X_i,i=1,\dots,N$ be sub-Gaussian random variables. Let $\sigma^2_i$ be the constant such that $\mathbb{P}(|X_i| > t ) \leq 2 \exp(-t^2/\sigma^2_i)$ for any $t>0$ and $i = 1,\ldots, N$. Let $\tilde\sigma^2_N = \bigvee_{i=1}^N \sigma^2_i$. Then for any $t>0$,  $\mathbb{P}\left(\max_i|X_i|>\sqrt{\tilde\sigma_N^2\log{2N}+t}\right)\leq \exp(-t)$.
\begin{proof}
    Let $u = \sqrt{\tilde\sigma_N^2\log{2N}+t}$,
        $$\mathbb{P}\left(\max_i|X_i|>u\right) \leq \sum_i \mathbb{P}(|X_i|>u) \leq 2N \exp\br{-u^2/\tilde\sigma_N^2} = \exp(-t).$$ 
\end{proof}
\end{lemma}
Next, we construct a test that satisfies the Type I and Type II error bound on a specified sieve space.
\begin{lemma}
(Existence of tests) There exist test functions $\{\Phi_{n\pscale}\}$, subset $\cU_n,\Theta_n\subset\Theta$, and constant $K_1,K_2,c_1,c_2>0$ such that
    \begin{enumerate}
        \item[(a)] $\mathbb{E}_{\theta_0}\Phi_{n\pscale}\to 0$, as $n\to \infty$ and $p\to \infty$;
        \item[(b)] $\sup_{\theta\in \cU_n^c\cap \Theta_{n}} \mathbb{E}_{\theta}(1-\Phi_{n\pscale})\leq K_1e^{-c_1n\pscale}$;
        \item[(c)] $\Pi(\Theta_{n}^c)\leq K_2e^{-c_2n\pscale}$.
    \end{enumerate}
\end{lemma}

\begin{proof}
    Define the sieve space of $\theta$ as 
    $\Theta_n$, which be decomposed into product of the following parameter space:
    \begin{align*}
    \Theta_{n} &= \Theta_{\alpha,n}\times\prod_{k=1}^q\Theta_{\zeta,k,n} \times\prod_{i=1}^n\Theta_{\eta,i,n}\\
        \Theta_{\alpha,n} &= \br{\alpha\in \Theta_{\alpha}:\sup_{s\in \cR_1\cup\cR_{-1}}\|D^{\omega}\alpha(s)\|_{\infty}<\sqrt{n\pscale}, \|\omega\|_1\leq \rho}\\
        \Theta_{\zeta,k,n} & = \br{\zeta_k\in\Theta_\zeta: \sup_{s\in\cS}\|D^{\omega}\zeta_k(s)\|_{\infty}<n\pscale, \|\omega\|_1\leq \rho}, k=1,\dots,q\\
        \Theta_{\eta,i,n} & = \br{\eta_i\in\Theta_\eta: \sup_{s\in\cS}\|D^{\omega}\eta_i(s)\|_{\infty}<n\pscale, \|\omega\|_1\leq \rho}, i=1,\dots,n
    \end{align*}
    where $D^\omega f(s)$ stands for $(\partial^{\|\omega\|_1}/\partial^{\omega_1},\ldots,\partial^{\|\omega\|_1}/\partial^{\omega_d})f(s)$ for any $\omega = (\omega_1,\ldots, \omega_d)^\rT$ with $\omega_j (j = 1,\ldots, d)$ being positive intergers and $s\in \R^d$. 
    
    To show the conditions (a) and (b), we use Lemma 8.27(i) in \cite{vandervaart_2017}, by viewing $\mathbf{\mpdf}\sim N_{n\pscale}(\bfmu,\sigma^2 I)$, $\bfmu=\{\mu_i(s_j)\}_{i=1,j=1}^{n,\pscale}\in \R^{n\pscale}$. By Lemma 8.27(i), for any $\bfmu_1,\bfmu_0\in \R^{n\pscale}$, there exists $\Phi(\bfmu_1)$ such that for any $\bfmu$ where $\|\bfmu-\bfmu_1\|_2\leq \|\bfmu_0-\bfmu_1\|_2/2$,
    \begin{align*}
        \mathbb{E}_{\bfmu_0}\Phi(\bfmu_1)\vee \mathbb{E}_{\bfmu}\{1-\Phi(\bfmu_1)\}&\leq \exp\left\{-c_1\|\bfmu_0-\bfmu_1\|_2^2/\sigma_M^2\right\}
    \end{align*}

    Because the type II error in condition (b) does not depend on a single $\bfmu_1$, to remove the dependence on $\bfmu_1$, and to use a neighborhood $\mathcal{U}_n$ defined by the empirical norm as the distance metric instead of the Euclidean norm,
    we use the same technique as the one in Proposition 11 in \cite{vandervaart2011}. For any $r\geq 1$, any integer $j\geq 1$, define shells for $\bfmu$
    \begin{align*}
        \mathcal{C}_{j,r}&:=\br{\Theta_{n}: jr\leq \|\bfmu-\bfmu_0\|_2\leq (j+1)r}
    \end{align*}
    
    Denote $\cP(\mathcal{C}_{j,r},jr/2,\|\cdot\|_2)$ as the largest packing number of $\mathcal{C}_{j,r}$ with Euclidean distance $jr/2$, and denote the corresponding $jr/2$-separated set of $\mathcal{C}_{j,r}$ as $\cP_j$.
    Note that $\cP_j$ is also a $jr/2$-covering set of $\mathcal{C}_{j,r}$. Hence for any $\bfmu\in \mathcal{C}_{j,r}$, there exists $\bfmu_1\in \cP_j$ such that 
    \[\|\bfmu-\bfmu_1\|_2\leq \frac{jr}{2}\leq \frac{1}{2}\|\bfmu_1-\bfmu_0\|_2.\]

    Choose $\Phi_j=\max_{\bfmu_1\in \cP_j}\{\Phi(\bfmu_1)\}$, then for any $\bfmu\in \mathcal{C}_{j,r}$, conditioning on $\bfX,\bfC$,
    \begin{align*}
        \mathbb{E}_{\bfmu_0,\sigma_0}\Phi_j &\leq 2\cP(\mathcal{C}_{j,r},\frac{jr}{2},\|\cdot\|_2)\exp\br{-c_1[(jr)^2]/\sigma_M^2}\\
        \mathbb{E}_{\bfmu,\sigma}(1-\Phi_j) &\leq \exp\br{-c_1[(jr)^2]/\sigma_M^2}
    \end{align*}
    Denote $\cN(\Theta_{n},r,\|\cdot\|_{\infty})$ as the smallest covering number for the set $\Theta_{n}$ with radius $r$ and distance function $\|\cdot\|_{\infty}$. Now we need an upper bound on $\log\cN(\Theta_{n},r,\|\cdot\|_{\infty})$. 
    Note that by Lemma 2 in \cite{ghosal2006} and a similar approach in Lemma A1 in \cite{kang2018}, there exist constants $K_\alpha,K_\zeta,K_\eta$, such that $\log\cN(\Theta_{\alpha,n},r,\|\cdot\|_{\infty})\leq K_\alpha(n\pscale)^{d/(2\rho)}r^{-d/\rho}$, and  $\log\cN(\Theta_{\eta,i,n},r,\|\cdot\|_{\infty})\leq K_\eta(n\pscale)^{d/\rho}r^{-d/\rho}$, $\log\cN(\Theta_{\zeta,k,n},r,\|\cdot\|_{\infty})\leq K_\zeta(n\pscale)^{d/\rho}r^{-d/\rho}$. 
    Hence there exists constant $K_0$,
    \begin{align*}        &\log\cN(\Theta_{n},r,\|\cdot\|_{\infty}) \\
    &\leq \log\cN(\Theta_{\alpha,n},r,\|\cdot\|_{\infty}) + \sum_{k=1}^q \log\cN(\Theta_{\zeta,k,n},r,\|\cdot\|_{\infty})+\sum_{i=1}^n \log\cN(\Theta_{\eta,i,n},r,\|\cdot\|_{\infty})\\ 
        &\leq K_0n(n\pscale)^{d/\rho}r^{-d/\rho}
    \end{align*}
    
    Conditioning on $\sbr{\bfX,\bfC}$, denote \[\Theta_{n}^*:=\br{\bfmu\in\R^{n\pscale}: \mu_{ij} = \alpha(s_j)X_i + \sum_{k=1}^q \zeta_k(s_j)C_{i,k} + \eta_i(s_j), \theta\in \Theta_{n}}. \]

    Now we first show that conditioning on $\sbr{\bfX,\bfC}$, given $c^*_n=\max_i\br{|X_i|, \|\bfC_i\|_\infty}_i$,
    \[\log\cN(\Theta_{ n}^*,r/\sbr{4\sqrt{n\pscale}},\|\cdot\|_\infty)\leq \log\cN(\Theta_{ n},r/\sbr{4c^*_n\sqrt{n\pscale}},\|\cdot\|_\infty).\]
    
    Denote $\cS^*_{\mu,n}$ as a $(c^*_n r)$-covering set of $\Theta_{n}^*$ under $\|\cdot\|_\infty$. $\cS^*_{\mu,n}$ is constructed in the following way: for any $\bfmu\in \Theta_{n}^*$, there exists a corresponding  $\theta_{\mu} = \sbr{\alpha, \br{\zeta_k}_{k=1}^q, \br{\eta_i}_{i=1}^n }\in \Theta_{n}$ such that $\mu_{ij} = \alpha(s_j)X_i + \sum_{k=1}^q \zeta_k(s_j)C_{i,k} + \eta_i(s_j)$, hence there exists $\theta_{\mu,1}\in \cN_{\mu,n}$ where $\cN_{\mu,n}$ is the smallest covering set with cardinality $\cN(\Theta_{n},r,\|\cdot\|_\infty)$, and there exists corresponding $\bfmu_{1}\in \Theta_{n}^*$ given $\theta_1$. 
    \[|\mu_{1,ij} - \mu_{ij}|\leq |\sbr{\alpha(s_j)-\alpha_1(s_j)}X_i| + \sum_{k=1}^q |\sbr{\zeta_k(s_j)-\zeta_{1,k}(s_j)}C_{i,k}| + |\eta_i(s_j)-\eta_{1,i}(s_j)| \leq c^*_nr\]. 
    Hence $\cS^*_{\mu,n}$ can be constructed as a collection of all such $\bfmu_1$. Let $|\cS^*_{\mu,n}|$ be the cardinality of such $\cS^*_{\mu,n}$. By the construction of $\cS_{\mu,n}^*$, $|\cS_{\mu,n}^*| \leq \cN(\Theta_{n},r,\|\cdot\|_\infty)$.
    
    Since $\|\cdot\|_{2,n\pscale}\leq \|\cdot\|_\infty$, we have 
    \begin{align*}
        \log\cP(\Theta_{ n}^*,r/2,\|\cdot\|_2)
        &\leq \log\cN(\Theta_{ n}^*,r/4,\|\cdot\|_2)
        = \log\cN(\Theta_{n}^*,r/\sbr{4\sqrt{n\pscale} },\|\cdot\|_{2,n\pscale})\\
        &\leq \log\cN(\Theta_{ n}^*,r/\sbr{4\sqrt{n\pscale}},\|\cdot\|_\infty)
        \leq \log|\cS_{\mu,n}^*|\\
        &\leq \log\cN(\Theta_{ n},r/\sbr{4c^*_n\sqrt{n\pscale}},\|\cdot\|_\infty)\\
        &\leq K_0(4c^*_n)^{d/\rho}n(n\pscale)^{3d/(2\rho)}r^{-d/\rho}\numberthis \label{eq:covering}
     \end{align*}
     Denote event $A=\mbr{c^*_n<a\sqrt{\log\br{n}}}$ and $I_A$ be its indicator, where $a$ is an absolute constant, Lemma \ref{lem:max_subG_tail} implies that $\mathbb{P}(I_{A^c}) \to 0$ as $n\to\infty$, where $A^c$ denotes the complement of $A$. Hence given $A$, $\log\cP(\Theta_{ n}^*,r/2,\|\cdot\|_2) \leq K_a \sbr{\log n}^{d/(2\rho)}n(n\pscale)^{3d/(2\rho)}r^{-d/\rho}$.

    Then for any $\bfmu\in \cup_{j\geq 1}\mathcal{C}_{j,r},\sigma\in \cup_{j\geq 1} \mathcal{C}_{j,\epsilon}$, define $\Phi=\sum_{j\geq 1}\Phi_j I(\bfmu\in \mathcal{C}_{j,r})$, for some constants $K_2,K_3$, conditioning on $\bfX,\bfC$, 
    \begin{align*}
        \mathbb{E}_{\bfmu_0}&\Phi\leq \sum_{j\geq 1}2\cP(\mathcal{C}_{j,r},jr/2,\|\cdot\|_2)
        \exp\{-c_1[(jr)^2]/\sigma_M^2\}\\
        &\leq 2\cP(\Theta^*_{n},r/2,\|\cdot\|_2)\sum_{j\geq 1}\exp\{-c_1[j(r)^2]/\sigma_M^2\}\\
        &\leq 2\cP(\Theta^*_{n},r/2,\|\cdot\|_2)K_2\exp\left(-\frac{c_1r^2}{4\sigma_M^2}\right) \\
        &\leq
        K_3 \cP(\Theta^*_{n},r/2,\|\cdot\|_2) \exp\left(-\frac{c_1r^2}{\sigma_M^2}\right)
        \\
        \mathbb{E}_{\bfmu}(1-\Phi)&\leq \sum_{j\geq 1}\exp\{-c_1[(jr)^2]\sbr{2\sigma_M^2}^{-1}\}\\
        &\leq K_3 \exp\br{-\frac{c_1r^2 }{\sigma_M^2}}
    \end{align*}
    Choose $r=\sqrt{n\pscale}\epsilon$, for any $\epsilon>0$, we can choose $n,\pscale$ large enough such that $r>1$. By Proposition \ref{prop:iden}(b), $\cU_M^c\subset\cU^c_{M,1}$ almost surely, where
    \begin{align*}
        \cU^c_M &= \left\{\Theta: \|\alpha(s)-\alpha_0(s)\|^2_{2,\pscale} + \sum_{k=1}^{q}\|\zeta_{k}(s)-\zeta_{k,0}(s)\|^2_{\pscale}  + \frac{1}{n}\sum_{i=1}^n\|\eta_i(s)-\eta_i(s)\|^2_{2,\pscale} >\epsilon^2\right\}\\
        \cU^c_{M,1} &= \left\{\Theta: \|\bfmu-\bfmu_0\|_{2,n\pscale}>\sqrt{c_0}\epsilon\right\}
    \end{align*}

    Then for any $\theta\in \Theta_n\cap \cU_{M,1}^c$, note that $(\log n)^{d/(2\rho)}(n\pscale)^{d/\rho}<n^{d/(2\rho)}(n\pscale)^{d/\rho}<\pscale$ given Assumption \ref{asm:true_func}, $\rho>d+3/(2\tau)$. 
    \begin{align*}
         \mathbb{E}&\br{\mathbb{E}_{\bfmu_0,\sigma_0}\br{\Phi\mid \bfX,\bfC}}\leq \mathbb{E}_A\br{\cP(\Theta_{n}^*,\sqrt{n\pscale}c_0\epsilon/2,\|\cdot\|_2)}K_4\exp\{-c_1''n\pscale\epsilon^2\} + \mathbb{E}\br{I_{A^c}}\\
        &\leq K'\exp\left\{c'''_1\sbr{\log n}^{d/(2\rho)}n(n\pscale)^{d/\rho}\epsilon^{-d/\rho} -c_1''n\pscale\epsilon^2 \right\}\overset{\pscale\to\infty}{\to} 0\\
        \mathbb{E}&_{\Theta_n\cap \cU_M^c}(1-\Phi)\leq E_{\Theta_n\cap \cU_{M,1}^c}(1-\Phi)
        \leq K''\exp\{-c_2'n\pscale\epsilon^2\}
    \end{align*}
    To verify (c), $\Pi(\Theta_n^c)\leq \Pi(\Theta_{\alpha,n}^c) + \sum_{i=1}^n\Pi(\Theta_{\eta,i,n}^c) + \sum_{k=1}^q\Pi(\Theta_{\zeta,k,n}^c) $.
    Theorem 5 in \cite{ghosal2006} ensures that $\Pi(\Theta_{\eta,i,n}^c)\leq K_3e^{-c_3 (n\pscale)^2}, \Pi(\Theta_{\zeta,k,n}^c)\leq K_3e^{-c_3 (n\pscale)^2}$, Lemma 4 in \cite{kang2018} ensures that $\Pi(\Theta_{\alpha,n}^c)\leq K_{\alpha}e^{-c_\alpha n\pscale}$. 
    Hence 
    \begin{align*}
        \Pi(\Theta_n^c)&\leq K_{\alpha}e^{-c_\alpha n\pscale} + K_3e^{-\sbr{c_3 (n\pscale)^2-\log (n+q)}}  \\
        &\leq K_2e^{-c_2n\pscale}
    \end{align*}
\end{proof}
    The proof for Theorem \ref{thm:alpha_consistency} is complete. Note that this can be easily extended to the marginal consistency of $\alpha$ alone by conditioning on other parameters at the true value.

\subsection{Proof of Theorem \ref{thm:beta_consistency}}\label{supp:sec:proof_thm2}
Similar to Theorem \ref{thm:alpha_consistency}, we verify the conditions in Theorem A.1 in \cite{Ghosal2004}. 

Let $\theta_0$ denote the set of true parameters $\br{\beta_0,\gamma_0, \bfxi_{0}}$ that generate the outcome variable $Y_i$ given $\mathcal{M}_i$, $X_i$ and $\bfC_i$. Let $\theta = \sbr{\beta,\gamma, \bfxi}\in \Theta_\beta\times \R^{q+1}$ denote any parameter in the parameter space, where $\Theta_\beta$ is defined in Definition \ref{def:param_space}.

\begin{lemma}
(Prior positivity condition) Under model \eqref{eq:model1}, define $\Lambda_i(\theta_0,\theta)= \log\br{\pi(Y_i;\theta_0)/\pi(Y_i;\theta)}$, there exists a set $B\subset \Theta$ such that $\Pi(B)>0$ and for any $\theta\in B$:
\begin{itemize}
    \item[(a)] $\liminf_{n\to\infty}\Pi[\theta\in B: n^{-1}\sum_{i=1}^n \bE_{\theta_0}\{\Lambda_i(\theta_0,\theta )\}<\epsilon]>0$ for any $\epsilon>0$
    \item[(b)] $n^{-2}\sum_{i=1}^n \Var_{\theta_0}\{\Lambda_i(\theta_0,\theta )\} \to 0$
\end{itemize}

\end{lemma}

\begin{proof}
For one individual $i$, the density 
\[\pi_i(Y_i,\mcdf_i,X_i,\bfC_i;\theta) = \pi_i(Y_i|\mcdf_i,X_i,\bfC_i;\theta)\pi_i(\mcdf_i,X_i,\bfC_i).\]
Here, with the abbreviated notation $\tilde\bfgamma = (\gamma,\bfxi^\rT)^\rT\in \R^{q+1}$, and $\tilde\bfX_i = (X_i,\bfC_i^\rT)^\rT\in \R^{q+1}$. Hence given $\br{\tilde\bfX_i}_{i=1}^n$ and $\br{\mcdf_i(\vox_j)}_{i=1,j=1}^{n,p}$,  and denote $\bfmcdf_i = \br{\mcdf_i(\vox_j)}_{j=1}^p$, 
$$ Y_i  \overset{\ind}{\sim} N\left(\sum_{j=1}^{\pscale} \beta(\cvox_j)\mcdf_i(\vox_j) + \tilde\bfgamma^\rT \tilde{\bfX}_i,\sigma_Y^2\right).$$ 

Given the true parameters $\beta_0(s)$, $\tilde\bfgamma_0$, define the subset
\[B_{\delta} = \br{\Theta:\sup_{j}|\beta(\cvox_j)-\beta_0(\cvox_j)|^2\leq\delta, \|\tilde\bfgamma - \tilde\bfgamma_0\|_2^2\leq \delta }\]

If we denote the mean of $Y_i$ under true parameters as $\mu_{i,0}$, otherwise as $\mu_i$,
the log-likelihood ratio for $\theta_0 = (\beta_0(s),\tilde\bfgamma_0)$ versus $\theta = (\beta(s),\tilde\bfgamma)$ can be written as 
\begin{align*}
    \Lambda_{i}(D_{n,i};\theta_0,\theta) &= \log\br{\pi_i(Y_i;\beta_0(s),\tilde\bfgamma_0)} - \log\br{\pi_i(Y_i;\beta(s),\tilde\bfgamma)}\\
    &= -\frac{1}{2\sigma_{Y}^2} (Y_i-\mu_{i,0})^2  + \frac{1}{2\sigma_Y^2} (Y_i-\mu_i)^2
\end{align*}

Hence,
\begin{align*}
    K_{i,n}(\theta_0,\theta)&:=\bE_{\theta_0}(\Lambda_{i}(D_{n,i};\theta_0,\theta)) = \bE\br{\bE_{\theta_0}\sbr{\Lambda_i|\bfmcdf_i,\tilde\bfX_i}}\\
    &=\bE\br{\frac{1}{2\sigma_Y^2}(\mu_i-\mu_{i,0})^2}\\
    & \leq \bE\mbr{\frac{1}{2\sigma_Y^2}\br{ (\tilde\bfgamma - \tilde\bfgamma_0)^\rT \tilde\bfX_i + \sum_{j=1}^{\pscale}(\beta(\cvox_j)-\beta_0(\cvox_j))\mcdf_i(\vox_j)}^2}   
\end{align*}
Note that by equation \eqref{eq:model2-3}, given $\tilde\bfX_i$, $\mcdf_i(\vox_j)\sim N(\mu_{i}(\cvox_j)\leb(\vox_j), \sigma_M^2\leb(\vox_j))$ with its second moment as $\sigma_M^2\leb(\vox_j)-\sbr{\mu_{i}(\cvox_j)\leb(\vox_j)}^2$. 
When $ \leb(\vox_j) =1/p$, the second moment is $\sigma_M^2/p-\sbr{\mu_{i}(\cvox_j)}^2/p^2$, and its 4th moment is of the order $O(p^{-4})$. Hence $\mathbb{E}\br{\|\bfmcdf_i\|_2^2\big|\tilde\bfX_i}$ can be upper bounded by a constant, and so does $\Var\br{\|\bfmcdf_i\|_2^2\big|\tilde\bfX_i}$. For the finite dimensional vector $\tilde\bfX_i$ with finite 4-th moment (Assumption \ref{asm:alpha_iden}(a)), there is a finite bound $\bE\|\tilde\bfX_i\|_2^2<K_0$.

For any $(\tilde\bfgamma,\beta(s))\in B_\delta$,
\begin{align*}
    K_{i,n}(\theta_0,\theta)&\leq \frac{1}{2\sigma_Y^2} \bE\br{ \delta\|\tilde\bfX_i\|_2^2  \|\bfmcdf_i\|_{2}^2 }
\end{align*}

Hence we have $K_{i,n}(\theta_0,\theta)\leq \delta K' $ for some constant $K' > 0$.
Similarly, denote $Z_i = (Y_i-\mu_{i,0})/\sigma_{Y}$ as the standard normal variable under $H_0$,
\begin{align*}
    V_{i,n}(\theta_0,\theta)& = \Var\br{\bE_{\theta_0}(\Lambda_i\mid\tilde\bfX_i,\bfmcdf_i)} + \bE\br{\Var_{\theta_0}(\Lambda_i\mid\tilde\bfX_i,\bfmcdf_i)}\\
    \Var\br{\bE_{\theta_0}(\Lambda_i|\tilde\bfX_i,\bfmcdf_i)}=&\Var\br{\frac{1}{2\sigma_Y^2}(\mu_i-\mu_{i,0})^2}\\
     \leq&\frac{1}{4\sigma_{Y}^4}\Var\mbr{\br{(\tilde\bfgamma - \tilde\bfgamma_0)^\rT\tilde\bfX_i +  \sum_{j=1}^{\pscale}(\beta(\cvox_j)-\beta_0(\cvox_j))\mcdf_i(\vox_j)}^2} \\
    \leq& \frac{1}{\sigma_{Y}^4} \Var\mbr{\delta \|\tilde\bfX_i\|_2^2 + \delta \|\bfmcdf_i\|_{2}^2} \\
    \leq& \frac{1}{\sigma_{Y}^4} \Var\br{\delta \|\tilde\bfX_i\|_2^2 + \bE\sbr{ \delta \|\bfmcdf_i\|_{2}^2\bigg\vert\tilde\bfX_i}} + \\
    &\frac{1}{\sigma_{Y}^4}\bE\br{\Var\sbr{\delta\|\tilde\bfX_i\|_2^2 + \delta \|\bfmcdf_i\|_{2,\pscale}^2 \bigg\vert \tilde\bfX_i }}\\
    <&\infty 
\end{align*}
For the second term,
\begin{align*}
    \bE_{\theta_0}\br{\Var_{\theta_0}(\Lambda_i|\tilde\bfX_i,\bfmcdf_i)} =&
    \bE\mbr{\Var_{\theta_0}\br{\left. -\frac{1}{2}Z_i^2 + \frac{1}{2}\sbr{Z_i+\frac{\mu_{i,0}-\mu_i}{\sigma_{Y}}}^2 \right|\tilde\bfX_i,\bfmcdf_i}}\\
    =&\bE\mbr{ \Var_{\theta_0}\br{\left . \frac{\mu_{i,0}-\mu_i}{\sigma_Y}Z_i\right|\tilde\bfX_i,\bfmcdf_i}}\\
    =&\bE\br{\frac{1}{\sigma_Y^2}\sbr{\mu_i-\mu_{i,0}}^2}\\
    \leq& \frac{1}{\sigma_Y^2} \bE\sbr{\delta\|\tilde\bfX_i\|_2^2 + \delta\|\bfmcdf_i\|_{2}^2} <\infty
\end{align*}

Hence for any $\beta\in B_\delta$, \[\frac{1}{n^2}\sum_{i=1}^n V_{n,i}(\beta_0,\beta)\to 0\]

For any $0<\epsilon<\delta K'$, 
\begin{align*}
    &\Pi\sbr{ (\beta,\tilde\bfgamma,\sigma_Y)\in B_\delta: \frac{1}{n}\sum_{i=1}^n K_{n,i} <\epsilon}   \\
    &\geq\Pi\sbr{ \sup_{j} |\beta_0(\cvox_j)-\beta(\cvox_j)| < \sqrt{\epsilon/K'}, \|\tilde\bfgamma-\tilde{\bfgamma}_0\|_2^2 <\epsilon/K'}>0.
\end{align*}
The last inequality follows from Theorem 1 in \cite{kang2018} and the assumption that for any $\epsilon>0$, $\Pi(\|\tilde\bfgamma-\tilde\bfgamma_0\|_2^2<\epsilon)>0$.

\end{proof}

\noindent\textbf{Verifying the Existence of test condition}

To verify the existence of test condition, we need the basis expansion expression of model \eqref{eq:model1}.
Recall model \eqref{eq:model1}, we abbreviate the scalar and vector covariates and denote  $\tilde\bfgamma = (\gamma,\bfxi^\rT)^\rT\in \R^{q+1}$, $\tilde\bfX_i = (X_i,\bfC_i^\rT)^\rT\in \R^{q+1}$. Let $\tilde{\mcdf}_{i,l} = \sum_{j=1}^p \psi_l(\cvox_j)\mcdf_i(\vox_j)$, and define the $n\times L_n$ matrix $\bm{\tilde{\mcdf}}_n:=(\tilde{\mcdf}_{i,l})_{i=1,..,N, l=1,...,L_n}$. 
\begin{align*}
    Y_i & = \sum_{j=1}^{\pscale}\beta(\cvox_j)\mcdf_i(\vox_j) + \tilde\bfgamma^\rT \tilde\bfX_i   +\epsilon_i \\
    &= \sum_{j=1}^{\pscale}\br{\sum_{l=1}^{\infty}\theta_{\beta,l}\psi_l(\cvox_j) }M_i(\vox_j)+\tilde\bfgamma^\rT \tilde\bfX_i + \epsilon_i\\
    &=\sum_{l=1}^{\infty} \theta_{\beta,l}\sum_{j=1}^p \psi_l(\cvox_j)\mcdf_i(\vox_j) +\tilde\bfgamma^\rT \tilde\bfX_i+ +\epsilon_i\\
    &= \sum_{l=1}^{\infty} \theta_{\beta,l}\tilde{\mcdf}_{i,l} +\tilde\bfgamma^\rT \tilde\bfX_i +\epsilon_i\\
    &=(\bm{\tilde{\mcdf}}_n,\tilde\bfX_n) \sbr{\begin{array}{c}
         \bftheta_\beta \\
          \tilde\bfgamma
    \end{array}} + r_{L_n,i} +\epsilon_i \numberthis \label{eq:y_newform}
\end{align*}
The remainder term $r_{L_n,i} = \sum_{l=L_n}^\infty \theta_{\beta,l}\sum_{j=1}^p\psi_l(\cvox_j)\mcdf_i(\vox_j)$.

Before verifying the existence of test condition, we introduce the following lemma.

\begin{lemma} \label{lem:residual}Let independent residual terms $$r_{L_n,i} = \sum_{l=L_n}^\infty \theta_{\beta,l}\sum_{j=1}^p\psi_l(\cvox_j)\mcdf_i(\vox_j)$$ as defined in \eqref{eq:y_newform} across $i=1,\dots,n$. Denote the event $A_{L_n} = [|r_{L_n,i}|<t]$. Then for any given $i$, and for some sufficiently large positive constant $t$ ,  $\mathbb{P}[A_{L_n} i.o.] = 1$.
\end{lemma}
\begin{proof}
Denote the mean function in \eqref{eq:model2-3} of $\mcdf_i(\vox_j)$ as $\mu_{i}(\cvox_j)$.
Then $\mcdf_i(\vox_j) = p^{-1}\mu_{i}(\cvox_j) + p^{-1/2}Z_{i,j}$ where $Z_{i,j}$ is independent standard normal variable across $i=1,\dots,n, j=1,\dots,p$. 
Let $\Tilde{\mcdf}_{i,l} = \sum_{j=1}^p \mcdf_{i}(\vox_j)\psi_l(\cvox_j)$. Then
\begin{align*}
    r_{L_n,i} &= \sum_{l=L_n}^\infty \theta_{\beta,l}\Tilde{\mcdf}_{i,l}=\sum_{l=L_n}^\infty\theta_{\beta,l} \frac{1}{p}\sum_{j=1}^p \mu_{i}(\cvox_j)\psi_l(\cvox_j) + \sum_{l=L_n}^\infty \theta_{\beta,l}\frac{1}{\sqrt{p}}\sum_{j=1}^p\psi_l(\cvox_j)Z_{i,j}
\end{align*}
which implies that $r_{L_n,i}$ follows a normal distribution with mean 
\[\mu_{L_n,i,r} = \sum_{l=L_n}^\infty\theta_{\beta,l} \frac{1}{p}\sum_{j=1}^p \mu_{i}(\cvox_j)\psi_l(\cvox_j) \] 
and variance 
\[\sigma_{L_n,r}^2 = \frac{1}{p}\sum_{j=1}^p \sbr{\sum_{l=L_n}^\infty \theta_{\beta,l}\psi_l(\cvox_j)}^2.\]

Let $\theta_{M,i,l} = p^{-1}\sum_{j=1}^p \mu_{i}(\cvox_j)\psi_l(\cvox_j)$. Since $\sum_{l=L_n}^\infty \theta_{M,i,l}^2 \to 0$ for any $i$, and $\sum_{l=L_n}^\infty \theta_{\beta,l}^2 \to 0$ as $L_n \to\infty$, the mean $\mu_{L_n,i,r}\to 0$ as $n\to \infty$. 

Given the orthonormality of the basis, and denote $\beta_{L_n}(s) = \sum_{l=1}^{L_n}\theta_{\beta,l}\psi_l(s)$ as the finite basis smooth approximation of $\beta(s)$, write
\begin{align*}
   \sigma^2_{L_n,r} & = \int_\cS |\beta(s)-\beta_{L_n}(s)|^2 \dd \leb(s) + r_p=\sum_{l=L_n}^\infty \theta_{\beta,l}^2 + r_p,
\end{align*}
where the approximation error $r_p = \left|\int_\cS |\beta(s)-\beta_{L_n}(s)|^2 \dd \leb(s) - p^{-1}\sum_{j=1}^p |\beta(\cvox_j)-\beta_{L_n}(\cvox_j)|^2\right|$. From Definition \ref{def:param_space}(d) $r_p< K_\beta p^{-2/d}$, where $K_\beta>0$ is a constant. Hence $\sigma^2_{L_n,r}\to 0$ as $n\to \infty$.  

For large enough $n$, $\mu_{L_n,i,r}$ is bounded for all $i$. By the normal tail bound (Proposition 2.1.2 in \cite{vershynin2018high}), for $Z\sim N(0,1)$, $\mathbb{P}(Z>t) \leq \frac{1}{t\sqrt{2\pi}}\exp\br{-t^2/2}$.  Then we have
\begin{align*}
    \mathbb{P}(r_{L_n,i}>t) & 
    \leq \frac{\sigma_{L_n,r}}{t-\mu_{L_n,i,r}}\exp\br{-\frac{\sbr{t-\mu_{L_n,i,r}}^2}{2\sigma_{L_n,r}^2}}\leq a_n = C\sigma_{L_n,r}\exp(-c'/\sigma_{L_n,r}^2). \numberthis \label{eq:norm_tail}
\end{align*}

By Definition \ref{def:param_space}, $a_n\leq \exp(-c'n^{\nu_1\nu_2})<n^{-1}$, hence $\sum_{i=1}^n \mathbb{P}(r_{L_n,i}>t)<\infty$. 
\begin{align*}
 \mathbb{P}(A_{L_n}^c) &= \mathbb{P}(|r_{L_n,i}|> t) \leq \mathbb{P}(r_{L_n,i}>t) + \mathbb{P}(r_{L_n,i}< -t)
\end{align*}
For the $\mathbb{P}(r_{L_n,i}< -t)$ part, we only need to replace $t-\mu_{L_n,i,r}$ by $t+\mu_{L_n,i,r}$ in \eqref{eq:norm_tail}, and the same conclusion follows, $\sum_{i=1}^n \mathbb{P}(r_{L_n,i}<-t)<\infty$.
By Borel-Cantelli Lemma, we can draw the conclusion.

\end{proof}

\begin{lemma}\label{lem:beta_exist_test}(Existence of tests) There exist test functions $\Phi_{n}$, subsets $\cU_n,\Theta_{n}\subset\Theta$, and constant $K_1,K_2,c_1,c_2>0$ such that
    \begin{enumerate}
        \item[(a)] $\mathbb{E}_{\theta_0}\Phi_{n}\to 0$;
        \item[(b)] $\sup_{\theta\in \cU_n^c\cap \Theta_{n}} \mathbb{E}_{\theta}(1-\Phi_{n})\leq K_1e^{-c_1n}$;
        \item[(c)] $\Pi(\Theta_{n}^c)\leq K_2e^{-c_2n}$.
    \end{enumerate}
    
\end{lemma}
\begin{proof}
 To verify the existence of tests, we define the sieve space of $\beta$ as 
    \[\Theta_{\pscale,n}:=\left\{\beta\in\Theta_{\beta}: \sup_{s\in \cR_1\cup\cR_{-1}}\|D^{\omega}\beta(s)\|_{\infty}<\pscale^{1/(2d)},  \|\omega\|_1\leq \rho\right\}\]

     The construction of the test follows a similar idea as in Lemma 1 in \cite{armagan2013posterior}. For any $\epsilon > 0$, denote $$\cU^c = \{\beta\in \Theta_\beta,\tilde\bfgamma\in\Theta_{\tilde\bfgamma}: \|\beta-\beta_0\|_{2,\pscale} + \|\tilde\bfgamma - \tilde\bfgamma_0\|_2>\epsilon\}.$$

Following the notations and new formulation of model \eqref{eq:model1} in \eqref{eq:y_newform} under the basis decomposition, we create the test as follows. Denote $\bftheta_\beta=\sbr{\theta_{\beta,1},\dots,\theta_{\beta,L_n}}^\rT $, $\bftheta_w = (\bftheta_\beta^\top, \tilde\bfgamma^\top)^\top$    as the vector of parameters.

For any $\epsilon>0$, to test the hypothesis 
\[H_0: \{\beta(s), \tilde\bfgamma\} = \{\beta_0(s), \tilde\bfgamma_0\},\quad \mbox{v.s.} \quad H_1:  \{\beta(s), \tilde\bfgamma\} \in \cU^c.\] 
Define test function 
\[\Phi_n = I\br{ \left\| \sbr{\tilde\bfW_n^\rT \tilde\bfW_n}^{-1}\tilde\bfW_n^\rT \bfY - \bm{\theta_w^0} \right\|_2 >\frac{\epsilon}{2}},\]
where $\bfY = (Y_1,\ldots, Y_n)^\rT\in\R^n$. Let $\bm Z\sim N(0,I_{n})$ be a standard normal vector. As defined in the main text above Assumption 3, $\tilde\bfW_n = \sbr{\bm{\mathcal{\tilde M}}_n,\tilde\bfX}\in \R^{n\times(L_n+1+q)}$, and $\bm{\theta_w^0}$ denotes the true value of $\bm{\theta_w}$.

Let $\bm R_n = (r_{L_n,1},\ldots,r_{L_n,n})^\rT\in \R^n$ be the remainder term. Then under $H_0$,
\[\sbr{\bfY - \tilde\bfW_n \bm{\theta_w^0}} = \bm R_n^0 + \bm Z\sigma_Y.\] Let $H:=\sbr{\tilde\bfW_n^\rT \tilde\bfW_n}^{-1}\tilde\bfW_n^\rT$.
\[H \bfY - \bm{\theta_w^0} = H \sbr{\bfY - \tilde\bfW_n \bm{\theta_w^0}} + H \tilde\bfW_n \bm{\theta_w^0} -\bm{\theta_w^0} =H \sbr{\bfY - \tilde\bfW_n \bm{\theta_w^0}}   \]

Denote the singular value decomposition of $\tilde\bfW_n$ as $\tilde\bfW_n=U\Lambda V^\rT$ where $UU^\rT = I_n$, $VV^\rT = I_{L_n}$, $\Lambda$ is at most rank $L_n$, and the smallest singular value is $\sigma_{\min,n}$. Let $\sigma_{\min,n} := \sigma_{\min}(\tilde\bfW)$. For some positive constant $c_{\min}$, denote event $$\Sigma = [\sigma_{\min,n}>c_{\min}\sqrt{n}].$$

Let $\chi^2(a,b)$ denote the non-central $\chi^2$ distribution with non-central parameter $a$ and degree $b$. Then under event $\Sigma$,
\begin{align*}
   &\left\| H \sbr{\bfY - \tilde\bfW_n \bm{\theta_w^0}}  \right\|_2^2
   =\left\| H\sbr{\bm R_n^0 + \bm Z\sigma_Y}\right\|_2^2\\
   &= \sbr{\bm R_n^0 + \bm Z\sigma_Y}^\rT U \Lambda^{-2} U^\rT \sbr{\bm R_n^0 + \bm Z\sigma_Y}\\
   &\leq \sbr{\bm R_n^0 + \bm Z\sigma_Y}^\rT \sigma_{\min,n}^{-2} \sbr{\begin{array}{cc}
        I_{L_n}& 0  \\
        0 & 0_{n-L_n}
   \end{array} }\sbr{\bm R_n^0 + \bm Z\sigma_Y}\\
   &=\sigma_Y^2 \sigma_{\min,n}^{-2} \sbr{\bm R_n^0/\sigma_Y + \bm Z}^\rT \sbr{\begin{array}{cc}
        I_{L_n}& 0  \\
        0 & 0_{n-L_n}
   \end{array} }\sbr{\bm R_n^0/\sigma_Y + \bm Z} \sim \sigma_Y^2 \sigma_{\min,n}^{-2} \chi^2(L_n, u_n )
\end{align*}
$u_{n} = \frac{1}{\sigma_Y^2}\left\|\sbr{\begin{array}{cc}
        I_{L_n}& 0  \\
        0 & 0_{n-L_n}
   \end{array} }\bm R_n^0\right\|_2^2$ is the non-central parameter in the non-central $\chi^2$ distribution of order $L_n$. Each element in $R_n^0$ is the residual term $r_{L_n,i}$.

Several results are available for the upper bound of noncentral $\chi^2$ tail probability, here we use Theorem 7 in \cite{zhang2020non}, when $x>L_n+2u_{n,\sigma_Y}$, for some constant $c$,
\[\bP\left\{\chi^2(L_n, u_n ) - (L_n+u_n)>x \right\}<\exp\sbr{-cx}\]
Hence if $\epsilon^2/(4\sigma_Y^2)\sigma_{\min,n} > L_n+2u_{n,\sigma_Y}$, 
\begin{align*}
    \bE_{\theta_0}\br{\Phi_nI(\Sigma)}& \leq \bP\br{\sigma_Y^2\sigma_{\min,n}^{-2}\chi^2(L_n,u_{n,\sigma_Y}) > \frac{\epsilon^2}{4}}  = \bP\br{\chi^2(L_n,u_{n,\sigma_Y}) > \frac{\epsilon^2}{4\sigma_Y^2} \sigma_{\min,n}^2 }\\
    &\leq \exp\br{-c\sbr{\frac{\epsilon^2}{4\sigma_Y^2} \sigma_{\min,n}^2-L_n-u_{n,\sigma_Y}}}.
\end{align*}

By Lemma \ref{lem:residual}, for sufficiently large $n$, $|r_{L_n,i}|<c_0$ with probability 1.
Note that $L_n+u_{n,\sigma_Y}< \sbr{1 + c_0^2/\sigma_Y^2}L_n$, given $\sigma_{\min,n}>\sqrt{n}c_{\min}>0$, for sufficiently large $n$, there exists a constant $c'>0$ such that 
$\epsilon^2/(4\sigma_Y^2)\sigma_{\min,n}^2-L_n-u_{n,\sigma_Y} >c'n$.  Hence by Assumption \ref{asm:W},
  $\bE_{\beta_0}\br{\Phi_nI(\Sigma)}\leq  \exp\br{-c'n}$
and
$
    \bE_{\beta_0}(\Phi) = \bE_{\beta_0}\br{\Phi I(\Sigma)} + \bE_{\beta_0}\br{\Phi I(\Sigma^c)}\leq  \exp\br{-c'n} + \exp\br{-\tilde c n} \leq \exp\br{-\tilde c'n},
$ for $n > 2\log(2)/\tilde{c}'$, 
where $\tilde{c}' = \min\{\tilde{c},c'\}/2$. 

To find the upper bound of the Type II error, let 
$\tilde{r}_p = \int_{\mathcal{S}}\{\beta(s)-\beta_0(s)\}^2\lambda(d s) - \|\beta(s) - \beta_0(s)\|^2_{2,p}$ and 
$r_{L_n} = \sum_{l=L_n}^\infty \theta_{\beta,l}^2$. Then $\tilde{r}_p \to 0$ as $p\to \infty$ and $r_{L_n}\to 0$ as $n\to \infty$. 
Note that
\begin{align*}
    \int_{\mathcal{S}}\{\beta(s)-\beta_0(s)\}^2\lambda(d s)=\int_{\cS}\br{\sum_{l=1}^\infty(\theta_{\beta,l} - \theta_{\beta^0,l})\psi_l(s) }^2\lambda(\dd s)
    =\|\bm{\theta_\beta} - \bm{\theta_{\beta^0}}\|_2^2+r_{L_n},
\end{align*}
where $\bm{\theta_\beta},\bm{\theta_{\beta^0}}\in \R^{L_n}$. 
By $\|\bm{\theta_w} - \bm{\theta_{w}^0}\|_2^2=\|\bm{\theta_\beta} - \bm{\theta_{\beta^0}}\|_2^2+ \|\tilde\bfgamma - \tilde\bfgamma_0\|_2^2$, 
$$\|\beta(s)-\beta_0(s)\|_{2,\pscale}^2 + \|\tilde\bfgamma - \tilde\bfgamma_0\|_2^2  =\|\bm{\theta_w} - \bm{\theta_{w}^0}\|_2^2-\tilde{r}_{\pscale}+r_{L_n}.$$
For a sufficiently large $n$ and $p$, we have $\tilde{r}_p < \epsilon^2/16$ and $r_{L_n} < \epsilon^2/16$. Then $r_{L_n} - r_p < \epsilon^2/8$. Thus, when $\|\beta(s)-\beta_0(s)\|_{2,\pscale}^2 + \|\tilde\bfgamma - \tilde\bfgamma_0\|_2^2>\epsilon^2/2$, $\|\bm{\theta_w} - \bm{\theta_{w}^0}\|_2^2>3\epsilon^2/8$. 

Recall $$\cU^c = \{\beta\in \Theta_\beta,\tilde\bfgamma\in\Theta_{\tilde\bfgamma}: \|\beta-\beta_0\|_{\pscale} + \|\tilde\bfgamma - \tilde\bfgamma_0\|_2>\epsilon\}.$$
Define the sieve space $\Theta_{n}:=\Theta_{p,n}\times \Theta_{\bfgamma}$.
\begin{align*}
    \sup_{ \cU^c \cap \Theta_{n}} \bE_{\beta}(1-\Phi_n)I(\Sigma)&=\sup_{ \cU^c \cap \Theta_{n}} \bP\br{\left\|H\bfY - \bm{\theta_w^0}\right\|_2\leq \frac{\epsilon}{2}}\\ 
    &\leq \sup_{ \cU^c \cap \Theta_{n}} \bP \br{\left| \left\|H\bfY - \bm{\theta_w}\right\|_2 - \left\|\bm{\theta_w}- \bm{\theta_w^0}\right\|_2 \right| \leq\frac{\epsilon}{2}}\\
    &\leq \sup_{ \cU^c \cap \Theta_{n}} \bP \br{ \left\|H\bfY - \bm{\theta_w}\right\|_2 > -\frac{\epsilon}{2} +  \left\|\bm{\theta_w}- \bm{\theta_w^0}\right\|_2}\\
    &\leq \sup_{ \cU^c \cap \Theta_{n}} \bP \br{ \left\|H\bfY - \bm{\theta_w}\right\|_2 > c_1\epsilon },
\end{align*}
where $c_1=\sbr{\sqrt{3/8}-1/2}$.

Lastly, by Lemma 4 in \cite{kang2018}, for some constant $c_2$, $\Pi(\Theta_{n}^c)\leq K_2'e^{-c_2\pscale^{1/d}}\leq K_2e^{-c_2n}$ with Assumption \ref{asm:true_func}(b) that $\pscale \geq O(n^{\tau d})$.

\end{proof}

\subsection{Proof of Theorem \ref{thm:total_consistency}}
\begin{proof}
First we show that, conditioning on all other parameters, the joint posterior of $\alpha(s)$ and $\beta(s)$ can be factored into the marginal posteriors of $\alpha(s)$ and $\beta(s)$. Let $\bfD = \br{\bfY, \bfM, \bfX, \bfC}$.  For simplicity, we omit ``$(s)$" in $\alpha(s)$ and $\beta(s)$ in the following derivation.
\begin{align*}
    \Pi(\alpha,\beta \mid \bfD) &= \frac{\Pi(\bfD\mid \alpha,\beta)\pi(\alpha,\beta)}{\Pi(\bfD)}\\
    &=\frac{\Pi(\bfM,\bfY\mid \alpha,\beta, \bfX, \bfC)\pi(\alpha)\pi(\beta)\pi( \bfX, \bfC)}{\Pi(\bfY\mid \bfM, \bfX, \bfC)\Pi(\bfM \mid  \bfX, \bfC)\pi( \bfX, \bfC)}\\
    &=\frac{\Pi(\bfY \mid \bfM,\alpha,\beta, \bfX, \bfC)\Pi(\bfM\mid \alpha,\beta, \bfX, \bfC)\pi(\alpha)\pi(\beta)}{\Pi(\bfY\mid \bfM, \bfX, \bfC)\Pi(\bfM\mid \bfX, \bfC)}\\
    &=\frac{\Pi(\bfY\mid \bfM,\beta,\bfX, \bfC)\pi(\beta)}{\Pi(\bfY\mid \bfM,\bfX, \bfC)} \frac{\Pi(\bfM\mid\alpha,\bfX, \bfC)\pi(\alpha)}{\Pi(\bfM\mid \bfX, \bfC)}
    \\
    &=\Pi(\beta\mid \bfD)\Pi(\alpha\mid \bfD)
\end{align*}
Now,
\begin{align*}
    \Pi&\left( \|\alpha\beta-\alpha_0\beta_0\|_{1,\pscale}>\epsilon\mid \bfD\right)\\
    &=\Pi\left( \|(\beta-\beta_0)(\alpha-\alpha_0)+\alpha_0(\beta-\beta_0)+\beta_0(\alpha-\alpha_0)\|_{1,\pscale}>\epsilon\mid \bfD\right)\\
    &\leq \Pi\left( \|(\beta-\beta_0)(\alpha-\alpha_0)\|_{1,\pscale}+\|\alpha_0(\beta-\beta_0)\|_{1,\pscale}+\|\beta_0(\alpha-\alpha_0)\|_{1,\pscale}>\epsilon\mid \bfD\right)\\
    &\leq \Pi\left( \|(\beta-\beta_0)(\alpha-\alpha_0)\|_{1,\pscale}>\epsilon\mid \bfD\right)+
    \Pi\left( \|\beta_0(\alpha-\alpha_0)\|_{1,\pscale}>\epsilon\mid \bfD\right)+\\
    &\qquad\Pi\left( \|\alpha_0(\beta-\beta_0)\|_{1,\pscale}>\epsilon\mid \bfD\right)\numberthis\label{eq:26}
\end{align*}
Given that both $\alpha_0$ and $\beta_0$ are defined on a compact set $\cS\in \R^d$ (Definition \ref{def:param_space}), there exists $K>0$ such that $\|\alpha_0\|_\infty\leq K$ and $\|\beta_0\|_\infty\leq K$, by Theorem \ref{thm:alpha_consistency}, \ref{thm:beta_consistency}, and the norm inequality $\|\cdot\|_{1,p}\leq \|\cdot\|_{2,p}$, the last two terms in (\ref{eq:26}) goes to 0 in $P^{(n)}_{\alpha_0,\beta_0}$-probability as $n\to \infty$.

For any $\delta>0$,
\begin{align*}
    \Pi&\left( \|(\beta-\beta_0)(\alpha-\alpha_0)\|_{1,\pscale}>\epsilon\mid \bfD\right)\\
    &\leq \Pi\left( \|\beta-\beta_0\|_{2,\pscale}\|\alpha-\alpha_0\|_{2,\pscale}>\epsilon \mid \bfD\right)\\
    &\leq \Pi\left( \|\beta-\beta_0\|_{2,\pscale}\|\alpha-\alpha_0\|_{2,\pscale}>\epsilon\mid \bfD, \|\alpha-\alpha_0\|_{2,\pscale}>\delta\right)\Pi\left( \|\alpha-\alpha_0\|_{2,\pscale}>\delta \mid \bfD\right) +\\
    &\qquad \Pi\left( \|\beta-\beta_0\|_{2,\pscale}\|\alpha-\alpha_0\|_{2,\pscale}>\epsilon\mid \bfD, \|\alpha-\alpha_0\|_{2,\pscale}<\delta\right)\Pi\left( \|\alpha-\alpha_0\|_{2,\pscale}<\delta \mid \bfD\right) \\
    &\leq \Pi\left( \|\alpha-\alpha_0\|_{2,\pscale}>\delta \mid \bfD\right) + 
    \Pi\left( \|\beta-\beta_0\|_{2,\pscale}\delta>\epsilon \mid \bfD, \|\alpha-\alpha_0\|_{2,\pscale}<\delta\right)\\
    &=\Pi\left( \|\alpha-\alpha_0\|_{2,\pscale}>\delta \mid \bfD\right) + 
    \Pi\left( \|\beta-\beta_0\|_{2,\pscale}\delta>\epsilon\mid \bfD\right).
\end{align*}
As $n\to\infty$, $\Pi\left( \|\alpha-\alpha_0\|_{2,\pscale}>\delta \mid \bfD\right) \to 0$ and
    $\Pi\left( \|\beta-\beta_0\|_{2,\pscale}\delta>\epsilon\mid \bfD\right)\to 0$ in $P^{(n)}_{\alpha_0,\beta_0}$-probability,  which implies that $\Pi\left( \|(\beta-\beta_0)(\alpha-\alpha_0)\|_{1,\pscale}>\epsilon\mid \bfD\right)\to 0$

\end{proof}

\subsection{Proof of Corollary \ref{corol:sign_consistency}}
\begin{proof}

The proof of the sign consistency is similar to Theorem 3 in \cite{kang2018}.

To show Corollary \ref{corol:sign_consistency}, for simplicity, denote $\cE(s):=\alpha(s)\beta(s)$ and $\cE_0(s):=\alpha_0(s)\beta_0(s),~\forall s\in \cS$ as the true function of the total effect.
Since both $\alpha(s)$ and $\beta(s)$ satisfy Definition \ref{def:sparse_param}, we use the notations 
\[\cR^{f}_i:=\Big\{s\in\cS: \sgn \{f(s)\}=i\Big\},~f\in\{\alpha,\beta\},~i\in\{-1,0,1\},\]
and by Definition \ref{def:sparse_param}, $\cR^{\alpha}_{\pm 1},\cR^{\beta}_{\pm 1}$ are open sets.
Define  $\cR^{\cE}_1 = \Big(\cR^{\alpha}_1\cap\cR^{\beta}_1\Big)\cup\Big(\cR^{\alpha}_{-1}\cap\cR^{\beta}_{-1}\Big)$, 
$\cR^{\cE}_{-1} = \Big(\cR^{\alpha}_{-1}\cap\cR^{\beta}_1\Big)\cup\Big(\cR^{\alpha}_{1}\cap\cR^{\beta}_{-1}\Big)$,
$\cR^{\cE}_0 = \cS-(\cR^{\cE}_1\cup\cR^{\cE}_{-1})$, $\cR^{\cE}_{\pm 1}$ are open sets. To show $\cR^{\cE}_0$ has nonempty interior, if we denote $\bar A:=S-A$ as the complementary set of $A$ in $S$, we only need to show \[\left(\overline{\cR_1^\alpha\cup\cR^\alpha_{-1}}\right)\cup \left(\overline{\cR_1^\beta\cup\cR^\beta_{-1}}\right)
\subseteq R_0^\cE\]
where the LHS has nonempty interior by the Definition \ref{def:sparse_param}. $\cR_0^\cE = \overline{\cR_1^\cE} \cap \overline{\cR_{-1}^\cE}$,
\begin{align*}
    \overline{\cR_1^\cE} &= \overline{\Big(\cR^{\alpha}_1\cap\cR^{\beta}_1\Big)}\cap\overline{\Big(\cR^{\alpha}_{-1}\cap\cR^{\beta}_{-1}\Big)}\\
    &= \left(\overline{\cR_1^\alpha\cup\cR^\alpha_{-1}}\right)\cup \left(\overline{\cR_1^\beta\cup\cR^\beta_{-1}}\right) \cup
    \left(\overline{\cR_1^\beta\cup\cR^\alpha_{-1}}\right)\cup \left(\overline{\cR_1^\alpha\cup\cR^\beta_{-1}}\right)
\end{align*}
Similarly we can show $\left(\overline{\cR_1^\alpha\cup\cR^\alpha_{-1}}\right)\cup \left(\overline{\cR_1^\beta\cup\cR^\beta_{-1}}\right)\subseteq \overline{\cR_{-1}^\cE}$, hence $\cR_0^\cE$ has nonempty interior. The parameter space of $\cE$, $\Theta_\cE$ satisfies Definition \ref{def:sparse_param}.

Now denote $\cS_0=\{s\in\cS:\cE_0(s)=0\}$, $\cS_+=\{s\in\cS:\cE_0(s)>0\}$, $\cS_-=\{s\in\cS:\cE_0(s)<0\}$. 
Notice that $\cR^{\cE_0}_{\pm 1}\subseteq\cS_{\pm}$, and $\cS_{0}\subseteq \cR^{\cE_0}_0$. The key difference is that $\cS_{0,\pm}$ are not necessarily open sets.

For any $\cA\subseteq\cS$ and any integer $m\geq 1$, let $Q_{\pscale}$ be the discrete measure that assigns $1/p$ mass to each fixed design points in $\br{s_j}_{j=1}^p$,  define \[\cF_m(\cA):=\left\{\cE\in\Theta_\cE: \int_{\cA}|\cE(s)-\cE_0(s)|d Q_{\pscale}(s)<\frac{1}{m}\right\}.\]
Note that for any $\cA\subseteq\cB\subseteq\cS$, we have $\cF_m(\cS)\subseteq\cF_m(\cB)\subseteq\cF_m(\cA)$.

\[\Pi(\cF_{m}(\cS_0)\mid \bfD)\geq \Pi(\cF_{m}(\cS)\mid \bfD)\to 1,  \ \mbox{as}\  n\to\infty.\]
By the monotone continuity of probability measure,
\[\Pi\{ \cE(s)=\cE_0(s)=0\mid \bfD\}=\Pi\{ \cE(s)=0, s\in\cS_0\mid \bfD\}=\lim_{m\to \infty}\Pi\{ \cF_m(\cS_0)\mid \bfs\} = 1,\ \mbox{as}\ n\to\infty.\]

Now to show the consistency of the positive sign, for any small $\delta>0$, denote $\cS_{+}^\delta:=\{s\in\cS: \cE_0(s)>\delta\}$. Because $\mathcal{E}_0$ is a continuous function, its preimage $\mathcal{E}^{-1}((\delta,\infty))$ supported on $\R^d$ is also an open set, $\cS_{+}^\delta = \cR_{1}^{\mathcal{E}} \cap \mathcal{E}^{-1}((\delta,\infty))$ hence is also an open set.

For any $s_0\in \cS_+^{\delta}$, we can find a small open ball $B(s_0,r_1)=\br{s: \|s-s_0\|_2< r_1}\subseteq  \cS_+^{\delta}$. By the continuity of $\mathcal{E}$ and $\mathcal{E}_0$, for any large $m$, there exists $r_2>0$ such that $\|s-s_0\|_2< r_2$ implies $|\mathcal{E}(s) - \mathcal{E}(s_0)|<1/m$. Let $r = \min\br{r_1,r_2}$. 

For any open subset $B$ in $\cS$, Definition \ref{def:param_space}(d) implies that for any large $m$, there exists $N_m$ such that for any $n>N_m$,
\[ \left| \int_{B}|\mathcal{E} - \mathcal{E}_0| Q_p(\dd s) - \int_{B}|\mathcal{E} - \mathcal{E}_0|\leb(\dd s) \right| <\frac{V_m}{2m},\]
 where we denote $V_m = \leb\{B(s_0,r)\}\to 0$ as $m\to\infty$.

Hence for any small $\delta>0$, notice that 
\begin{align*}
    &\frac{1}{V_m}\int_{B(s_0,r)} |\cE(s) - \cE_0(s)|\leb(\dd s) <\frac{1}{m}\\
    \Rightarrow\quad& \frac{1}{V_m}\int_{B(s_0,r)} \cE(s)\leb(\dd s) > \frac{1}{V_m}\int_{B(s_0,r)} \cE_0(s)\leb(\dd s) -\frac{1}{m}\\
    \Rightarrow\quad & \frac{1}{V_m}\int_{B(s_0,r)} \cE(s)\leb(\dd s) > \delta - \frac{1}{m}\\
    \Rightarrow\quad &\exists s_1\in B(s_0,r),~s.t. ~\cE(s_1) > \delta - \frac{1}{m}\\
    \Rightarrow\quad & \cE(s_0) +\frac{1}{m}> \delta - \frac{1}{m}, \forall s_0\in S_+^\delta.
\end{align*}
Hence we have 
\begin{align*}
    &\Pi\br{\forall s_0\in\cS_+^\delta,\cE(s_0)>0\mid \bfD} \geq \Pi\br{\forall s_0\in\cS_+^\delta,\cE(s_0)\geq \delta \mid \bfD}\\
    =&\lim_{m\to\infty}\Pi\br{\forall s_0\in\cS_+^\delta,~ \cE(s_0)>\delta - \frac{2}{m}\Big|\bfD}\\
    \geq &\lim_{m\to\infty} \Pi\br{\int_{B(s_0,r)} |\cE(s) - \cE(s_0)|\leb(\dd s) <\frac{V_m}{m} \Big | \bfD}\\
    \geq &\lim_{m\to\infty} \Pi\br{\int_{B(s_0,r)} |\cE(s) - \cE(s_0)|\dd Q_p(s) <\frac{V_m}{2m} \Big | \bfD} = 1, 
\end{align*}

The proof for the consistency of the negative sign is similar to the positive sign.

\end{proof}

\section{Example for Assumption \ref{asm:W}} \label{subsec:verify}

In this section, we give an example that demonstrates the generative model \eqref{eq:model2} satisfies Assumption \ref{asm:W} under some stronger assumptions.

\begin{assumption}\label{asm:M_cond} When viewing the mediator model~\eqref{eq:model2} as the true generative model of $\tilde\bfW_n$, assume 
\begin{enumerate}
    \item for any $s\in\cS$,
         $\sum_{i=1}^n X_i\epsilon_{M,i}(s) = 0$ and $\sum_{i=1}^n C_{k,i}\epsilon_{M,i}(s) = 0, ~k=1,\dots,q $, with probability one;
    \item for the chosen basis $\br{\psi_l(s)}_{l=1}^\infty$, the individual effects $\eta_i(s)$ can be viewed as one realization of the random Gaussian process $\eta_i\sim \cGP(0,\sigma_\eta \kappa)$, and can be decomposed as $\eta_i(s) = \sum_{l=1}^\infty \theta_{\eta,i,l}\psi_l(s)$ where $\theta_{\eta,i,l}\overset{\ind}{\sim}\normal(0,\sigma_{\eta}^2 \lambda_l)$;
\end{enumerate}
\end{assumption}

\begin{proposition}\label{prop:rand_mat}
Under Assumption \ref{asm:M_cond}, the least singular value of $\tilde\bfW_n$ satisfies 
\[0 < c_{\min} < \liminf_{n\to\infty}\sigma_{\min}(\tilde\bfW_n)/\sqrt{n}\]
with probability $1-\exp(-\tilde c n)$ for some constant $\tilde c ,c_{\min}>0$.
\end{proposition}

Recall the notations in \eqref{eq:y_newform}, $\tilde\bfW_n = (\bm{\tilde{\mcdf}}_n,\tilde\bfX) \in \R^{n\times(L_n+1+q)}$, and $\tilde\bfX_i = (X_i,\bfC_i^\rT)^\rT\in \R^{q+2}$.

The proof of Proposition \ref{prop:rand_mat} needs to show that the least singular value of $\tilde\bfW_n$, denoted as $\sigma_{\min}(\tilde\bfW_n)$ satisfies that
\[ \mathbb{P}\sbr{\sigma_{\min}(\tilde\bfW_n)<c\sqrt{n}\mid \bfX,\bfC}\leq e^{-c'n}\]

\begin{proof}

Given \eqref{eq:model2-3} for $\mcdf(\vox)$ and $\leb(\vox_j)=\frac{1}{p}$, we can write
\begin{align*}
    \tilde{\mcdf}_{i,l} = \tilde\theta_{\alpha,l}X_i + \sum_{k=1}^q \tilde\theta_{\zeta,k,l}C_{i,k} + \theta_{\eta,i,l} + \tilde\varepsilon_{i,l}
\end{align*}
where $\tilde\varepsilon_{i,l}\sim N\{0,(\sigma_M^2/p)\sum_{j=1}^p \psi_l(\cvox_j)^2\}$, and each $\tilde\theta_{\alpha,l} = \langle \alpha,\psi_l\rangle_{\pscale}$, $\tilde\theta_{\zeta,k,l} = \langle \zeta_k,\psi_l\rangle_{\pscale}$. Hence we can write
\begin{align*}
    \bm{\tilde\mcdf}_n = \tilde\bfX \bftheta_{M}+\bfTheta_E
\end{align*}
Here, $\bftheta_M = \sbr{\begin{array}{cc}
     \tilde\theta_{\alpha,1},\dots,\tilde\theta_{\alpha,L_n}\\
     \tilde\theta_{\zeta_1,1},\dots,\tilde\theta_{\zeta_1,L_n}\\
    \dots\\
    \tilde\theta_{\zeta_q,1},\dots,\tilde\theta_{\zeta_q,L_n}
\end{array}
}\in \R^{(q+1)\times L_n}$, 
$(\bfTheta_E)_{i,l} =\langle\eta_i,\psi_l\rangle_{\pscale} + \tilde\varepsilon_{i,l}$, .

By Assumption \ref{asm:alpha_iden}(c) and Assumption \ref{asm:M_cond}.1, we have that $\bfTheta_E^\rT\tilde\bfX =\bfzero$. Denote $\bfA_n = \tilde\bfX^\rT\tilde\bfX$, then
\begin{align*}
    \tilde\bfW_n^\rT\tilde\bfW_n &= \sbr{\begin{array}{cc}
        \bm{\tilde\mcdf}_n^\rT\bm{\tilde\mcdf}_n & \bm{\tilde\mcdf}_n^\rT \tilde\bfX \\
         \tilde\bfX^\rT\bm{\tilde\mcdf}_n &  \tilde\bfX^\rT \tilde\bfX
    \end{array}}
    =\sbr{\begin{array}{cc}
       \sbr{\tilde\bfX \bftheta_{M}+\bfTheta_E}^\rT \sbr{\tilde\bfX \bftheta_{M}+\bfTheta_E}  & \sbr{\tilde\bfX \bftheta_{M}+\bfTheta_E}^\rT \tilde\bfX \\
        \tilde\bfX^\rT\sbr{\tilde\bfX \bftheta_{M}+\bfTheta_E} & \tilde\bfX^\rT\tilde\bfX 
    \end{array}}\\
    &=\sbr{\begin{array}{cc}
        \bftheta_M^\rT \bfA_n \bftheta_M + \bfTheta_E^\rT\bfTheta_E & \bftheta_M^\rT \bfA_n \\
        \bfA_n\bftheta_M & \bfA_n 
    \end{array}}.
\end{align*}
Furthermore,
\begin{align*}\sbr{\tilde\bfW_n^\rT\tilde\bfW_n}^{-1} = \sbr{\begin{array}{cc}
        \sbr{ \bfTheta_E^\rT\bfTheta_E}^{-1} &  -\sbr{ \bfTheta_E^\rT\bfTheta_E}^{-1}\bftheta_M^\rT \\
         -\bftheta_M\sbr{ \bfTheta_E^\rT\bfTheta_E}^{-1}& \bfA_n^{-1} + \bftheta_M \sbr{ \bfTheta_E^\rT\bfTheta_E}^{-1} \bftheta_M^\rT
    \end{array}}.
\end{align*}
This implies that the Schur complement of $\bfA_n$ in $\sbr{\tilde\bfW_n^\rT\tilde\bfW_n}^{-1}$ is $\sbr{ \bfTheta_E^\rT\bfTheta_E}^{-1}$.
Denote $\|\cdot\|$ as the operator norm.
Notice that $\frac{1}{\sigma^2_{\min}\sbr{\tilde\bfW_n}} = \|\tilde\bfW_n^{-1}\|^2 = \nbr{\sbr{\tilde\bfW_n^\rT\tilde\bfW_n}^{-1}}$. By Lemma \ref{lem:M_least}, $\sigma_{\min}\sbr{\Theta_E}$ has a lower bound $c\sqrt{n}$ with probability $1-e^{-c'n}$.
\begin{align*}
    \nbr{\sbr{\tilde\bfW_n^\rT\tilde\bfW_n}^{-1}}^2&\leq \nbr{\sbr{ \bfTheta_E^\rT\bfTheta_E}^{-1}}^2 + 2\nbr{\sbr{ \bfTheta_E^\rT\bfTheta_E}^{-1}\bftheta_M^\rT}^2 + \nbr{\bfA_n^{-1} + \bftheta_M \sbr{ \bfTheta_E^\rT\bfTheta_E}^{-1}\bftheta_M^\rT}^2\\
    &\leq \frac{1}{\sigma^4_{\min}\sbr{\bfTheta_E}}\sbr{1+\nbr{\bftheta_M}^2}^2 + \nbr{\bfA_n^{-1}}^2
\end{align*}
Note that $\sum_{l=1}^\infty\theta_{\alpha,l}^2<\infty$ and $\sum_{l=1}^\infty\theta_{\zeta_k,l}^2<\infty,k=1,..,q$, hence $\nbr{\bftheta_M}$ is bounded by a constant. With Assumption \ref{asm:alpha_iden}.(a), $\sigma_{\min}(\bfA_n)>n$. Hence with probability $1-e^{-c'n}$,
\begin{align*}
    \frac{1}{\sigma^4_{\min}\sbr{\tilde\bfW_n}}\leq C \frac{1}{\sigma^4_{\min}\sbr{\bfTheta_E}} + \frac{1}{n^2} \leq \frac{C'}{n^2}
\end{align*}
Hence $\sigma_{\min}\sbr{\tilde\bfW_n}>c\sqrt{n}$ with probability $1-e^{-c'n}$.
\end{proof}

\begin{lemma}\label{lem:M_least}
Under model \eqref{eq:model2},
\[\bm{\tilde\mcdf}_n = \tilde\bfX \bftheta_{M}+\bfTheta_E\]
Then under assumptions \ref{asm:true_func}-\ref{asm:M_cond},  the smallest singular value $\sigma_{\min}\sbr{\bfTheta_E}$ satisfies that, for some $c_1,c_2>0$,
\begin{equation}\label{eq:singular_ineq}
    \mathbb{P}\left\{ \sigma_{\min}\sbr{\bfTheta_E}<c_1  \sqrt{n} \left|\bm{X},\bm{C}\right.\right\} \leq e^{-c_2n}
\end{equation}

\end{lemma}

\begin{proof}

To show \eqref{eq:singular_ineq}, we can write 
\begin{align*}
   \bm{ \Theta}_E & = \bm{\tilde \Theta}_{\eta} + \bm{\tilde\Theta}_{E} + \bm{R}_p.
\end{align*}
To unpack each matrix, we give the $(i,l)$th element in each matrix: $\sbr{\bm{\tilde \Theta}_{\eta}}_{i,l} = \int_{\cS} \eta_i(s)\psi_l(s)\leb(\dd s)$, $\sbr{\bm{\tilde\Theta}_{E}}_{i,l} \sim N(0,\sigma_M^2\int_{\cS}\psi_l(s)^2\leb(\dd s)) $.
Note that we view $\eta_i(s)$ as independent copies of Gaussian processes, and by Assumption \ref{asm:M_cond}(b), $\sbr{\bm{\tilde \Theta}_{\eta}}_{i,l}\sim N(0,\lambda_l \sigma_\eta^2)$.

The remainder term $\bm{R}_p$ is the approximation error between the continuous integrals and their fixed grid approximation. Denote the fixed grid approximations as $\sbr{\bm{ \Theta}^*_{\eta}}_{i,l} = \frac{1}{p}\sum_{j=1}^p \eta_i(\cvox_j)\psi_l(\cvox_j)$, $\sbr{\bm{\Theta}^*_{E}}_{i,l} \sim N(0,\frac{1}{p}\sum_{j=1}^p\psi_l^2(\cvox_j))$, and $\bm{R}_p = \br{\bm{ \Theta}^*_{\eta} - \bm{ \tilde \Theta}_{\eta}} + \br{\bm{\Theta}^*_{E} - \bm{\tilde \Theta}_{E}}$, and $|\sbr{\bm{R}_p}_{i,l}|\leq K p^{-2/d}$ almost surely for all $i,l$,

We need to show 
\begin{enumerate}
    \item[(i)] $\sigma_{\min}\sbr{\bm{\tilde \Theta}_E}$ has a lower bound $c\sqrt{n}$ with probability $1-e^{-\tilde{c}n}$.
    \item[(ii)] $\sigma_{\min}\sbr{\bm{\tilde \Theta}_E + \bm{\tilde \Theta}_\eta}$ has a lower bound $c\sqrt{n}$ with probability $1-e^{-\tilde{c}n}$.
    \item[(iii)] Adding the error term $\bm{R}_p$ does not change this lower bound.
\end{enumerate}

To show (i), let $\bm{Z}$ be an $L_n\times n$ dimensional random matrix where the entries are i.i.d standard normal variables. Then by Theorem 1 in \cite{rudelson2009smallest}, 
\[\mathbb{P}\br{ \sigma_{\min}\sbr{\bm{Z}}<\epsilon \sbr{ \sqrt{n}-\sqrt{L_n-1} }  } \leq \sbr{C\epsilon}^{n-L_n+1}+e^{-cn}\]
Because we have $L_n=o(n)$ (Assumption \ref{asm:M_cond}.3), hence we use a relaxed lower bound, for some $c_0,c'_0>0$,
\[\mathbb{P}\left( \sigma_{\min}\sbr{\bm{Z}}<c_0\sqrt{n}   \right) \leq e^{-c'_0n}.\]

Because $\psi_l$ forms an orthonormal basis, $\int_\cS\psi_l^2(s)\leb(\dd s) = 1$, $\bm{\tilde \Theta}_E = \sigma_M Z$.

To show (ii), note $\bm{\tilde \Theta}_\eta = \sigma_\eta \Lambda Z$. $\Lambda$ is the diagonal matrix with element $\lambda_l$. $\bm{\tilde \Theta}_\eta + \bm{\tilde \Theta}_E = D_E Z$ where $D_E$ is a diagonal matrix with $l$th element $\sqrt{\sigma_\eta^2\lambda_l +\sigma_M^2}$. For any $x\in \R^{L_n}$,
\begin{align*}
    \sigma_{\min}\sbr{D_E Z} &= \min_{\|x\|_2=1} \| Z^\rT D_E^\rT x\|_2 =  \min_{\|x\|_2=1} \frac{\| Z^\rT D_E^\rT x\|_2}{\|D_E^\rT x\|_2} \|D_E^\rT x\|_2 \geq \min_{\|y\|_2=1} \| Z^\rT y\|_2 \min_{\|x\|_2=1} \|D_E^\rT x\|_2 \\
    &= \sigma_{\min}\sbr{Z} \sigma_{\min}\sbr{D_E}
\end{align*}
Hence $\sigma_{\min} \sbr{\bm{\tilde \Theta}_\eta + \bm{\tilde \Theta}_E}= \sigma_{\min}\sbr{D_E Z} \geq  \sigma_{\min}\sbr{Z} \sigma_{\min}\sbr{D_E} \geq \sqrt{\sigma_\eta^2 \lambda_{L_n} + \sigma_M^2} \sigma_{\min, n}\sbr{Z}$. Since $\lambda_{L_n}\to 0$ as $n\to \infty$, $\sigma_M^2$ is the leading term.

To show (iii), by Weyl's inequality, $\sigma_{\min}\sbr{ \bm{\tilde \Theta}_\eta + \bm{\tilde \Theta}_E  + \bm{R}_p } \geq \sigma_{\min}\sbr{ \bm{\tilde \Theta}_\eta + \bm{\tilde \Theta}_E  } -\sigma_{\max}\sbr{ \bm{R}_p }$. Since we have $\max_{i,l} |\sbr{\bm{R}_p}| \leq K p^{-2/d}$, by Assumption \ref{asm:true_func} and \ref{asm:M_cond}, $\sigma_{\max}\sbr{\bm{R}_p} \leq K\sqrt{n\times L_n}p^{-2/d}\leq n^{\frac{\nu_1+1}{2} - 2\tau}\to 0$ as $n\to\infty$ (Assumption \ref{asm:true_func}).
\end{proof}


\section{Approximation of the mediator model \eqref{eq:model2}}\label{supp_sec:approx_mediator}

In model \eqref{eq:model1}, we use the discretized $\mcdf(\vox_j)$. Regarding $\mcdf_i(\vox)$, the general definition $\mcdf_i(\vox) = \int_{\vox} M_i(s)\leb(\dd s)$ is consistent with Equation \eqref{eq:model2-3}, by plugging in the M-regression in \eqref{eq:model2},
    \begin{align*}
        \mcdf_i(\vox) &= \int_{\vox} M_i(s)\leb(\dd s) \overset{(*)}{=} \int_{\vox} M_i(\omega, \dd s) \\
        &= \int_{\vox} \alpha(\cvox) X_i + \bfzeta^\top(\cvox)\bfC_{i} + \eta_i(\cvox) \leb(\dd s) + \int_{\vox} \epsilon_{M,i}(\omega,\dd s) \\
        &\overset{(**)}{=} \br{\alpha(\cvox_j) X_i + \bfzeta^\top(\cvox_j)\bfC_{i} + \eta_i(\cvox_j)}\leb(\vox_j)  + \epsilon_{M,i}(\vox_j).
    \end{align*}
    This is the detailed derivation from the general definition $\mcdf_i(\vox) = \int_{\vox} M_i(s)\leb(\dd s)$ to Equation \eqref{eq:model2-3}. 
    Note that we slightly abuse the notation $\int_{\vox} M_i(s)\leb(\dd s)$ to represent integration w.r.t. the Lebesgue measure and keep the random part fixed when integrating w.r.t. a random function $M_i(s)$. 
    The full expression should include the random part $\omega$ like in equation $(*)$. The last equation $(**)$ uses the approximation for any $s\in\vox_j,\alpha(s)\equiv \alpha(\cvox_j)$, similar for $\bfzeta,\eta_i$. This approximation is consistent with the image data because we could only estimate the functional parameters upto the given image resolution.

\section{Additional Results for Simulation}\label{sec_supp:simulation}

\subsection{Simulation Settings for PTG and CorS}\label{sec:PTG_CorS_Settings}

\textbf{Product Threshold Gaussian prior (PTG)}~\citep{song2020ptg} constructs prior distribution of the bivariate vector 
$\{\beta(s_j),\alpha(s_j)\}$ for each location $s_j$ by thresholding a bivariate Gaussian latent vector $\{\tilde\beta(s_j),\tilde\alpha(s_j)\}\sim \normal_2(0,\bfSigma)$ and their product. i.e.
\begin{align*}
\beta(s_j) &= \Tilde{\beta}(s_j) \max\left\{ I(|\tilde\beta(s_j)|>\lambda_1),I(|\tilde\beta(s_j)\tilde\alpha(s_j)|>\lambda_0) \right\},\\
    \alpha(s_j) &= \Tilde{\alpha}(s_j) \max\left\{ I(|\tilde\alpha(s_j)|>\lambda_2),I(|\tilde\beta(s_j)\tilde\alpha(s_j)|>\lambda_0) \right\}.
\end{align*}
PTG model uses the threshold parameters $\lambda_1,\lambda_2$ and $\lambda_0$ to control the sparsity in $\beta(s_j)$, $\alpha(s_j)$ and the indirect effect $\beta(s_j)\alpha(s_j)$ respectively, and \cite{song2020ptg} directly set $\bfSigma = \diag\br{\sigma_\beta^2,\sigma_\alpha^2}$. However, the spatial correlation in spatially-varying coefficients among different locations $s_j$ is not taken into consideration. Hence we anticipate this method to be less suitable for spatially correlated applications such as brain imaging. 
This method has been implemented in the \texttt{R} package \texttt{bama}~\citep{bama}. We set $\lambda_1 = \lambda_2 = \lambda_0 = 0.01$. 
A total number of 1500 MCMC iterations are performed with 1000 burnins.

\textbf{Correlated Selection model} \citep[CorS]{song2020cors} adopts a mixture model with four components to specify different sparsity patterns of $\alpha(s_j)$ and $\beta(s_j)$ and incorporate  the spatial correlations into prior specifications of mixing weights. 
\begin{align*}
    [\beta(s_j),\alpha(s_j)]^\rT &\sim \pi_{1}(s_j)\normal_2(0,\bfV_1)+\pi_{2}(s_j)\normal_2(0,\bfV_2)+\pi_{3}(s_j)\normal_2(0,\bfV_3)+\pi_{4}(s_j)\bfdelta_0,
\end{align*}
and a membership variable $\gamma(s_j)\in \{1,2,3,4\}$, where $\gamma(s_j)=1$ indicates $\beta(s_j)\alpha(s_j)\neq 0$, $\gamma(s_j)=2$ indicates $\beta(s_j)\neq 0,\alpha(s_j) = 0$, $\gamma(s_j)=3$ indicates $\beta(s_j)= 0,\alpha(s_j) \neq 0$, and $\gamma(s_j)=4$ indicates $\beta(s_j)=\alpha(s_j) \neq 0$.  When $\gamma(s_j)=1$, $\bfV_1$ is assigned an inverse Wishart prior. When $\gamma(s_j)=2$ or 3, $\bfV_2$ or $\bfV_3$ only contains $\sigma_\beta^2$ or $\sigma_\alpha^2$ on the diagonal and 0 otherwise.

Each $\gamma(s_j)$ is assumed to follow a multinomial distribution with probability $$\boldsymbol{\pi}(s_j) =\{\pi_1(s_j),\pi_2(s_j),\pi_3(s_j),\pi_4(s_j)\}^\top$$ with $\sum_{k=1}^4 \pi_k(s_j) = 1$. For each $m = 1,2,3$, let $\bfpi_m = \br{\pi_m(s_1),\dots,\pi_m(s_j)}\in \R^p$. $\logit(\bfpi_m)$ is assumed to follow a multivariate normal prior with a pre-specified covariance matrix $\sigma_m^2\bfD\in \R^{p\times p}$, independently for each $m=1,2,3$. Hence $\bfD$ is used to reflect the mediator-wise correlation.

We anticipate this method to have good performance in the spatially correlated data application.
We use the GitHub implementation of this method (\url{https://github.com/yanys7/Correlated_GMM_Mediation.git}). In the simulation study, we set the initial values for all $\alpha(s)$ and $\beta(s)$ to be 0.5, the initial values for $\{\pi_k(s_j),k=1,2,3,4\}$ to be 0.25, the 2 by 2 scale matrix in Inverse-wishart prior for $\bfV_1$ to be $[1,0.5;0.5,1]$, and the $\pscale\times \pscale$ matrix $\bfD$ to be estimated from the input image correlations. A total number of 2000 MCMC iterations are performed with 1000 burn-ins.

\subsection{Simulation Comparison with BI-GMRF \citep{wang2023high}}\label{sec:compare_BIGMRF}

In addition to the comparisons shown in Section 5.1 in the main text, we also compare BIMA with a more recent Bayesian approach BI-GMRF \citep{wang2023high} that can handle high-dimensional image data. The BI-GMRF approach uses Ising prior to account for sparsity, and Gaussian Markov Random Field to account for the spatial correlation. However, this method requires careful tuning of 7 hyper-parameters through cross-validation, otherwise, the estimation of $\beta$ and $\gamma$ cannot converge to a reasonable value. The current publicly available implementation of BI-GMRF \url{https://github.com/jadexq/BI-GMRF/tree/main} does not provide a user-friendly tuning procedure, hence we are only able to test on the exact simulated data given in the GitHub page of BI-GMRF. We compare the butterfly-shaped pattern and provide the comparison of the posterior mean and ROC curve of NIE in this section. 

The data set given by BI-GMRF contains $n=500, p=64\times 64$ samples. In our implementation, we use the same Mat\'ern kernel as used in the high-dimensional simulation setting in the main text. $\sthresh_\beta = 0.5$, $\sthresh_\alpha = 0.2$. The true pattern of $\beta$ contains binary values of 0 and 0.2, and the true $\alpha$ contains binary values of 0 and -0.2. 

Note this is not a favorable testing case for BIMA, because the true signal pattern has sharp changes and discontinuous jumps on the boundaries, which does not match the prior belief that brain signals are more likely to be spatially varying smooth transitions. We use the default  Mat\'ern kernel same as in our other high-dimensional simulation patterns, whereas for BI-GMRF, hyper-parameters are carefully tuned through cross-validation, and every time the data and true signal change, the hyper-parameters must be tuned again to achieve convergence.

Nonetheless, without careful tuning of the Gaussian kernel or other parameters such as $\sthresh$, BIMA can achieve comparable performance with BI-GMRF in terms of the ROC curve. If we compare Figure 3 in \cite{wang2023high} with Figure \ref{fig:Butterfly_compare} below, BIMA is the only method that has a comparable performance with BI-GMRF among all other methods compared in \cite{wang2023high}, and BIMA can achieve this performance without complex tuning procedure. Although the peripheral noises shown in Figure \ref{fig:Butterfly_compare} can potentially be reduced or removed by choosing a more appropriate kernel or thresholding parameter $\sthresh$.

\begin{figure}
    \centering
    \includegraphics[width=0.7\linewidth]{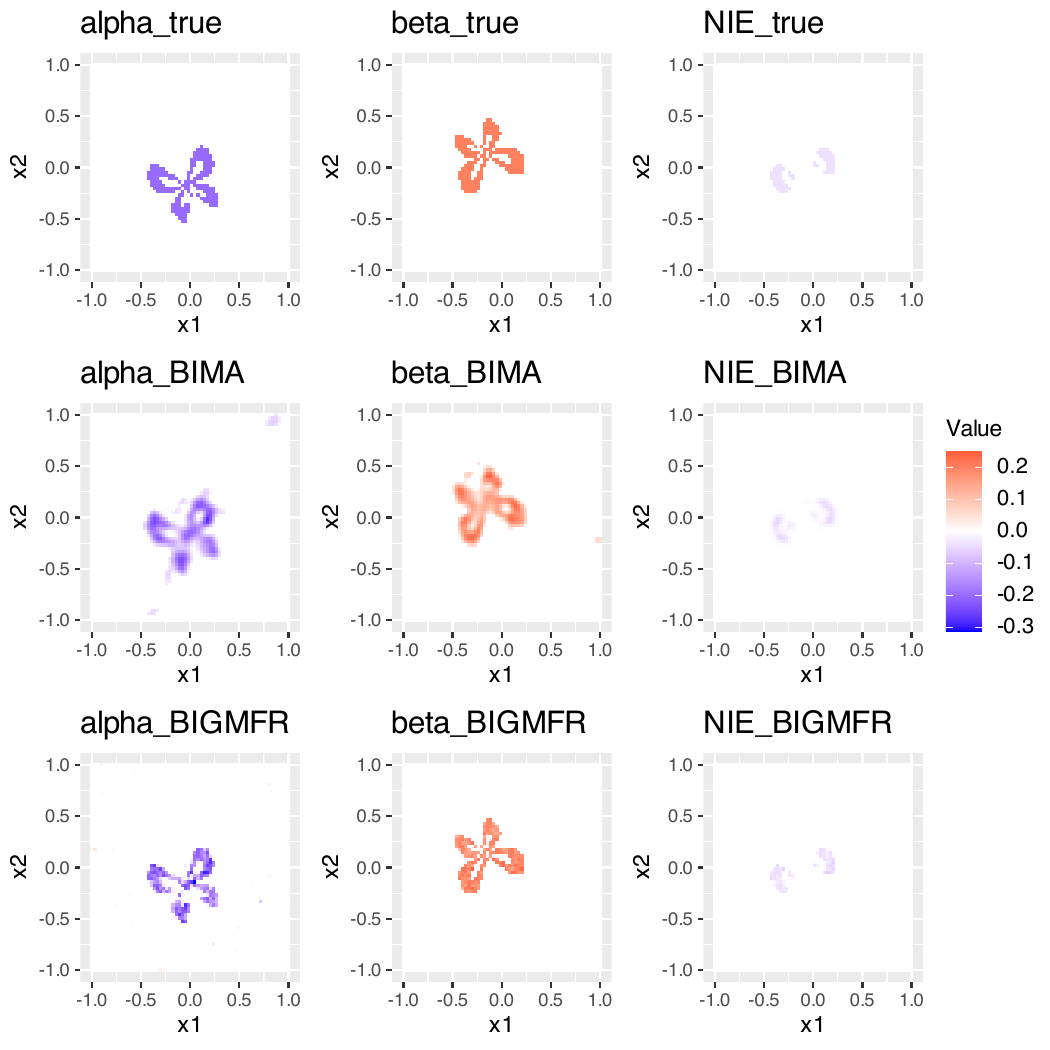}
    \caption{Posterior mean of $\alpha$, $\beta$, and TIE, estimated by BIMA and BI-GMRF. $n=500, p=64\times 64$.}
    \label{fig:Butterfly_compare}
\end{figure}

\begin{figure}
    \centering
    \includegraphics[width=0.5\linewidth]{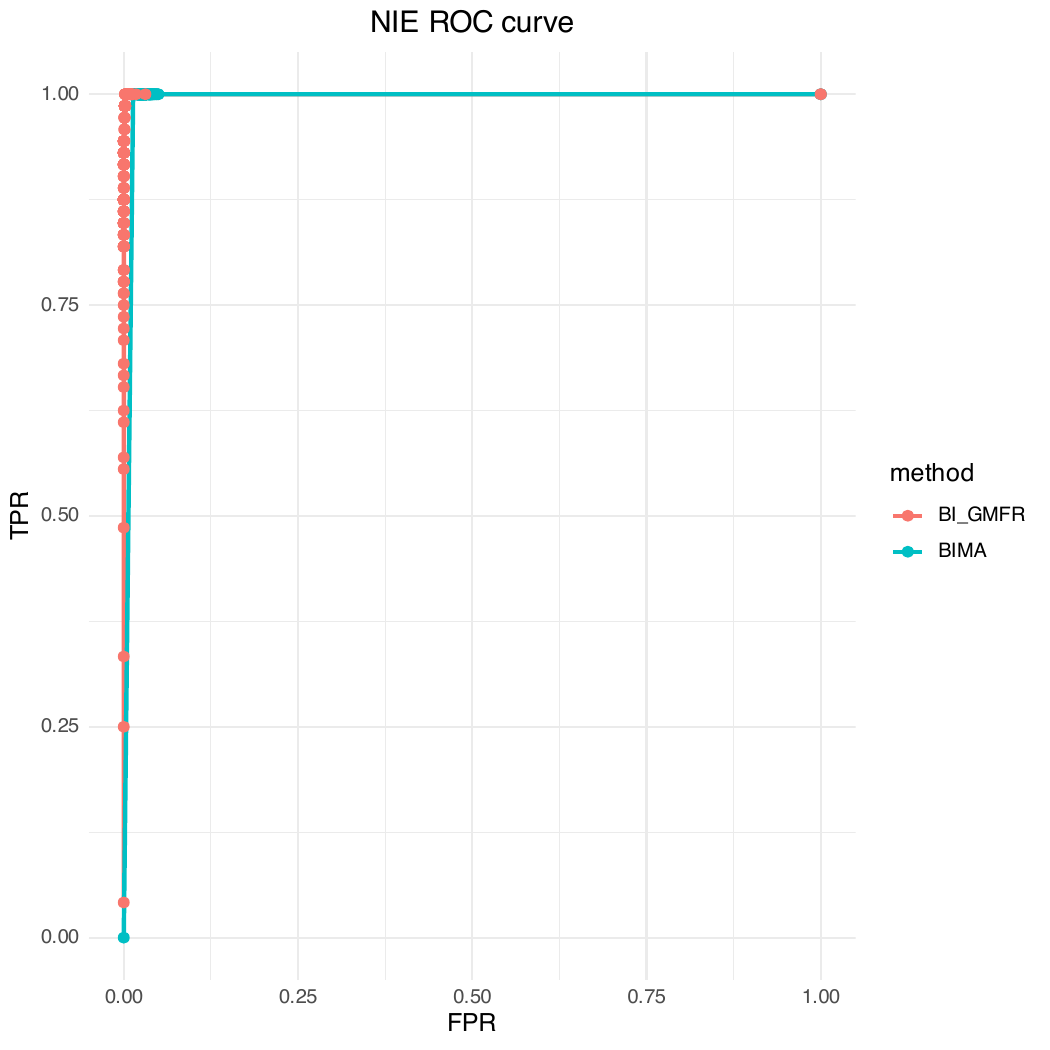}
    \caption{ROC curve of NIE for BIMA and BI-GMRF.}
    \label{fig:Butterfly_ROC}
\end{figure}

Due to the lack of user-friendly implementation of BI-GMRF, we are unable to use BI-GMRF to perform simulation with other data cases, or real data comparison.

\subsection{Simulation with nonsmooth patterns}\label{sec_supp:sim_nonsmooth}

\begin{figure}[ht!]
    \centering
    \includegraphics[width=0.8\linewidth]{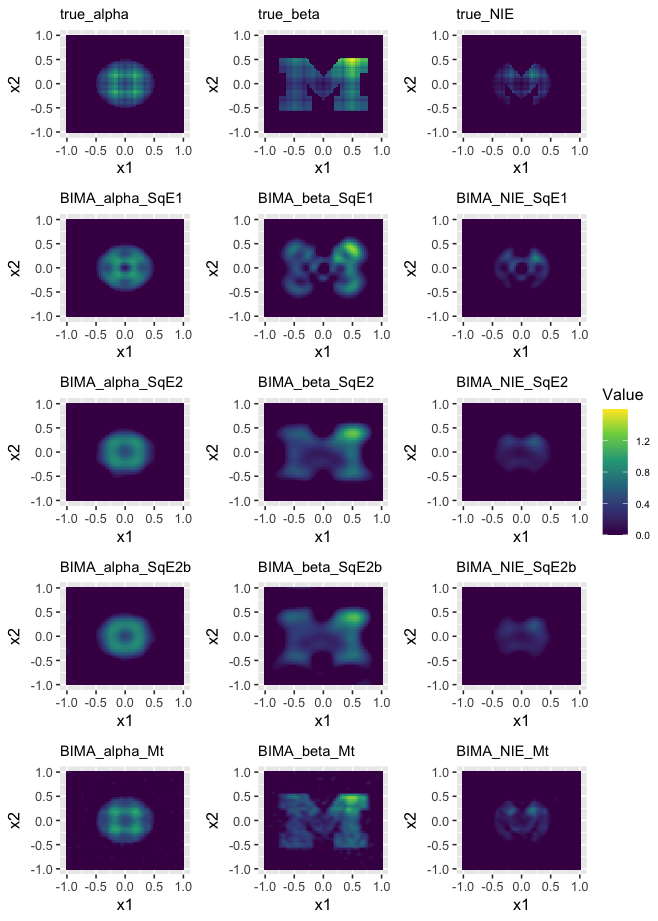}
    \caption{Simulation with nonsmooth patterns. Rows from top to bottom: true signals, BIMA with squared-exponential kernel ($a=0.01,b=10$), BIMA with square-exponential kernel ($a=0.001,b=10$), BIMA with square-exponential kernel ($a=0.01,b=1$), BIMA with Mat\'ern kernel, same smoothness as the kernel used in the high-dimensional simulation in Section \ref{sec:simulation}.}
    \label{fig:nonsmooth}
\end{figure}

This section includes a simulation study for the extremely non-smooth patterns and experiments with different kernels to check the robustness of our model. Figure \ref{fig:nonsmooth} provides the visualization of the true signals and the posterior means of $\alpha,\beta,\cE$.

In this simulation, the true signals are designed to be non-smooth 2D surfaces with sharply changing patterns. The active area of $\alpha$ is a circle, with the signal decaying to 0 on the edge of the circle; the active area of $\beta$ is M-shaped, with the top right corner having the largest effect size. Both $\alpha$ and $\beta$ images are initially created as smooth surfaces, but we multiply the original surface by a 2D Weierstrass function (see Section \ref{sec:app_RDA_sim3}, $a=0.5,b=3,k_{\max}=20$), so that the surfaces of both $\alpha$ and $\beta$ show the checkerboard type of irregular, nonsmooth patterns. Not only is the surface irregular, but the resulting $\cE$ also has very weak patterns and smoothly decays to 0. The result in Figure \ref{fig:nonsmooth} provides a comparison for the BIMA model with three different kernel specifications. The second row is the squared-exponential kernel used in Section \ref{subsec:comparison}, $\kappa(s,s'; a, b) = \mathrm{cor}\{\beta(s),\beta(s')\} =  \exp\{-a (s^2 + {s'}^2) - b \|s-s'\|^2 \}$ with
 $a=0.01$ and $b=10$. The third row is the squared-exponential kernel with $a=0.001,b=10$, the fourth row with $a=0.01,b=1$. The last row is the same Mat\'ern kernel used in Section \ref{subsec:comp_sens}.
 
Figure \ref{fig:nonsmooth} shows that for this type of highly irregular patterns, the squared exponential kernels are overly smooth. However, the Mat\'ern kernel can still give close estimates of the functional parameters to a certain extent. Further careful tuning of the thresholding parameters $\sthresh_\alpha,\sthresh_\beta$ could help reduce the background noise in the Mat\'ern kernel result. 

Table \ref{tb:sim_diff_ker} shows the variable selection accuracy averaged overal 100 replicated studies, for active $\cE(s)$ in terms of the FDR, Power, and Accuracy. When determining the active voxels, we apply the same threshold (PIP$>0.999$) on the PIP for all GP kernels; additionally, we set a threshold on the effect size and only deem a voxel to be active if the estimated NIE is greater than 0.05. Table \ref{tb:sim_diff_ker} indicates that when the FDR is below 0.2, the Mat\'ern kernel has the highest power.

\begin{table}[ht]
\centering
\caption{Simulation results on the NIE $\cE(s)$ selection accuracy among three different GP kernels, averaged over 100 replicated studies.}
\label{tb:sim_diff_ker}
\begin{tabular}{lrrr}
\hline
\toprule
 & FDR & Power & ACC \\
 \hline
SqE1 ($a=0.01,b=10$) & 0.141 & 0.886 & 0.965 \\ 
SqE2 ($a=0.001,b=10$) & 0.200 & 0.942 & 0.960 \\ 
SqE3 ($a=0.01,b=1$) & 0.200 & 0.943 & 0.960\\
Mat\'ern &0.181 & 0.971 & 0.967 \\ 
\bottomrule
\end{tabular}
\end{table}

\section{Additional Results for ABCD study}\label{supp_sec:additional_rda}

\subsection{Summary statistics of the ABCD data}

Table \ref{tb:summary_ABCD} provides the summary statistics for the ABCD data covariates and the outcome, stratified by the parent education level. Continuous variables including g-score and age are reported in means and standard deviations. Categorical variables including gender, race and ethnicity, and income level are reported in the number of observations in each category.

\begin{table}[ht!]
\centering
\caption{Summary statistics of the ABCD data stratified by Parent Degree. Mean (standard deviation) are reported for g-Score and Age. Counts are reported for Gender, Income, Race and Ethnicity}
\label{tb:summary_ABCD}
\begin{tabular}{rrrr}
\toprule
Parent degree              & Bachelor or higher & No bachelor  & Overall      \\
\midrule
g-Score                    & 0.47 (0.77)        & -0.15 (0.80) & 0.27 (0.83)  \\
Age                        & 10.09 (0.61)       & 10.01 (0.63) & 10.06 (0.62) \\

\multicolumn{4}{l}{Gender}\\
Female& 611                & 281          & 892 \\
Male& 635                &  334         & 969\\[2mm]
\multicolumn{4}{l}{Race and Ethnicity}          \\
Asian              & 30                 & 3            & 33           \\
Black               & 47                 & 84           & 131          \\
Hispanic            & 151                & 216          & 367          \\
White               & 924                & 254          & 1178         \\
Other               & 94                 & 58           & 152          \\[2mm]
\multicolumn{4}{l}{Income}\\
 \textless 50K       & 98                 & 336          & 434          \\
50$\sim$100K        &  375                & 213          & 558           \\
\textgreater{}=100K &    773                & 66           & 839    \\[2mm]
\multicolumn{1}{l}{Total} & 1246               &    615      & 1861         \\
\bottomrule
\end{tabular}

\end{table}

\subsection{Scatter plots for the posterior mean}\label{sec:app_scatter_NIE}

\begin{figure}[ht]
    \centering
    \includegraphics[width=0.5\linewidth]{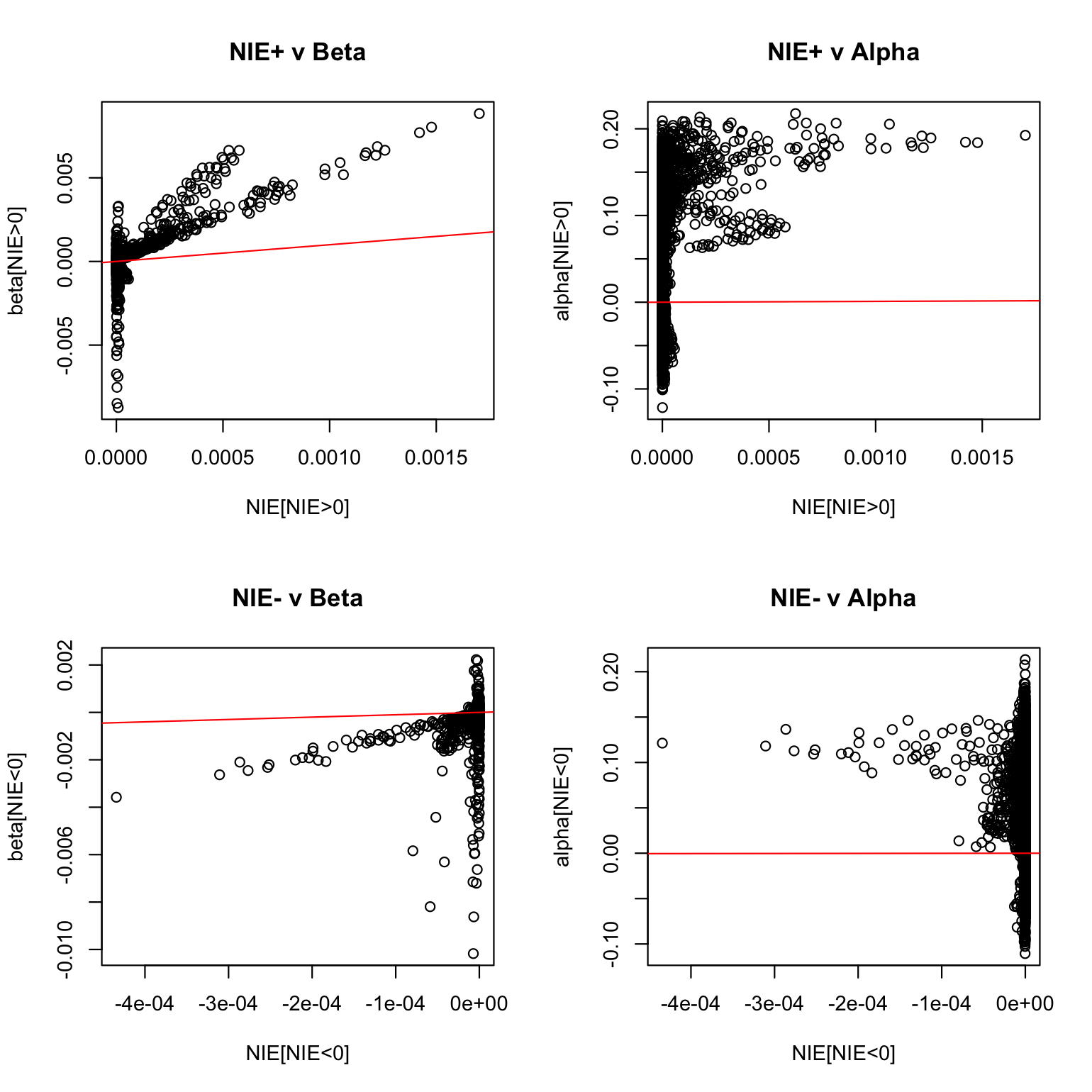}
    \caption{Scatter plot of posterior mean of NIE, stratified by the sign, and compared to the posterior mean of $\alpha$ and $\beta$ on all voxels. Each dot represents one voxel.}
    \label{fig:scatter_NIE}
\end{figure}

Figure \ref{fig:scatter_NIE} provides scatter plots on the posterior mean of NIE against that of $\beta$ and $\alpha$ on all voxels, stratified by the sign of NIE. Each dot is one voxel location. These plots show that most of the large positive effects of NIE consist of both positive $\alpha$ and $\beta$, instead of negative effects of $\alpha$ and $\beta$. The negative effects of NIE are very small and negligible. This helps us to have a further understanding of the mediation mechanism: higher parental education leads to positive effects on the children's working memory, and stronger working memory signal in children's fMRI image leads to higher IQ score, i.e. both $\alpha$ and $\beta$ are positive. Only this positive mediation pathway can lead to positive effect of higher parental education on children's IQ score.

\subsection{Sensitivity Analysis on Hyper-parameters in ABCD Data Analysis} \label{subsec_supp:sensitivity}

In this section, we provide details on data preprocessing and selecting the kernel parameters and the prior parameter $\lambda$ in both models \eqref{eq:model1} and \eqref{eq:model2}.

To get an appropriate kernel for the real data, we choose the Mat\'ern kernel parameters based on the smoothness of the image mediators. The input images are standardized across subject.
To get parameters in the Mat\'ern kernel function as defined in \eqref{eq:matern}, we tune $(\rho,u)$ on a grid in the following way:
First, the empirical sample correlations of the image predictors are computed, then the parameters $(\rho,u)$ are obtained using grid search so that the estimated correlation from the kernel function can best align with the empirical correlation computed from the image mediators. The kernel parameters are chosen region-by-region. We refer to this set of kernel parameters as the optimal kernel.

\begin{table}[h]
\centering
\caption{Predictive MSE for different kernels}
\label{tb:compare_kernel}
\begin{tabular}{llllll}
\toprule
    & Optimal Kernel & 90\% of $\rho$ & $u=1,\rho=15$ & $u=0.2,\rho=80$ & 110\% of $\rho$ \\\midrule
Test MSE & 0.515          & 0.516          & 0.547         & 0.539           & 0.507            \\
\bottomrule
\end{tabular}
\end{table}

To test and compare the performances of different kernels, we split the data into 50\% as training data and 50\% as testing data. Because the performance of different kernels can be directly compared through testing MSE using the outcome model \eqref{eq:model1} , we conduct a sensitivity analysis using model \eqref{eq:model1} to select an appropriate  set of kernel parameters.
The optimal kernel is obtained in the aforementioned way. To test the sensitivity of the kernel, we fix $u$ to be the same as the optimal $u$, but change $\rho$ to be 90\% and 110\% of the optimal $\rho$. Another 2 kernels where $u,\rho$ are constant across different regions are also included in the comparison. The comparison result is in Table \ref{tb:compare_kernel}. Based on Table \ref{tb:compare_kernel}, the case 110\% of the optimal $\rho$ seems to give a slightly better prediction performance, hence we choose this kernel for model \eqref{eq:model1}. The kernel in model \eqref{eq:model2} remains to be the optimal kernel we choose.

\begin{table}[h]
\centering
\begin{tabular}{crrrr}
\toprule
$\sthresh$  & 0.01 & 0.05& 0.07 & 0.1 \\\midrule
Training MSE & 0.0003 & 0.3621 & 0.4043 & 0.4693 \\
Test MSE     & 1.8444 & 0.5079 & 0.5120 & 0.5254 \\ 
\bottomrule
\end{tabular}
\caption{Training and test MSE for model \eqref{eq:model1} under different prior thresholding parameter $\sthresh$ for the coefficient $\beta(s)$.}
\label{tb:compare_lambda_Y}
\end{table}

We use the same 2-fold cross validation method to select an appropriate value of $\sthresh$ in the prior of $\beta(s)$. Based on Table \ref{tb:compare_lambda_Y}, if we select a very small $\sthresh = 0.01$, there is severe overfitting issue; if $\sthresh$ gets too large, the testing accuracy also decreases. Hence based on this 2-fold testing result, $\sthresh=0.05$ appears to be the most appropriate thresholding parameter. The running time for fitting model \eqref{eq:model1} based on 50\% of the data is only within 1 hour, so this testing procedure under the current data scale is not very computationally expansive.

\begin{table}[h]
\centering
\begin{tabular}{lllll}
\toprule
Value of $\sthresh$         & 0.05     & 0.08     & 0.1      & 0.5      \\\midrule
Averaged test MSE & 1.008132 & 1.008075 & 1.007796 & 1.007751 \\
\midrule
Value of $\sthresh$         & 1        & 1.5      & 1.7      & 2.0      \\\midrule
Averaged test MSE & 1.007740 & 1.007611 & 1.007532 & 1.007711\\
\bottomrule
\end{tabular}
\caption{Averaged testing MSE over all voxels under different value of $\sthresh$ for model \eqref{eq:model2}.}
\label{tb:compare_lambda_M}
\end{table}
A similar sensitivity analysis is conducted for model \eqref{eq:model2} to select $\sthresh$ in the prior of $\alpha(s)$. Estimating the individual effect $\{\eta_i(s)\}_{i=1}^N$ can be very time-consuming, hence the individual effects are set to 0 only for the sensitivity analysis.
From table \ref{tb:compare_lambda_M}, the difference in the testing MSE among different values of $\sthresh$ is very small. Hence we choose $\sthresh = 0.1$ conservatively to be able to include more activation voxels without compromising the predictive ability.

\subsection{Discussion on computational details}

\subsubsection{MALA initial values}

As discussed in section 4.3 in the main text, we can use Gibbs sampler to fit the outcome and mediator model first, and then use the posterior mean of $\beta$ and $\alpha$ as the initial value for MALA algorithm. In the real data analysis, for the mediator model \eqref{eq:model2}, we directly use the posterior mean of $\bftheta_{\alpha}$ as the initial value for $\bftheta_{\alpha}$ in the MALA algorithm. For the outcome model \eqref{eq:model1}, we use the Lasso regression to estimate $\beta$ first, then add $\sthresh$ to locations where $\beta(s)>0$, and subtract $\sthresh$ from $\beta(s)$ when $\beta(s)<0$, to get a hard-thresholded version of the latent GP $\tilde\beta$. The last step is to use basis on $\tilde\beta$ to get the initial values for $\bftheta_\beta$ in MALA.

\subsubsection{Convergence of the ABCD data analysis}\label{sec:app_conv_plot}

\begin{figure}[ht]
    \centering
    \includegraphics[width=0.7\linewidth]{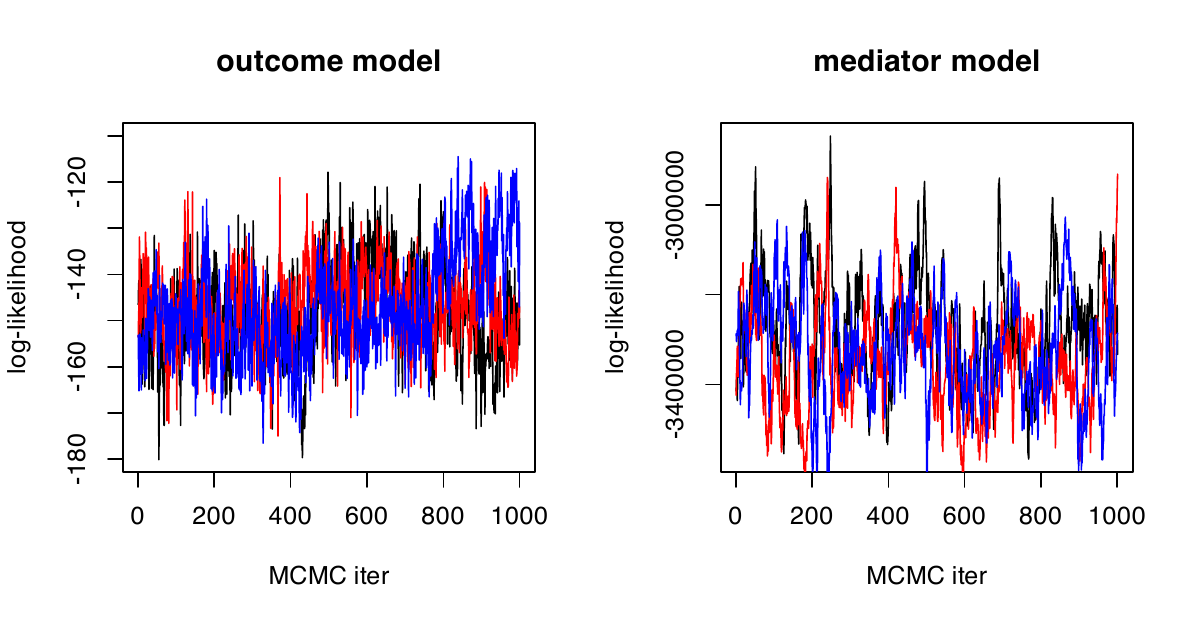}
    \caption{Trace plot of the log-likelihood after burn-in iterations of the outcome and mediator model. Colors indicate different chains.}
    \label{fig:conv_plot}
\end{figure}

To check the chain mixing, we provide the following trace plots on the log-likelihood of the outcome model \eqref{eq:model1} and the mediator model \eqref{eq:model2} in Figure \ref{fig:conv_plot}. The trace plots for both models are based on the thinned sample.

In addition, we also run 3 repeated experiments to compute the Gelman Rubin (GR) Diagnostic Statistics. The GR estimate for the log-likelihood of the outcome model is 1.05 with upper confidence interval 1.12, and the GR estimate for the mediator model is 1.11 with upper confidence interval 1.35. This indicates good mixing for both models.

\begin{figure}[ht]
    \centering
    \includegraphics[width=\linewidth]{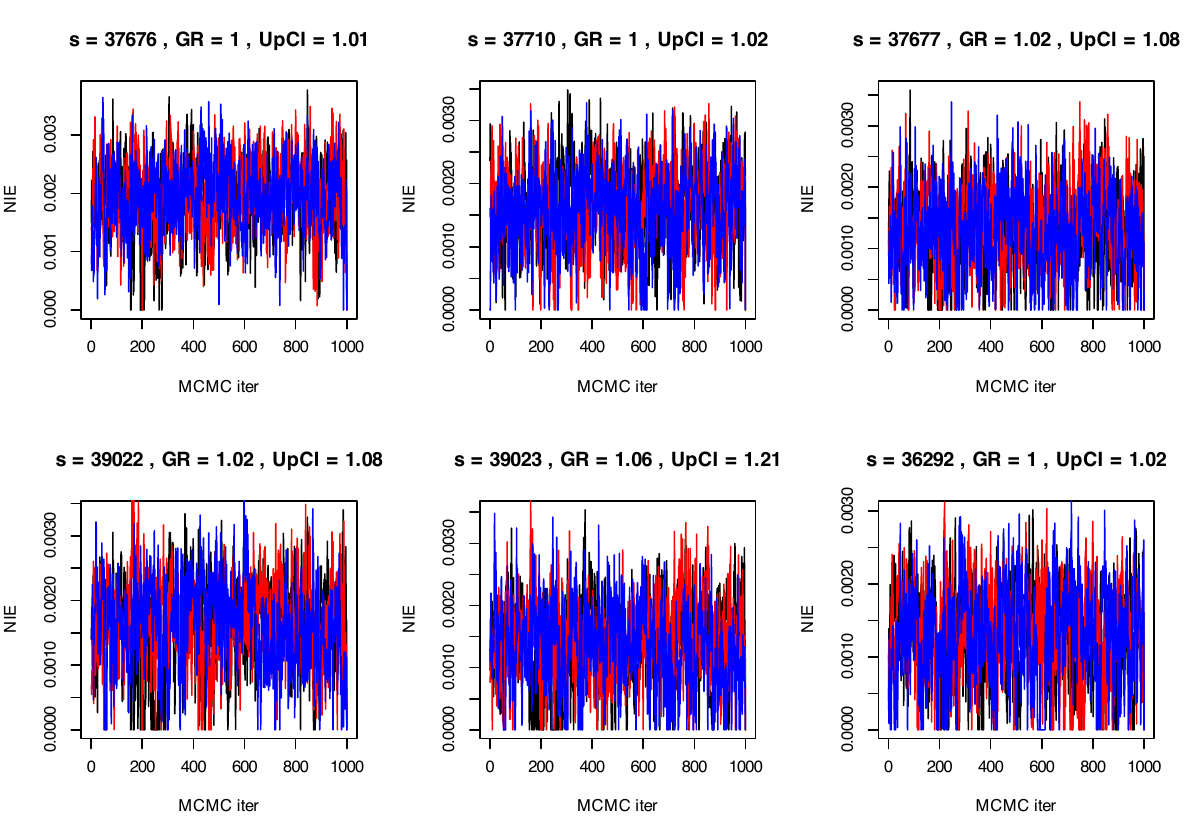}
    \caption{Trace plot of the NIE $\mathcal{E}(s)$ on a few selected voxels $s$ with the highest PIP, after burn-in iterations of the outcome and mediator model. Colors indicate different chains. The GR point estimates and Upper CI are reported in the plot titles.}
    \label{fig:NIE_conv_plot}
\end{figure}

Figure \ref{fig:NIE_conv_plot} shows trace-plots of NIE $\mathcal{E}(s)$ on a few selected voxels $s$ with the highest PIP. The GR test statistics and Upper CIs are reported in the title. All trace-plots indicate good mixing.

\subsection{Check regression residuals} \label{supp:sec:residual}
This section provides model fitting assessement on \eqref{eq:model1} and \eqref{eq:model2}.

\begin{figure}[ht!]
    \centering
    \includegraphics[width=0.7\linewidth]{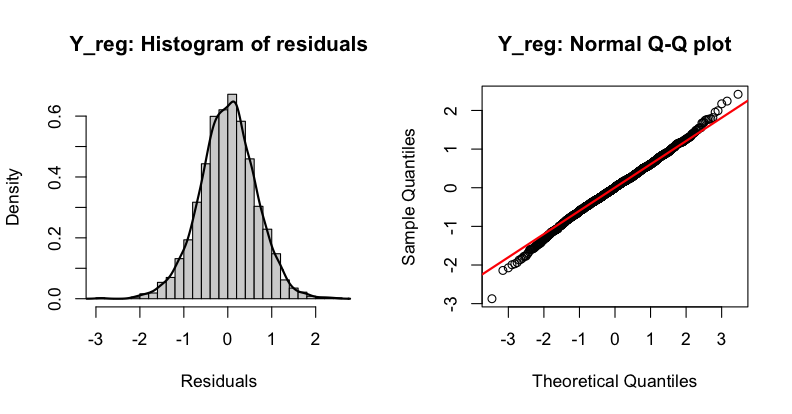}
    \caption{Residual checks of the outcome model \eqref{eq:model1}.}
    \label{fig:Y_reg_res}
\end{figure}

In particular, we perform normality checks on the regression residuals of the real data analysis. For the outcome model \eqref{eq:model1}, Kolmogorov–Smirnov (K-S) test yields a p-value of 0.5306, indicating  no evidence against the normality of the residual. See Figure \ref{fig:Y_reg_res} for the residual plots. 

\begin{figure}
    \centering
    \includegraphics[width=0.7\linewidth]{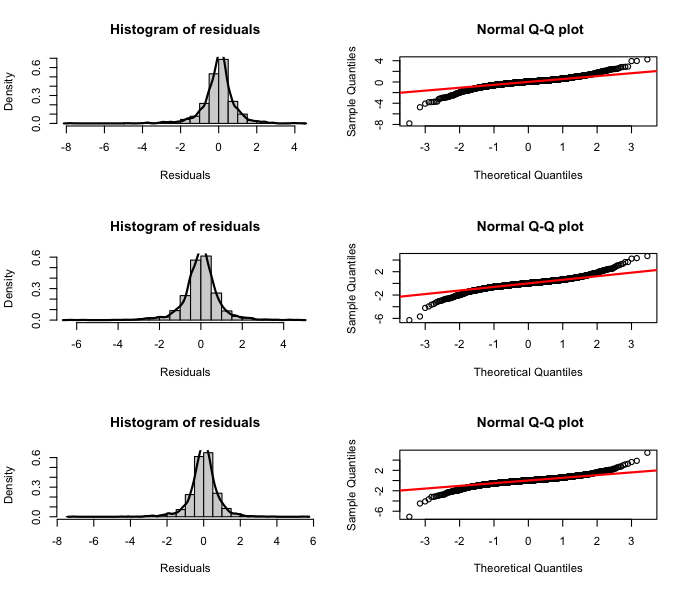}
    \caption{Residual plots for the 3 locations with the smallest p-values that reject the normality.}
    \label{fig:M_3residual}
\end{figure}

     For the mediator model \eqref{eq:model2},  the residuals  are spatially-varying across all $p=47,636$ locations, making residual checks are more challenging.  Moreover, with a large sample size, K-S test can be overly sensitive to minor deviations. To address this,  we report both  raw K-S test results and  the result after trimming a small proportion of subjects in each tail, with the Benjamini-Hochberg (BH) adjustment for multiple comparisons. For the untrimed residuals, 14,690 out of 47,636 (31\%) locations pass the normality check at the 0.05 level. For locations where normality is rejected,  Q–Q plots (Figure \ref{fig:M_3residual}) show that deviations were mainly due to a small proportion of extreme residuals. After removing the most extreme 0.5\% of observations in each tail, 36,754 (77\%) locations pass the test. With 2.5\% triming, 46,829(98\%) locations pass.

\subsection{Sensitivity Analysis (SA) for Vector-valued Unobserved Confounders}\label{supp:sec:SA}

BIMA model with unobserved confounders can be described using the following model \eqref{eq:U_y} - \eqref{eq:U}, where the unobserved confounder $\bfU_i\sim G(u;\theta_u)$ is assumed to follow a distribution $G(u;\theta_u)$ with parameter $\theta_u$, and $\bfrho^{\top}_y$ and $\bfrho^\top_m(\cvox_j)$ are the unobserved confounding effect on the outcome and the mediator respectively. The general sensitivity analysis for BIMA is to assign varying fixed values to $\bfrho^{\top}_y$ and $\bfrho^\top_m(\cvox_j)$, and jointly model \eqref{eq:U_y} - \eqref{eq:U} based on the Bayesian priors in BIMA, with $\bfU_i$ updated jointly conditional on \eqref{eq:U_y} and \eqref{eq:U_m}. 

\begin{align}
Y_i &= \sum_{j=1}^{p} \beta(\cvox_j) \mcdf_{i}(\vox_j)  + \gamma X_i + \bfxi^{\top} \bfC_i +\ \bfrho^{\top}_y \bfU_i + \epsilon_{Y,i},\quad \epsilon_{Y,i}\overset{\iid}{\sim} \normal(0,\sigma_Y^2), \label{eq:U_y}\\
\mpdf_{i}(\cvox_j) &= \alpha(\cvox_j) X_i + \bfzeta^\top(\cvox_j)\bfC_{i} +  \bfrho^\top_m(\cvox_j)\bfU_{i} + \eta_i(\cvox_j)+ \epsilon_{M,i}(\cvox_j),\quad \epsilon_{M,i}(\cvox_j)\overset{\iid}{\sim} \mN(0,\sigma_M^2)\label{eq:U_m} \\
\bfU_{i}&\sim G(u;\theta_U) \label{eq:U}
\end{align}

However, in this most general SA framework, the identifiability of the joint model \eqref{eq:U_y} - \eqref{eq:U} is not guaranteed. Hence we follow a simpler SA model proposed by \cite{Dorie2016-om} where $\bfU_i$ is a single binary variable, and propose the following SA algorithm for BIMA. We use $\bftheta_{\text{BIMA}}$ to denote all parameters used in the BIMA model. Algorithm \ref{algo:sa} provides details of the SA procedure. Note that the intuition in Step 7 in Algorithm \ref{algo:sa} behind the sampling of $U_i$ is that, for the unknown $U_i$, drawing samples conditional on model \eqref{eq:U_y} - \eqref{eq:U_m} is equivalent to guessing the most likely 0 or 1 assignment to $U_i$ based on the known fixed effects $\rho_m$ and $\rho_y$.

\begin{algorithm}[ht]
\caption{SA algorithm for BIMA with a single binary unobserved confounder}
\label{algo:sa}
\begin{algorithmic}[1]
\State Based on the joint model \eqref{eq:U_y} - \eqref{eq:U} where $U_i\sim \text{Ber}(p_u)$, and assume a spatial constant effect of $\rho_m(s_j)$ such that $\rho_m(s_j)\equiv\rho_m$ for all $j$.
\State Specify a combination of fixed values $\rho_y$ and $\rho_m$.
\For {each combination of fixed  $\rho_y$ and $\rho_m$}:
\State Initialize $U_i$ to 0. Initialize all other parameters in BIMA.
    \For {each MCMC iteration $t$}:
    \State Draw all BIMA parameters $\bftheta_{\text{BIMA}}$ based on \eqref{eq:U_y} - \eqref{eq:U_m} with fixed $\rho_y$ and $\rho_m$.
    \State Draw $U_i$ independently for each $i$ conditional on the joint model \eqref{eq:U_y} - \eqref{eq:U} 
    \begin{align*}
        &U_i \sim \text{Ber}(\pi_{u1}/\sbr{\pi_{u0}+\pi_{u1}})\\
        &\pi_{u1} = p_y(Y_i|U_i=1,\bftheta_{\text{BIMA}}) p_m(\bfM_i|U_i=1,\bftheta_{\text{BIMA}})p_u\\
        &\pi_{u0} = p_y(Y_i|U_i=0,\bftheta_{\text{BIMA}}) p_m(\bfM_i|U_i=0,\bftheta_{\text{BIMA}})(1-p_u)
    \end{align*}
    where $p_y$ and $p_m$ are the density function given in \eqref{eq:U_y} and \eqref{eq:U_m} respectively.
    \EndFor
    \State Output the posterior distribution of $\bftheta_{\text{BIMA}}$.
\EndFor

\end{algorithmic}
\end{algorithm}

We perform this sensitivity analysis on the ABCD study data. The potential binary unobserved confounder could be whether or not children's nutrient supply or exercise intensity is sufficient, etc.  We provide the following result on the NIE and NDE under varying values of $\rho_y$ and $\rho_m$. We choose $\rho_y$ to be in $(0.01,0.1,1)$, and $\rho_m$  to be in $(0.001,0.01,0.1)$ since the standardized $\bfM_i$ has small values of signal intensity on most voxels. 

\begin{figure}[ht]
    \centering
    \centering
\begin{subfigure}{.5\textwidth}
  \centering
  \includegraphics[width=\linewidth]{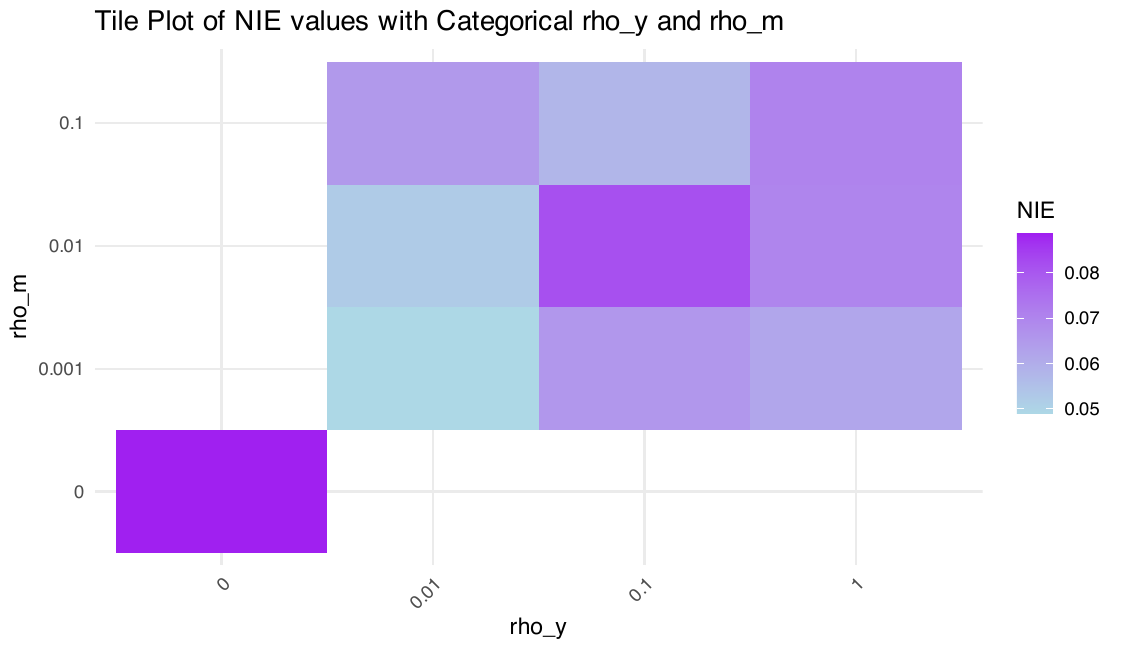}
    \caption{Posterior mean of NIE under varying $\rho_y$ and $\rho_m$}
    \label{fig:SA_NIE}
\end{subfigure}%
\begin{subfigure}{.5\textwidth}
  \centering
  \includegraphics[width=\linewidth]{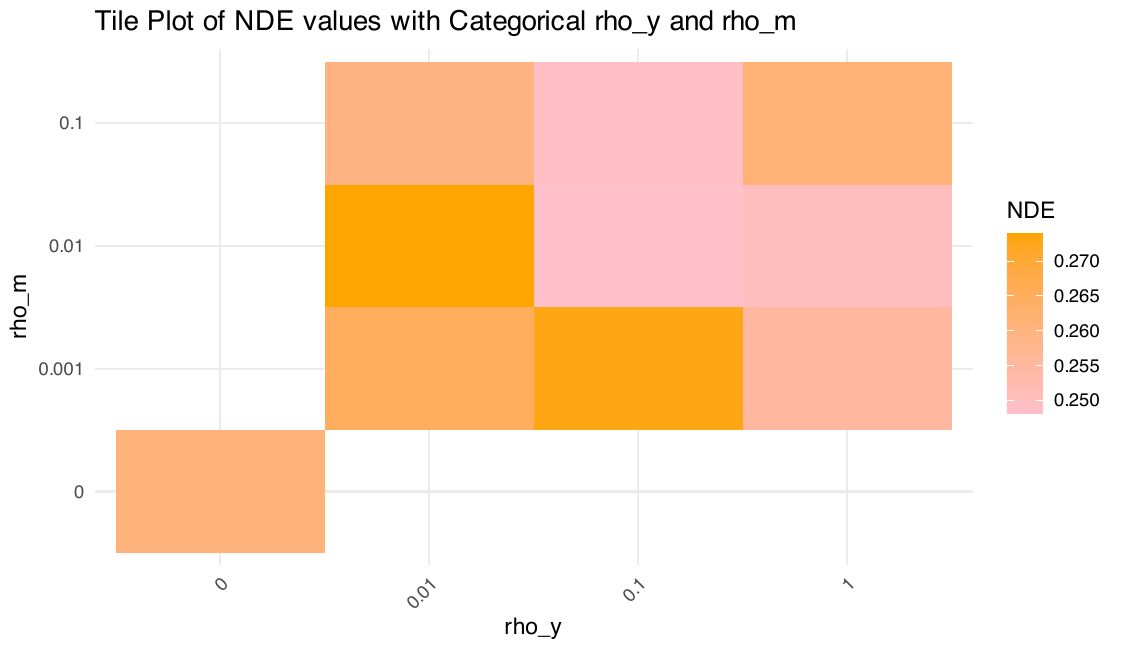}
    \caption{Posterior mean of NDE under varying $\rho_y$ and $\rho_m$}
    \label{fig:SA_NDE}
\end{subfigure}
\caption{Sensitivity analysis result.}
\label{fig:SA}
    
\end{figure}

The result based on Figure \ref{fig:SA}
shows that under varying levels of $\rho_y$ and $\rho_m$, the NIE and NDE vary within a relatively small range. 

In addition to the simple case of the binary unobserved confounder, we are currently working on a more general follow-up approach to account for unobserved confounders.  We note that accounting for unobserved confounders in high-dimensional mediation problems for more general scenarios is still an open research area and will be a valuable future direction, but more general ways to treat this issue are beyond the scope of this work.

\subsection{Causal Assumptions in the context of ABCD study}\label{sec:app_causal_interpretation}
This section provides a more detailed interpretation and discussion of the causal assumptions in Section 2.4 in the context of ABCD study.

The SUTVA assumption \citep{rubin1980randomization} states that one individual’s exposure assignment does not affect the outcome of others. In the context of ABCD study, this means one child's parental education level does not affect the cognitive ability of other children, which is a natural assumption. This is a reasonable and natural assumption in this setting, as parental education is unlikely to have cross-effects between unrelated individuals. In functional mediation analysis, this assumption is standard and has been employed in several recent works, including \cite{wang2023high}, \cite{song2020cors, song2020ptg, song2020shrinkage}, and \cite{jiang2023causal}. Also, one of the foundational papers in functional mediation analysis by \cite{lindquist2012functional} adopts the same causal framework when using temporal brain data as mediators. 

The interpretation for this assumption [A4] $Y_{i,(x,\bfm)} \perp \bfM_{i,(x')}\mid \bfC_i$ under the context of ABCD study is that: if there were two identical children (in terms of observed confounders) except for their parental education levels, one child's working memory $\bfM_{i,(x')}$ should be independent of the other child's cognitive ability $Y_{i,(x,\bfm)}$. This is a reasonable assumption given the context of the ABCD study. The discussion in \cite{andrews2021insights} concerns the unmeasured confounder that has cross-world impact, i.e. there is an unmeasured $U$ correlated with both $\bfM_{i,(x')}$ and $Y_{i,(x,\bfm)}$. In the ABCD study context, this means for the aforementioned identical children, there is an unobserved confounder that specifically impacts one child's working memory $\bfM_{i,(x')}$ and the other child's cognitive ability $Y_{i,(x,\bfm)}$, while does not influence the first child's cognitive ability $Y_{i,(x',\bfm)}$ or the second child's working memory $\bfM_{i,(x)}$. Although this is not entirely impossible, it is not a major concern of the ABCD study. 

Below, we provide a step-by-step derivation of NIE and NDE shown in \eqref{eq:NIE} and \eqref{eq:NDE} based on the causal assumptions listed in Section \ref{subsec:causal_mediation} and the linear functional structural equation models \eqref{eq:model1} and \eqref{eq:model2}: When the exposure/treatment takes on different values $x,x'$, conditional on the observed confounders $\bfC=\bfc$,
    \begin{align*}
        \text{ATE}(x,x') &= \bE\br{Y_{i,\br{x,\bfM_{i,(x)}}} - Y_{i,\br{x',\bfM_{i,(x')}}} \mid \bfC_i = \bfc}\\
        &=\underbrace{\bE\br{Y_{i,\br{x,\bfM_{i,(x)}}} - Y_{i,\br{x,\bfM_{i,(x')}}} \mid \bfC_i = \bfc}}_{\text{NIE(x,x')}} +\underbrace{\bE\br{Y_{i,\br{x,\bfM_{i,(x')}}} - Y_{i,\br{x',\bfM_{i,(x')}}} \mid \bfC_i = \bfc}}_{\text{NDE}(x,x')}
    \end{align*}

    The causal assumptions [A1]-[A4] (copied below) are used to equate the estimable averaged outcomes to the underlying potential outcomes. 
    
    \noindent\textbf{Causal Assumptions:} For any $i$, $x$ and $\bfm$, we assume: \textbf{[A1]} $Y_{i,(x,\bfm)}\perp X_i\mid \bfC_i$, \textbf{[A2]} $Y_{i,(x,\bfm)}\perp \bfM_i\mid \{\bfC_i,X_i\}$, \textbf{[A3]} $\bfM_{i,(x)}\perp X_i\mid \bfC_i$, \textbf{[A4]} $Y_{i,(x,\bfm)} \perp \bfM_{i,(x')}\mid \bfC_i$.
    
    We simplify the notations for clarity and drop the index $i$, and use small brackets to indicate the exogenous values $x,\bfm$. We follow a similar derivation as in Equation (21) in \cite{lindquist2012functional} and start from the observable averaged outcomes in the LHS (estimable based on the structural equation models),
    \begin{align*}
        \bE\br{Y\mid X=x,\bfM(X)=\bfm, \bfC} &\overset{\text{[A3]}}{=} \bE\br{Y(x,\bfm) \mid X=x,\bfM(x)=\bfm, \bfC}\\
        &\overset{\text{[A1]}}{=} \bE\br{Y(x,\bfm)\mid \bfM(x)=\bfm,\bfC}\\
        &\overset{\text{[A2]}}{=} \bE\br{Y(x,\bfm)\mid \bfC}
    \end{align*}
    Additionally, if we denote $\bfm' = \bfm(x')$, the endogenous mediator value if the exposure takes value $x'$,
    \begin{align*}
        \bE\br{Y(x,\bfm')\mid \bfC}&\overset{\text{[A4]}}{=}\bE\br{Y(x,\bfm')\mid \bfM(x')=\bfm', \bfC}\\
        &\overset{\text{[A1,A3]}}{=} \bE\br{Y\mid X=x, \bfM(x')=\bfm', \bfC}
    \end{align*}
    The above derivations hold regardless of any model assumptions. They are derived only based on the causal assumptions and the dependence structure in Figure \ref{fig:structure}.
    
    Now we can derive the NDE and NIE in estimable forms, and when we plug in the structural equation models, we can obtain the NIE and NDE in terms of the model parameters.
    \begin{align*}
        \text{NDE}(x,x') &= \bE\br{Y\sbr{x,\bfm'} - Y\sbr{x',\bfm'} \mid \bfC}\\
        &=\bE\br{Y\mid X=x, \bfM(x')=\bfm', \bfC} - \bE\br{Y\mid X=x',\bfM(X)=\bfm', \bfC}\\
        &\overset{(*)}{=} \bE\br{ \gamma X_i + \bfxi^{\top} \bfC_i \mid X_i=x,\bfC_i=\bfc } - \bE\br{ \gamma X_i + \bfxi^{\top} \bfC_i \mid X_i=x',\bfC_i=\bfc } \\
        &=\gamma (x-x')
    \end{align*}
    Here, $(*)$ uses the structural equation model \eqref{eq:model1}.
    \begin{align*}
        \text{NIE}(x,x')&=\bE\br{Y\sbr{x,\bfm} - Y\sbr{x,\bfm'} \mid \bfC }\\
        &=\bE\br{Y\mid X=x, \bfM(x)=\bfm, \bfC} - \bE\br{Y\mid X=x,\bfM(x')=\bfm', \bfC}\\
        &= \bE\mbr{\sum_{j=1}^p \beta(s_j) \br{ \mcdf_{i,(x)}(\vox_j)  - \mcdf_{i,(x')}(\vox_j) }  \mid \bfC_i=\bfc}\\
        &\overset{(**)}{=} \sum_{j=1}^p \beta(s_j) \br{ \alpha(s_j)x \leb(\vox_j)  - \alpha(s_j)x' \leb(\vox_j) }\\
        &= \sum_{j=1}^p \beta(\cvox_j)\alpha(\cvox_j)\leb(\vox_j)(x-x')
    \end{align*}
    Here, the third equation uses the structural equation \eqref{eq:model1}; $(**)$ uses the structural equation \eqref{eq:model2}, and the approximation \eqref{eq:model2-3}.

\subsection{ Results of ABCD study when using MUA and glmnet}\label{sec:RDA_MUA_glmnet}

\begin{figure}
\centering
\begin{tabular}{c}
Positive TIE (color range $[10^{-5}, 10^{-3}]$)\\
\includegraphics[width=\textwidth]{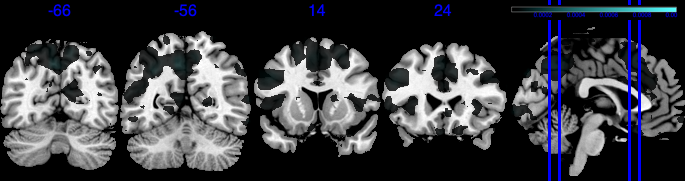}\\
Negative TIE (color range $[-10^{-4}, -10^{-5}]$)\\
\includegraphics[width=\textwidth]{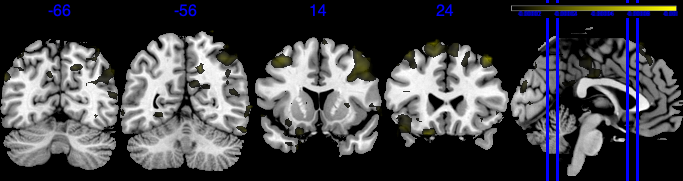}

\end{tabular}

\caption{Point estimate of NIE given by MUA+glmnet.}
\label{fig:TIE_MUA_glmnet}
\end{figure}

Due to the lack of appropriate methods for the ABCD study analysis, we use the Mass Univariate Analysis for the mediator model, and ridge regression for the outcome model. The point estimation is given by Figure \ref{fig:TIE_MUA_glmnet}.

Comparing Figure \ref{fig:TIE_MUA_glmnet}  and \ref{fig:total_mean} in the main text, we can see that the active areas selected by BIMA is a subset of the areas selected by MUA+glmnet. This suggests that the selected regions might contain active mediation effect even under the most naive method. However, the result given by MUA+glmnet has much smaller NIE effects, almost all close to 0, whereas BIMA can identify fewer key areas that contain significant effects, which can be cross-validated by the posterior inclusion probability.

\subsection{ Real Data Scale Simulation I}\label{sec:app_RDA_sim}

\begin{figure}
\centering
\begin{tabular}{c}
Region 59 \textit{left superior parietal}, region 67 \textit{left precuneus}, region 68 \textit{right precuneus}.\\
\includegraphics[width=\textwidth]{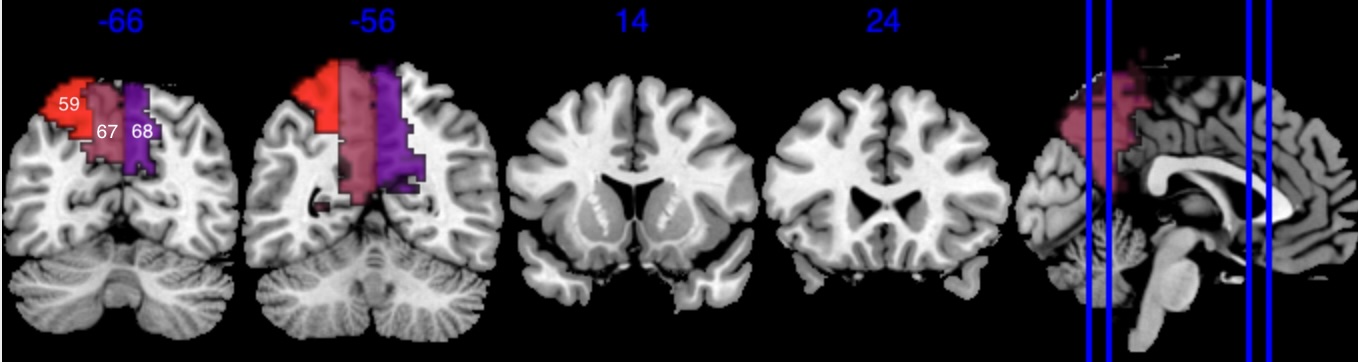}\\
True $\alpha(s)$ (color range $[0.01, 0.8]$)\\
\includegraphics[width=\textwidth]{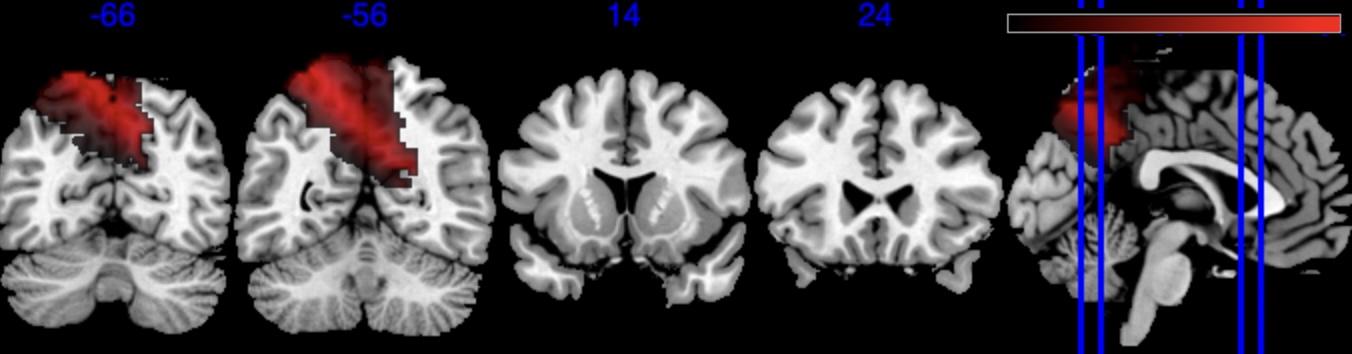}\\
True $\beta(s)$ (color range $[0.01, 0.8]$)\\
\includegraphics[width=\textwidth]{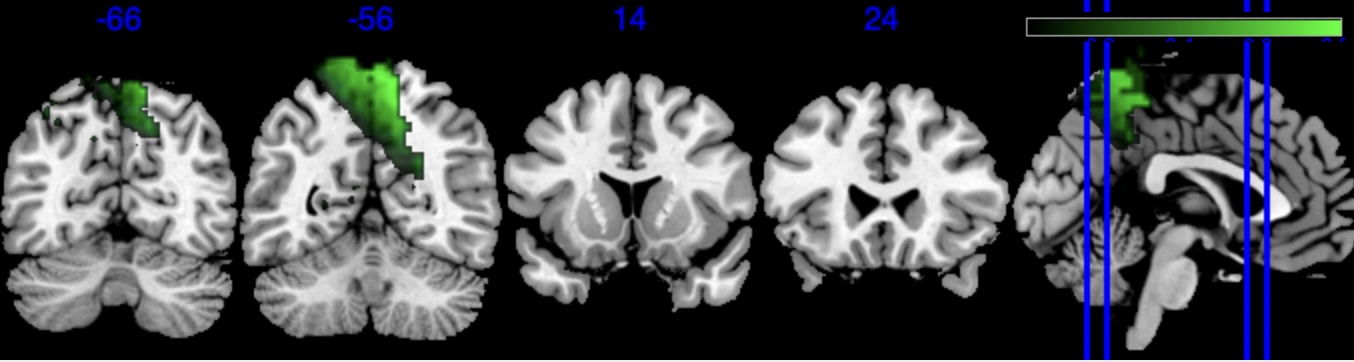}\\
True NIE $\cE(s)$ (color range $[0.01, 0.2]$)\\
\includegraphics[width=\textwidth]{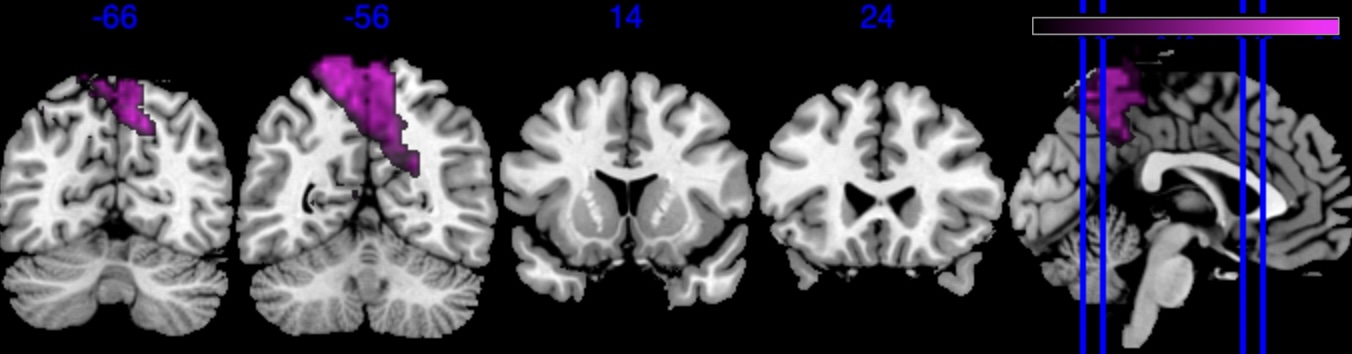}

\end{tabular}

\caption{True signals for the real data scale simulation I.}
\label{fig:true_signal_RDAsim}
\end{figure}

In this section, we provide a real data scale simulation based on the brain structure. The true $\alpha$ and $\beta$ are generated within 3 contiguous regions 59, 67, 68, as shown in Figure \ref{fig:true_signal_RDAsim}. The true signals are generated as spatially smooth functions across all three regions with no region-level independence structure. The true signal also contains small negative values (between -0.02 to 0) to match what we observe from the real data analysis that most large effect signals are positive signals. Due to the very small scale of the negative signals, we mainly focus on the positive signal with large effects. The dimension and sample size are the same with the real data. We only use the generated artificial signals to create synthetic fMRI data and outcome data. Due to the computational challenge, we only run this simulation once, and compare the result with the MUA+glmnet result.

We use this simulation to validate that, although we assume a prior region-wise independence structure to due computational convenience, which may induce discontinuous jumps across regions boundaries, but if the true signal is indeed a smooth function across regions, given sufficient sample size, the posterior will also be smooth across region boundaries with no obvious discontinuous jumps induced by the prior.

\begin{figure}
\centering
\begin{tabular}{c}
BIMA NIE $\cE(s)$ (color range $[0.01, 0.2]$)\\
\includegraphics[width=\textwidth]{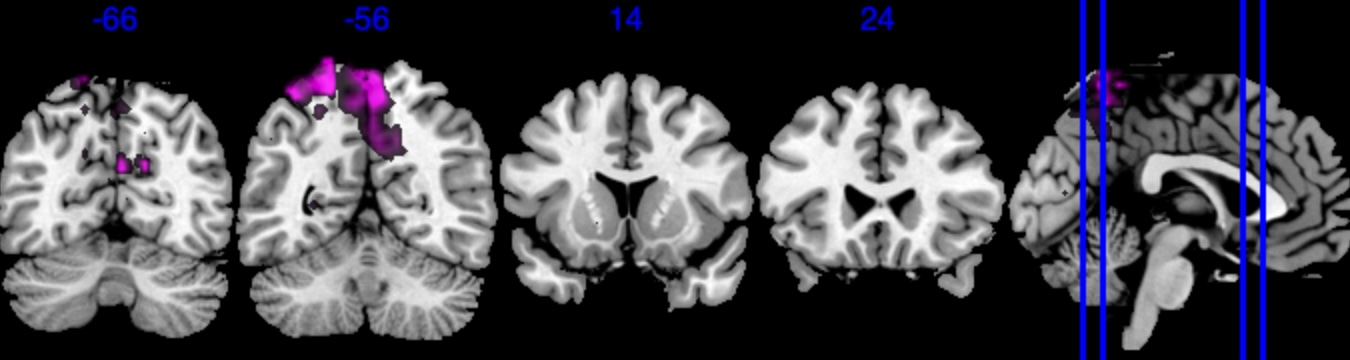}\\
MUA+glmnet NIE $\cE(s)$ (color range $[0.01, 0.2]$)\\
\includegraphics[width=\textwidth]
{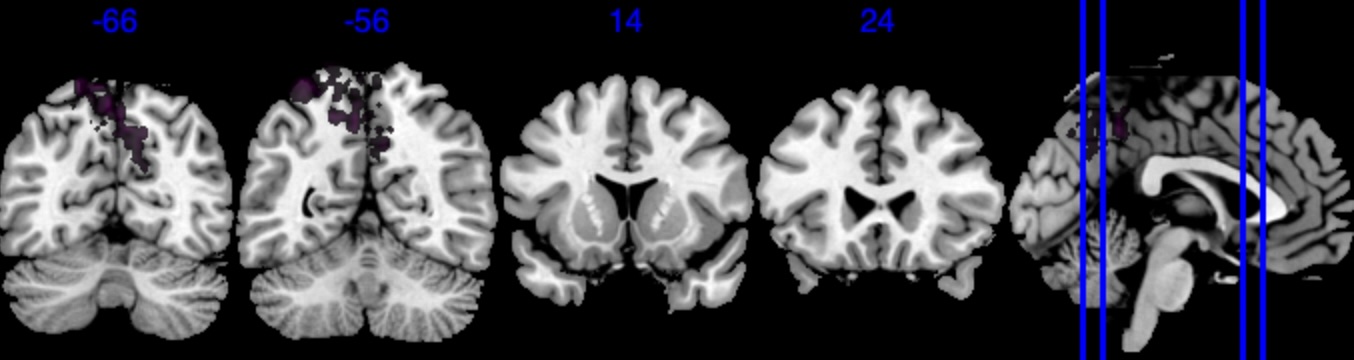}
\end{tabular}

\caption{Estimated signals for the real data scale simulation I.}
\label{fig:estimated_NIE}
\end{figure}

\begin{figure}
    \centering
    \includegraphics[width=\linewidth]{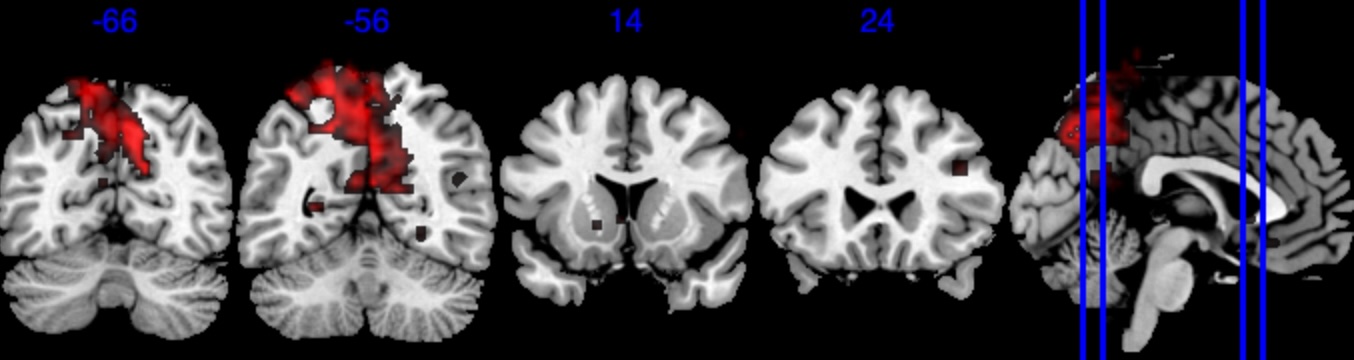}
    \caption{BIMA posterior mean of $\alpha(s)$}
    \label{fig:BIMA_alpha}
\end{figure}

The true signals are generated based on exponential squared kernel across the whole area spanning over region 59, 67, 68. Hence the true signals have spatial smoothness over the three regions, and have no discontinuity jumps across regions boundaries. In the BIMA estimation, we use the same kernel as the real data analysis, where a region-wise independent prior kernel structure is imposed. Based on the result shown in Figure \ref{fig:estimated_NIE}, we do not see obvious discontinuity jumps across region boundaries in the posterior mean of NIE $\cE(s)$. In theory, $\alpha(s)$ should be most influenced by the region-wise independence prior since the influence of other regions can only contribute to the $\alpha$ posterior through the noise variance. If we compare the estimation of $\alpha(s)$ (Figure \ref{fig:BIMA_alpha}) by BIMA with the true signal, although there are some small effects that are not picked up by BIMA, we do not see obvious jumps in the estimated values of $\alpha$ across region boundaries either. 

In addition, we provide the variable selection result in terms of the FDR, TPR, and Accuracy for the baseline (MUA+glmnet), BIMA selection by thresholding PIP, and BIMA selection by Morris FDR control method \citep{meyer2015bayesian} in Table \ref{tb:morris_RDAscale_sim}.

Because our true signal is generated using smooth functions that decay to 0, we set a threshold on the true $\cE_0$ such that only when $|\cE_0(s)|>0.01$, the corresponding voxel $s$ is active.

The first row in Table \ref{tb:morris_RDAscale_sim} is the selection made by the baseline method MUA+glmnet. Because the elastic net method does not come with valid p-values for variable selection, we choose to use a 0.01 cutoff on the effect size, i.e. if $|\cE(s_j)|>0.01$, $s_j$ is viewed as an active voxel.

The second row in Table \ref{tb:morris_RDAscale_sim} is the BIMA selection based on the criteria where $PIP(s_j)>0.99$ and the effect size $|\cE(s_j)|>0.01$. The effect size constraint is to stay consistent since we threshold the truth $|\cE_0(s)|>0.01$ for comparison.

The last row in Table \ref{tb:morris_RDAscale_sim} is the BIMA selection based on the Morris FDR control method \citep{meyer2015bayesian}, where the threshold on the size of $\cE(s)$ is set to be 0.01 (i.e. $\delta=0.01$ in the notation of \cite{meyer2015bayesian} where the definition of $P_{\text{BFDR}}$ is introduced), and the target FDR is set to be 0.01.

\begin{table}[ht!]
\centering
\begin{tabular}{llll}
\toprule
            & FDR   & TPR   & ACC   \\
MUA+glmnet  & 0.385 & 0.355 & 0.974 \\
BIMA-PIP    & 0.267 & 0.496 & 0.979 \\
BIMA-Morris & 0.341 & 0.597 & 0.978\\
\bottomrule
\end{tabular}
\caption{Comparison of signal selection accuracy in the real data scale simulation I.}
\label{tb:morris_RDAscale_sim}
\end{table}

Based on the results in Table \ref{tb:morris_RDAscale_sim}, none of these methods can achieve the target FDR. In fact, high-dimensional Bayesian FDR control is still an active research area. Both Morris FDR selection method, and cutoff on PIP are among some commonly used thresholding methods to select the active signals. However, based on our experience and evidence from this Table \ref{tb:morris_RDAscale_sim}, Morris selection tends to over-select active voxels and has a higher FDR than directly putting a cutoff on the PIP.

\subsection{ Real Data Scale Simulation II}\label{sec:app_RDA_sim2}

In Real Data Scale Simulation II, we provide a very low SNR case where we generate the simulated data based on the posterior mean of all parameters obtained from the real data analysis in \ref{sec:realdata}. This simulation provides a similar SNR to the real data analysis, where the generated true NIE is between -0.002 to 0.004, and the real data analysis gives a posterior mean of NIE between -0.0004 to 0.0017. This simulation study aims to present a range of cutoffs on PIP that can achieve a reasonable FDR and Power in terms of detecting true NIE signals. As shown in Table \ref{tb:top10}, the scale of NIE is very small. When computing the FDR and Power, we define a voxel $s$ being active if the true NIE satisfies $|\cE(s)|>5e-5$. 

\begin{table}[ht!]
\centering
\begin{tabular}{llll}
\toprule
Cutoff & FDR  & Power & ACC  \\
0.1   & 0.60 & 0.54  & 0.97 \\
0.2   & 0.43 & 0.47  & 0.98 \\
0.3   & 0.32 & 0.43  & 0.98 \\
0.4   & 0.25 & 0.37  & 0.98 \\
0.5   & 0.20 & 0.34  & 0.98 \\
0.6   & 0.15 & 0.30  & 0.98 \\
0.7   & 0.10 & 0.27  & 0.98 \\
0.8   & 0.04 & 0.25  & 0.98 \\
0.9   & 0.03 & 0.19  & 0.98 \\
\bottomrule
\end{tabular}
\caption{A range of cutoff on PIP with the corresponding FDR, Power, and Accuracy.}
\label{tb:sim_RDA_pip}
\end{table}

\begin{figure}
    \centering
    \includegraphics[width=0.7\linewidth]{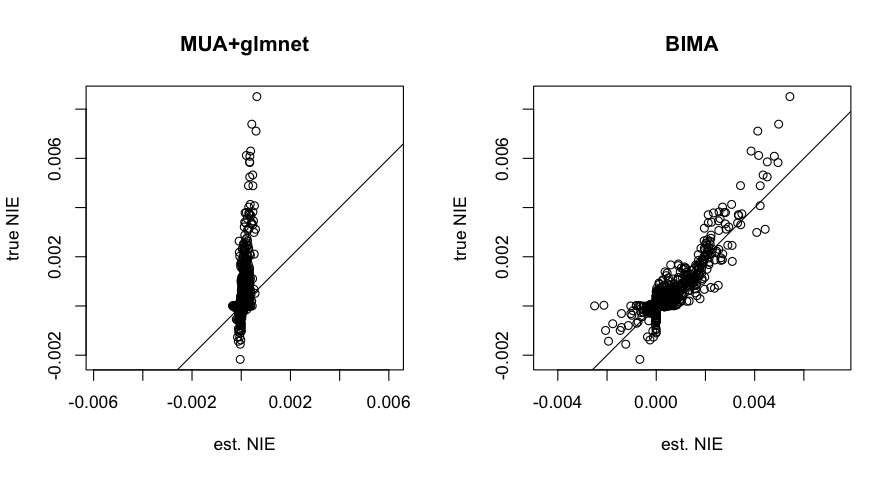}
    \caption{Scatter plot of true NIE and estimated NIE by BIMA and the baseline method MUA+glmnet.}
    \label{fig:sim_RDA4}
\end{figure}

We present the estimated NIE in Figure \ref{fig:sim_RDA4}, and the range of PIP cutoffs with corresponding FDR, Power and Accuracy in Table \ref{tb:sim_RDA_pip}. Based on these results, BIMA can identify very small signals relatively accurately even in this low SNR case, whereas the baseline method MUA+glmnet cannot provide a reasonable estimation. From the result in Table \ref{tb:sim_RDA_pip}, choosing the cutoff $\text{PIP}>0.5$ seems reasonable with FDR at 0.2 and power greater than 0.3.

\subsection{Real Data Scale Simulation III} \label{sec:app_RDA_sim3}

In order to further understand STGP prior's performance under irregular and nonsmooth signal patterns, we conduct Real Data Scale Simulation III, where the true signals $\alpha$ and $\beta$ are generated from a continuous but non-differentiable function with Weierstrass signals. For 3D voxel location $s\in \R^3$, define 
\[
f_w(s;a,b,k_{\text{max}}) = \sum_{i=1}^3 \sum_{k=0}^{k_{\max}} a^k \cos\sbr{2\pi b^k(s_i+0.5)} -3 \sum_{k=0}^{k_{\max}} a^k \cos\sbr{\pi b^k}
\]

\begin{figure}
    \centering
    \includegraphics[width=0.5\linewidth]{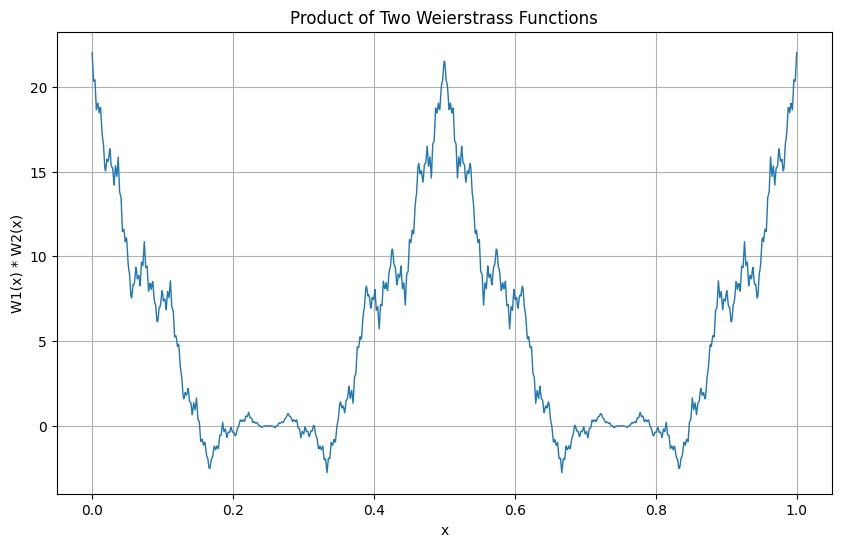}
    \caption{Product of 1-dimensional Weierstrass functions: $W_{\text{1d}}(s;a,b,k_{\text{max}}) = \sum_{k=0}^{k_{\max}} a^k \cos\sbr{2\pi b^k(s+0.5)}$. The figure shows the product of $W_{\text{1d}}(s;a=0.5,b=3,k_{\text{max}}=20)$ and $W_{\text{1d}}(s;a=1,b=1,k_{\text{max}}=10)$}.
    \label{fig:weierstrass}
\end{figure}

We generate a non-smooth $\alpha(s) = f_w(s;a=0.5,b=3,k_{\text{max}}=20)$ and a slightly more smooth $\beta(s) = f_w(s;a=1,b=1,k_{\text{max}}=10)$. The 1-dimensional product of these two Weierstrass functions is shown in Figure \ref{fig:weierstrass}.

This signal pattern is outside the RKHS of any reasonable kernel for neuroimaging signals. We only generate the signals on three regions in Figure \ref{fig:true_signal_RDAsim}, regions 59, 67, and 68. Because this simulation only studies the impact of non-smooth patterns, we restrict the analysis to these three regions as well. 

\begin{figure}
\centering
\begin{tabular}{c}
True NIE (color range $[0.01, 0.5]$ from black to red)\\
\includegraphics[width=\textwidth]{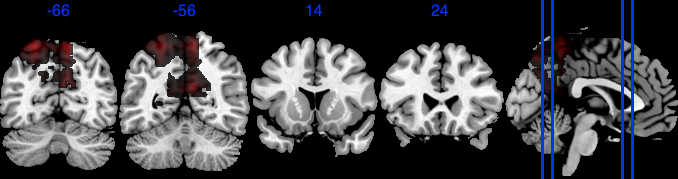}\\
BIMA est. NIE (color range $[0.01, 0.5]$ from black to red)\\
\includegraphics[width=\textwidth]{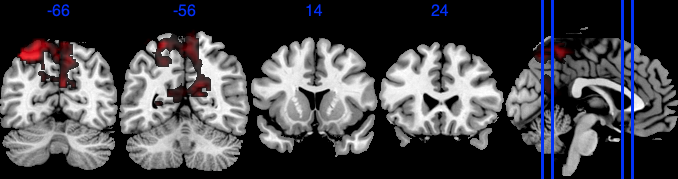}\\
MUA+glmnet est. NIE (color range $[0.01, 0.5]$ from black to red)\\
\includegraphics[width=\textwidth]{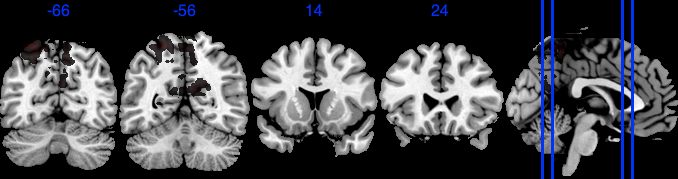}\\

\end{tabular}

\caption{True and estimated NIE for real data scale simulation III.}
\label{fig:RDAsim3}
\end{figure}

Table \ref{tb:morris_RDAscale_sim3} provides the selection results for the baseline MUA+glmnet, and BIMA results with two different selection criteria. Both Table \ref{tb:morris_RDAscale_sim3} and Figure \ref{fig:RDAsim3} show that even in this extreme case with very non-smooth patterns, BIMA can still detect most signals within higher power than the baseline method, and there is no obvious discontinuous jumps for BIMA in Figure \ref{fig:RDAsim3}. Although as shown in Figure \ref{fig:RDAsim3}, when the true signal is outside of the RKHS of smooth functions, the scale of BIMA estimates is off from the truth, and the signal boundary BIMA can identify is only restricted to signals with large effects, whereas smaller effects on the signal boundaries cannot be picked up.

\begin{table}[ht!]
\centering
\begin{tabular}{llll}
\toprule
            & FDR   & TPR   & ACC   \\
MUA+glmnet & 0.025 & 0.584 & 0.637 \\
BIMA: PIP$>0.1$       & 0.074 & 0.723 & 0.718 \\
BIMA: Morris target FDR =0.1    & 0.089 & 0.902 & 0.843 \\
\bottomrule
\end{tabular}
\caption{Comparison of signal selection accuracy in the real data scale simulation III.}
\label{tb:morris_RDAscale_sim3}
\end{table}

\end{document}